\documentclass[11pt,letterpaper]{article}
\usepackage{jheppub}
\usepackage{subcaption,multirow}
\usepackage{verbatim}
\pdfoutput=1 
\usepackage[T1]{fontenc} 
\usepackage[normalem]{ulem}
\usepackage{color} 
\usepackage{amsfonts,amssymb,amsmath} 
\usepackage{mathtools} 
\usepackage{slashed,braket}       
\usepackage{bm}  
\usepackage{hyperref} 
\usepackage{parskip} 
\usepackage{empheq}
\usepackage{mathrsfs}
\usepackage[utf8]{inputenc}
\usepackage{bbold}
\usepackage{booktabs}
\usepackage{perpage} 
\usepackage{slashed}
\usepackage{float}
\usepackage{color}
\usepackage{braket}
\usepackage[table,xcdraw]{xcolor}
\usepackage[makeroom]{cancel}
\usepackage{empheq}
\usepackage{gensymb}
\usepackage{overpic} 
\usepackage{graphicx}
\usepackage{feynmp-auto,tikz}
\usepackage{tikz-feynman} 
\usepackage{slashed} 
 \definecolor{myred}{rgb}{0.804688, 0.09375, 0.117188}
\usetikzlibrary{matrix}
\DeclareMathAlphabet\mathbfcal{OMS}{cmsy}{b}{n}
\def\lsim{\:\raisebox{-0.5ex}{$\stackrel{\textstyle<}{\sim}$}\:}
\def\gsim{\:\raisebox{-0.5ex}{$\stackrel{\textstyle>}{\sim}$}\:}

\newcommand{\plmu}{\partial_{\mu}}

\newcommand{\dlmu}{D_{\mu}}

\newcommand{\colvec}[1]{\left(\begin{array}{c}#1\end{array}\right)}
\newcommand{\colmatt}[1]{\left(\begin{array}{cc}#1\end{array}\right)}
\newcommand{\colmatth}[1]{\left(\begin{array}{ccc}#1\end{array}\right)}

\def\b{\beta}

\def\f{\phi}

\def\h{\eta}

\def\lpar#1#2#3#4{\rlap{\raise#3\hbox{$\hskip#4#1\left\{\mbox{\phantom{\rule[0mm]{0mm}{#2}}}\right.$}}}
\def\rpar#1#2#3#4{\rlap{\raise#3\hbox{$\hskip#4\left\}#1\mbox{\phantom{\rule[0mm]{0mm}{#2}}}\right.$}}}
\renewcommand{\subsubsection}[1]{\addtocounter{subsubsection}{1}
\par\nobreak
\medskip
\nobreak
\noindent{\it \thesubsubsection.  #1 }
\par\nobreak\medskip\nobreak}

\title{TASI Lectures on Non-Supersymmetric BSM Models}\label{ra_NonSUSY BSM}

\author[a]{Csaba Cs\'aki}
\author[a]{Salvator Lombardo}
\author[a]{Ofri Telem}

\emailAdd{csaki@cornell.edu}
\emailAdd{sdl88@cornell.edu}
\emailAdd{t10ofrit@gmail.com}

\affiliation[a]{Laboratory for Elementary Particle Physics, Cornell University, Ithaca, NY 14853, USA}

\abstract{These lectures provide a self-contained introduction to the essential aspects of non-supersymmetric beyond the Standard Model (BSM) physics for beginning graduate students who are already familiar with quantum field theory. After a detailed review of the physical meaning of the hierarchy problem, we introduce the key ingredients of the physics of Goldstone bosons necessary for many non-supersymmetric new physics models. Next we discuss the concept of collective symmetry breaking and present the main elements leading to little Higgs/composite Higgs models. We then turn to extra dimensional theories. After covering some of the basics of extra dimensional physics, we describe warped extra dimensions and explain how the AdS/CFT correspondence leads to realistic RS models and the holographic minimal composite Higgs model. 
}

\begin{document}
\tikzset{node style ge/.style={circle}}
\tikzset{node /.style={Latin Modern Math}}

\maketitle	

\section{Introduction: The Hierarchy Problem and Directions for Solving it\label{sec:intro}}

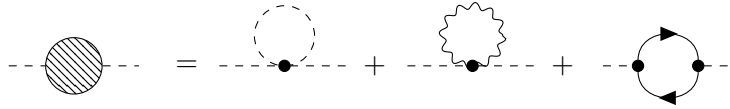
\begin{figure}
\begin{center}
\begin{tikzpicture}[baseline=(A.west)]
\begin{feynman}[]
\vertex (a){};
\vertex[right=1cm of a,blob] (b){};
\vertex[right=1cm of b] (c){};
\vertex[right=0.8cm of c] (d){};
\vertex[right=1cm of d,dot] (e){};
\vertex[right=1cm of e] (f){};
\vertex[right=0.5cm of f] (g){};
\vertex[right=1cm of g,dot] (h){};
\vertex[right=1cm of h] (i){};
\vertex[right=0.6cm of i] (j){};
\vertex[right=0.6cm of j,dot] (k){};
\vertex[right=0.8cm of k,dot] (l){};
\vertex[right=0.6cm of l] (m){};

\diagram* {
 (a) --[scalar] (b)--[scalar] (c),
 (d) --[scalar] (e)--[scalar] (f),
 (g) --[scalar] (h)--[scalar] (i),
(j) -- [scalar] (k),
(l) -- [scalar] (m),
(k) -- [half left, fermion] (l),
(l) -- [half left, fermion] (k),
};
\path (e)--++(90:0.4) coordinate (AA);
\draw (AA) [scalar]circle(0.4);
\path (h)--++(90:0.4) coordinate (BB);
\draw (BB) [photon]circle(0.4);
\node[align=center,black,opacity=1] (A) at (2.5,0) {$=$};
\node[align=center,black,opacity=1] (B) at (5.0,0) {$+$};
\node[align=center,black,opacity=1] (C) at (7.5,0) {$+$};
\end{feynman}
\end{tikzpicture}
\end{center}
\caption{The one loop corrections to the Higgs mass parameter in the SM. All three diagrams are quadratically divergent, leading to the hierarchy problem. \label{fig:Higgsloops}}
\end{figure}

The Standard Model (SM) of particle physics is an extremely successful theory: it is capable of reproducing the results of all experiments we have produced to date. Nevertheless most particle theorists believe that the SM is not the final theory, and that there should be physics beyond the SM (BSM) and that physics should not lie too far from the currently probed energy levels. The main reason for this lies in the hierarchy problem: the Higgs field responsible for electroweak symmetry breaking (EWSB) in the SM is quadratically sensitive to high scales. Formally, this appears first as a quadratic divergence in the one-loop contributions to the Higgs mass. For a Higgs potential of the form 
\begin{equation}
V(H) = -\mu^2 |H|^2 +\lambda |H|^4 
\end{equation}
the loops of Fig.~\ref{fig:Higgsloops} will contribute $-\mu^2 \to -\mu^2 +\delta \mu^2$ where
\begin{equation}
\delta \mu^2 = \frac{\Lambda^2}{32\pi^2} \left[ -6y_t^2 +\frac{1}{4} (9 g^2 +3 g'^2) + 6 \lambda \right]
\label{eq:quaddiv}
\end{equation}
where $\Lambda$ is the cutoff of the theory (for simplicity assumed to be universal for the various loops for now), $y_t$ is the top Yukawa coupling, $g,g'$ are the SU(2) and U(1) gauge couplings and $\lambda$ is the Higgs self-coupling. The minimum of the Higgs potential is at 
\[ \langle H \rangle = \left( \begin{array}{c}  0 \\ \frac{v}{\sqrt{2}} \end{array} \right) , \ \ v^2 = \frac{\mu^2}{\lambda} \]
and from the measured values of the $W,\,Z$ masses we know $v = 246$ GeV. Similarly the physical Higgs mass is 
\[ m_h = \sqrt{2\lambda} v = 125 \ {\rm  GeV} \] 
which implies $\lambda =0.13  \sim 1/8$. These measured values of $v, \lambda$ are the results of the full quantum corrected potential, which is supposed to include the quadratically sensitive shift to the mass parameter $\mu$. If $\Lambda \gg $ TeV we would find $\delta \mu^2 \gg \mu^2$, giving rise to the so-called hierarchy problem. The bare potential must be tuned to cancel off the quantum corrections in order to get the correct physical mass parameter, and the problem is worse the higher the cutoff. In particular a cutoff $\Lambda \sim 10$ TeV gives rise to the so-called little hierarchy problem, while a cutoff all the way at the Planck scale $\Lambda \sim M_{Pl} \sim 10^{19}$ GeV would give rise to the big hierarchy problem.

As we have seen, the hierarchy problem is the quadratic sensitivity of the Higgs mass (and the Higgs VEV) to {\it new physics}. The cutoff $\Lambda$ is a physical mass threshold where there must be new degrees of freedom to explain why the low-energy effective field theory seizes to be the correct description at this scale. As a corollary, any new mass scale (\textit{e.g.} new particles at a high scale) will feed into the Higgs potential at some point. There are several important points that we should clarify here regarding the hierarchy problem, which often causes misunderstandings.

\begin{itemize}
\item In the above discussion we have been somewhat cavalier with the cut-off scale $\Lambda^2$. One might worry (and indeed many people do!) that the hierarchy problem is merely an artifact of using a crude cut-off regulator. However, those understanding effective theories well realize quickly that the hierarchy problem is not at all about various regularization schemes. As in any good effective theory, $\Lambda$ in our calculations is merely standing in for the physical mass threshold at which new heavy particles appear. You can think of $\Lambda$ as literally the mass of a new heavy particle ($m_{NP}$), and the ``quadratically divergent" contributions to the Higgs mass parameter simply as log-divergent or finite contribution from the heavy particle which are proportional to $m_{NP}^2$. Moreover, these contributions contain an imaginary part from the new particle going on-shell, which is physical and cannot be removed by regulation scheme. Thus using dimensional regularization (a scheme where power law divergences are simply regulated to zero) is really not a solution of the hierarchy problem.

\item The hierarchy problem is really the sensitivity to new scales. If there is no new scale there really is no hierarchy problem. However most physicists believe that there are at least two issues that will force us to extend the SM: the appearance of quantum gravity around the Planck scale and the appearance of a Landau pole in the hypercharge gauge coupling at exponentially large scales. 

\item  For a while it was popular to play with the idea that the terms in Eq.~(\ref{eq:quaddiv}) actually cancel each other. This used to be known as the ``Veltman condition", which would have singled out a very particular value for the Higgs mass. However we can easily see that even if the mass had turned out to be the magical value (which it did not) this would not have solved the hierarchy problem. As we discussed in Eq.~(\ref{eq:quaddiv}) $\Lambda$ is merely a stand-in for the mass of a heavy particle that will ultimately regulate these loops. However this can numerically be different for the three diagrams, thus one should really be talking about the gauge cut-off scale $\Lambda_g$, the fermion cut-off scale $\Lambda_f$ and the Higgs cut-off scale $\Lambda_H$, which could all be different by ${\cal O}(1)$ factors or even more. Thus it is not really meaningful to talk about a Veltman-like condition, unless some symmetry ensures that all these cut-off scales are equal. 

\item A simple way to phrase the hierarchy problem is the fact that the Higgs mass term 
$\mu^2 |H|^2$ is a relevant operator, which grows towards the IR. The Wilsonian formulation of the hierarchy problem then is that it is difficult to choose a RG trajectory which in the IR flows to the correct Higgs mass: most trajectories will miss a light physical Higgs mass, and an immense tuning is needed to hit the right Higgs mass parameter in the IR. Note, that the Higgs mass parameter is the only relevant operator in the SM. 

\item Finally, we should remark that the hierarchy problem is specific to elementary scalars. The reason is that fermions and gauge bosons have a new symmetry appearing in the Lagrangian when the mass goes to zero. For example for fermion masses in 4D one has a new chiral symmetry appearing in the $m\to 0$ limit, which will protect the fermion masses from large unsuppressed corrections, and ensure that the correction is proportional to the mass itself: $\Delta m_e \propto m_e \log \frac{\Lambda}{m_e}$. Similarly, for gauge bosons there is an unbroken gauge symmetry appearing in the $M_W\to 0$ limit, which will ensure $\Delta M_W^2 \propto M_W^2 \log \frac{\Lambda}{M_W}$. 

\end{itemize}

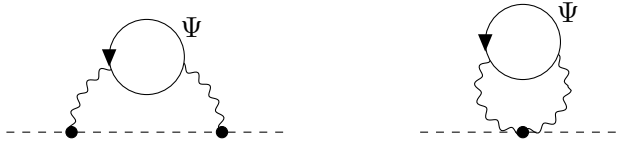
\begin{figure}
\vspace*{-0.5cm}
\begin{center}
\begin{tikzpicture}[baseline=(A.west)]
\begin{feynman}[]
\vertex (a){};
\vertex[right=1cm of a,dot] (b){};
\vertex[right=1cm of b] (cc){};
\vertex[above=1cm of cc] (dd){};
\vertex[right=2cm of b,dot] (c){};
\vertex[right=1cm of c] (d){};
\vertex[right=1.5cm of d] (e){};
\vertex[right=1.5cm of e,dot] (f){};
\vertex[right=1.5cm of f] (g){};
\vertex[above=1.2cm of f] (ff){};

\diagram* {
 (a) --[scalar] (b)--[scalar] (c)--[scalar] (d),
 (e) --[scalar] (f)--[scalar] (g),
};
\path (f)--++(90:0.6) coordinate (BB);
\draw (BB) [photon]circle(0.6);
\draw (c) [photon] arc (0:180:1);
\fill (dd) [white] circle (0.5);
\draw (dd) [fermion] circle (0.5);
\fill (ff) [white] circle (0.5);
\draw (ff) [fermion] circle (0.5);
\node[align=center,black,opacity=1] (CC) at (2.6,1.4) {$\Psi$};
\node[align=center,black,opacity=1] (CC) at (7.6,1.6) {$\Psi$};
\end{feynman}
\end{tikzpicture}
\end{center}
\vspace*{-0.25cm}
\caption{Corrections to the Higgs mass for the case when new heavy fermions charged under the SM are added. \label{fig:twoloopfermion}}
\end{figure}

The simplest demonstration of the the hierarchy problem would be to introduce yet another scalar $S$ (never mind for now that that scalar would have its own hierarchy problem). Introducing this scalar along with a quartic coupling with the Higgs 
\[ \lambda_S |H|^2 |S|^2 \]
will result in a loop correction for the $S$ particle giving rise to 
\begin{equation}
\delta \mu^2 = \frac{\lambda_S}{16\pi^2} \left[ \Lambda_{UV}^2 - m_S^2 \, \log \frac{\Lambda^2_{UV}}{m^2_S} + {\cal O} (m_S^2 )  \right] \ .
\end{equation}
We can see that even if we drop the $\Lambda_{UV}^2$ term there will be an explicit quadratic dependence on $m_S^2$ the mass of the new heavy particle, from log divergent or finite contributions, which is exactly the hierarchy problem. This dependence will be there irrespective of how one regulates this loop. One may wonder if the hierarchy problem can be avoided by not coupling the new physics directly to the Higgs scalar. One obvious example would be to use some heavy fermions that are charged under the SM but don't directly have a Yukawa coupling with the Higgs. While one loop corrections are in this case indeed avoided, the quadratic sensitivity to the Higgs mass will show up at two loops (see Fig.~\ref{fig:twoloopfermion}):
\[ \delta \mu^2  \propto \frac{g_{SM}^4}{(16\pi^2)^2} m_\Psi^2 \ . \]

By now we should be convinced that the hierarchy problem is a serious issue which should be resolved one way or another in a theory more complete than the SM. The leading approach toward solving it has been to assume that new physics actually shows up early: around the TeV scale rather than at the scales where it ultimately must show its face (like the Planck scale or the Landau scale) since the required fine-tuning is more severe the higher the energy scale the issue is addressed. We will see that the new TeV-scale physics can have a form that will make the Higgs insensitive to any further higher energy scales of new physics. The two most common choices for the new physics at the TeV scale that actually makes the Higgs insensitive to high scales are:
\begin{itemize}
\item Supersymmetry. In this case we introduce a fermion$\leftrightarrow$boson symmetry (``supersymmetry") which relates the SM Higgs to its fermionic partner. This symmetry will ensure that the chiral symmetry of the fermionic partner also protects the Higgs itself from quadratic sensitivity to high scales. Supersymmetry is covered in a separate lecture series by Howie Haber~\cite{HaberTASI2016}.
 
\item Composite Higgs \cite{Kaplan:1983fs} : there is no true elementary scalar, rather the Higgs is a bound state of some more fundamental, strongly-interacting fermions. This idea eliminates the largest part of the quadratic sensitivity as a form factor shuts off corrections to the Higgs mass above the compositeness scale $\Lambda$, thus effectively lowering the cutoff to $\Lambda \sim$ TeV.  It will be useful for the Higgs to be identified as a Goldstone boson to ensure that the Higgs is naturally lighter than the strong dynamics.
\begin{itemize} 
\item Goldstone's theorem~\cite{Goldstone}. If a global symmetry is spontaneously broken, massless scalars, ``Goldstone bosons" will appear, whose masses will be protected and remain vanishing by Goldstone's theorem. This will be the crucial idea used throughout these lectures. While Goldstone's theorem is a universal ingredient in many of these models, the actual implementation can be slightly different (though as we will see all of these models are actually related to each other).
\end{itemize}

\item Warped extra dimensions~\cite{RS}: in this case the variation of the fundamental energy scale along the extra dimension will lead to a solution to the hierarchy problem. As we will see using the AdS/CFT correspondence this picture is actually dual to that of a composite Higgs. Just as it was useful to have a Goldstone composite Higgs, it is important to have the extra dimensional analogue:
\begin{itemize}
\item Gauge - Higgs unification~\cite{GHUnif}: here the scalar is an extra dimensional component $A_5$ of the the gauge field. We will see that by the AdS/CFT correspondence this is the idea that the Higgs is identified with a Goldstone boson of a spontaneously broken global symmetry. The ultimately most successful and calculable models actually combine all of these ingredient into what is now known the holographic minimal composite Higgs model.
\end{itemize}
\end{itemize}

Besides the traditional supersymmetry or composite Higgs approach there are also more radical ideas for solving the hierarchy problem which we list here.
\begin{itemize}
\item Technicolor/Higgsless models~\cite{Technicolor,Higgsless}. In this case there is actually no Higgs particle. A condensate of the strong dynamics directly breaks the electroweak symmetry. While conceptually one of the most beautiful ideas, it is now clearly disfavored by the discovery of the SM-like Higgs boson. These models also had difficulty obtaining small corrections to the electroweak precision observables. Higgsless models were extra dimensional versions of technicolor using AdS/CFT, and are more under control as they are calculable. 

\item Large extra dimensions~\cite{ADD}. In these models the weak scale is actually the true fundamental scale (analogous to $M_{\text{Pl}}$) where gravity becomes strongly interacting, and thus there is no weak-Planck scale hierarchy problem to begin with. But it does predict interesting gravity-related phenomena at the TeV scale like production of mini black holes. The main drawback of such models is that the radius of the extra dimensions has to be exponentially larger than the fundamental length scale, which is hard to explain in a model with just one fundamental scale (that is the issue of radius stabilization is now equivalent to the original hierarchy problem).

\item Anthropic explanations in the Multiverse~\cite{Weinberg,Amherst}. A popular way out of the hierarchy problem is to speculate that we live in a multiverse of many universes, where the fundamental constants vary from one universe to the other. In most universes the Higgs mass would indeed be very large, but that would also result in a universe without chemistry and hence no life. It is no wonder then that we end up living in a universe where the Higgs mass is small and allows us to wonder about possible solutions to the hierarchy problem. While this approach may indeed be the correct one, we will likely never know. By definition the multiple universes can not be experimentally accessed. 

\item Relaxion~\cite{relaxion} type mechanisms. A very interesting recent idea is that while the Higgs mass parameter is currently very small, it has not always been like that in our Universe. A field called the relaxion has been continuously scanning the possible Higgs mass as the Universe expanded. When the Higgs mass square parameter switched sign, electroweak symmetry breaking happened, which triggered the end of the rolling of the relaxion and the scanning of the Higgs mass, leaving us stuck in a seemingly fine-tuned vacuum.

\item There are several other more exotic ideas for solving the hierarchy problem. For an excellent overview see~\cite{Craig}.

\end{itemize}

In these lectures we will be focusing on the composite Higgs (CH) solution: it is one of the simplest and most plausible ones, with very concrete predictions for the LHC or higher energy colliders. There are many other excellent reviews on CH models~\cite{Contino,PanicoWulzer, CT, CGT,BCS}.

\section{Goldstone Bosons \label{sec:GBs}}
 
\begin{figure}
\begin{center}
\includegraphics[scale=0.3]{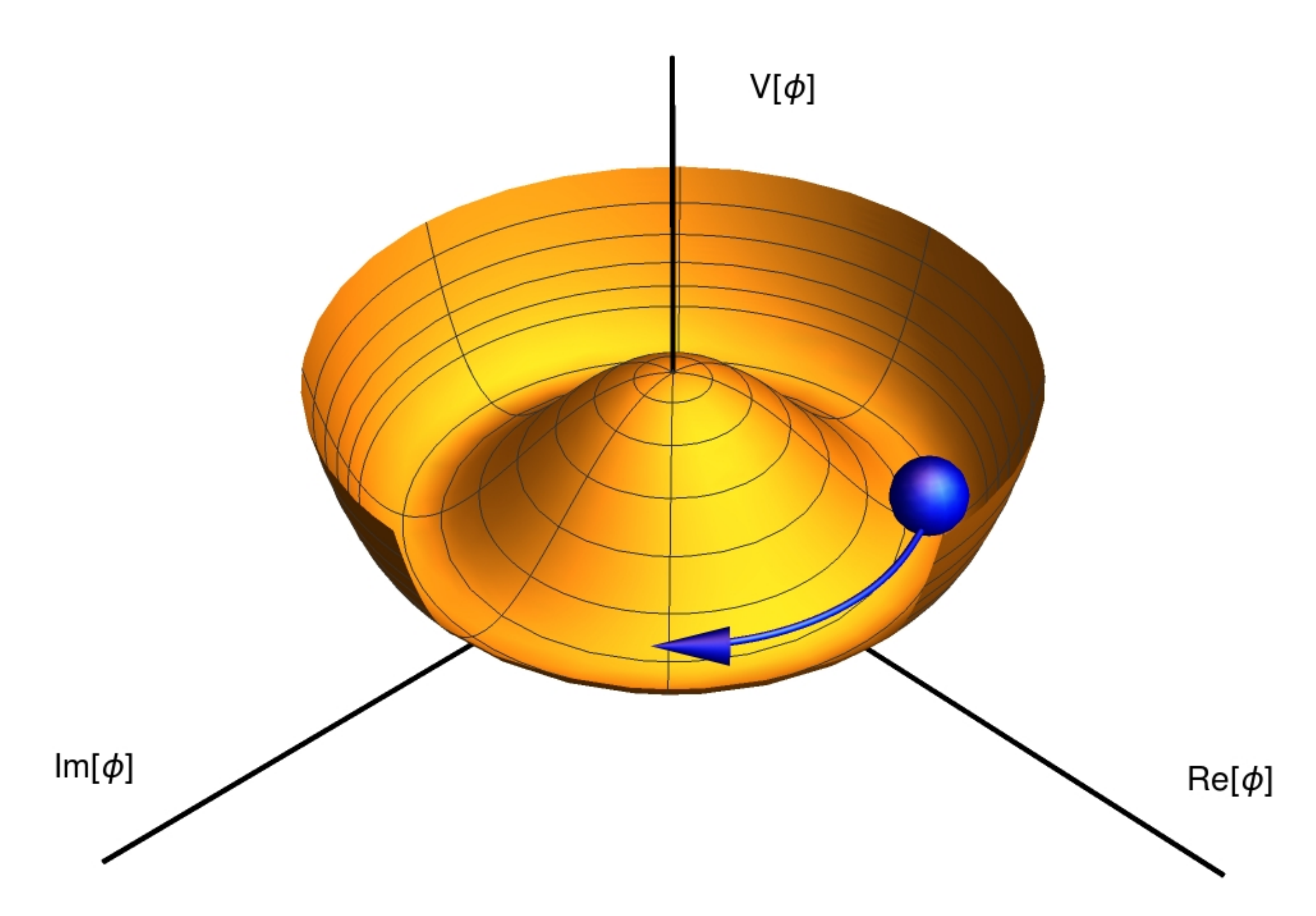}
\end{center}

\caption{Illustration of the Goldstone mode rolling at the valley at the bottom of the potential.\label{fig:GBs}}
\end{figure}

Throughout these lectures we will often be identifying the Higgs boson with Goldstone bosons of a spontaneously broken global symmetry. Thus it is important to first understand the properties of Goldstone bosons in detail. Goldstone's theorem tells us that whenever there is a spontaneously broken global symmetry there should be a corresponding massless scalar field, the Goldstone boson (GB), sometimes called the Nambu-Goldstone boson (NGB). The physical intuition is pretty simple: due to the global symmetry the minimum of the potential is either unique (in which case there is no spontaneous symmetry breaking) or degenerate (in which case there is spontaneous symmetry breaking). In this way, a spontaneously broken global symmetry ensures the presence of a degenerate valley at the bottom of the potential (see Fig.~\ref{fig:GBs}). The Goldstone bosons are the fields parametrizing the motion along this valley. More formally, we would say the the Goldstones span the coset $G/H$. This term is borrowed from group theory, where the coset $G/H$ marks the group G with the elements in H identified with the identity. By Goldstone's theorem, these fields are exactly massless. Indeed, because of the vacuum degeneracy, there should be zero energy cost move along the valley of the potential implying a vanishing mass term along this direction. Note that the unbroken generators of the original global symmetry annihilate the vacuum, while the broken generators are the ones generating the movement along the valley of inequivalent vacua. This simple observation forms the basis of writing effective GB Lagrangians, also known as chiral perturbation theory (since it was first developed for the theory of pions, which arise from the breaking of the $SU(2)_L\times SU(2)_R$ chiral symmetries of the strong interactions), or in its most powerful form as the Callan-Coleman-Wess-Zumino (CCWZ) formalism~\cite{CCWZ}. Here we will only discuss chiral perturbation theory which involves less formalism and is slightly more intuitive, but once the reader is familiar with that developing the full CCWZ formulation will be straightforward.

\subsection{Non-linear Goldstone fields}

The first important fact about the effective theory for the GB's is that it does not matter what the actual origin of symmetry breaking and the GB's actually is: there could be an elementary scalar developing a VEV, but there could equally well be some strong dynamics (like QCD) giving rise to a vacuum condensate. For the effective GB Lagrangian this doesn't matter. The heavy degrees of freedom (usually some radial modes, like the Higgs itself in the SM) are integrated out, and chiral perturbation theory fixes the effective Lagrangian for the GB's only. Where exactly the heavier degrees of freedom reside will depend on whether the theory is coupled strongly or weakly. As we will see for a strongly interacting theory the heavy degrees of freedom are all expected to lie at the cutoff scale of the theory. If the heavy degrees of freedom are weakly coupled, they can appear far below the cutoff. This will play an important role in composite Higgs models and we will discuss this issue in detail. 

Let us now explicitly identify the lowest term in the chiral Lagrangian. We assume that the global symmetry $G$ is spontaneously broken to $H$ by some VEV $\Sigma_0$ (the exact origin of $\Sigma_0$ does not matter). 
The NGBs are conveniently parameterized in the NGB matrix:
\begin{equation}
U_{\text{NGB}} = e^{i \Pi^a(x) T^a /f }\, ,
\end{equation}
where the index $a$ labels broken generators, $f$ is the pion decay constant determined by the magnitude of the VEV $\Sigma_0$, and the ratio $\Pi^a/f$ can be interpreted as the angle of transformation in the broken direction. 
This matrix acts on the VEV $\Sigma_0$ to rotate it along the broken directions:
\begin{equation}
\Sigma \,=\, U_{\text{NGB}} \, \left[\Sigma_0\right] \, .
\end{equation}
The exact way that $U_{\text{NGB}}$ acts on the VEV depends on it's representation under the global symmetry $G$. For example: 
\begin{eqnarray}
\Sigma \,=\, U_{\text{NGB}} \, \left[\Sigma_0\right] \,&=&\,U_{\text{NGB}} \, \Sigma_0~~~~~~~~~~~~~~~\text{for } \Sigma_0\, \text{in the fundamental}\,,\nonumber\\
\Sigma \,=\, U_{\text{NGB}} \, \left[\Sigma_0\right] \,&=&\,U^{\dagger}_{\text{NGB}} \, \Sigma_0 \,U_{\text{NGB}}~~~~~~~\text{for } \Sigma_0\, \text{in the adjoint}\,.
\end{eqnarray}

The simplest example that can illustrate this is QCD and the chiral Lagrangian. The QCD Lagrangian in terms of quarks is of course well-known:
\begin{equation}
{\cal L}_{QCD} =-\frac{1}{4} G^a_{\mu\nu}  G^{a\, \mu\nu} +\sum_i \bar{q}_i (i \slashed{D} -m_q) q_i \ .
\end{equation} 
The most important aspect for us from this Lagrangian is that in the $m_q\to 0$ limit the theory has a chiral global symmetry $G\,=\,SU(3)_L\times SU(3)_R$ (the classical global symmetry is even larger $U(3)_L \times U(3)_R$, one of the two additional $U(1)$'s is baryon number, while the axial $U(1)$ is anomalous). While the QCD Lagrangian is usually written in terms of 4-component Dirac fermions, for the purpose of understanding the symmetries it is better to think of it in terms of 2 component Weyl spinors\footnote{For a review of 2 component spinors, see \cite{Dreiner:2008tw}.}. A Dirac spinor can be written as
\[ \Psi = \left( \begin{array}{c} \chi \\ \bar{\psi} \end{array} \right) \]
where $\chi$ is a left handed (LH) and $\bar{\psi}$ a right-handed (RH) 2-component Weyl spinor. Since the conjugate of a RH spinor is a LH one we can immediately see that $\chi , \psi$ are both two-component LH spinors, with $\chi$ transforming as $3$ and $\psi$ as $\bar{3}$ of $SU(3)_{QCD}$. In this language the fermionic Lagrangian can be written as 
\begin{equation}
\bar{\Psi} (i \slashed{D} -m )\Psi = i \psi_{\dot{\alpha}}^\dagger \bar{\sigma}^{\mu \dot{\alpha} \alpha} D_\mu \psi_\alpha +i \chi_{\dot{\alpha}}^\dagger \bar{\sigma}^{\mu \dot{\alpha} \alpha} D_\mu \chi_\alpha - m (\psi^\alpha \chi_\alpha +h.c. )
\end{equation}
In this last form the appearance of the chiral symmetries in the $m\to 0$ limit is pretty straightforward to see: since it is only the mass term connection $\chi$ and $\psi$ in the massless limit we have the independent rotations 
\[ \psi \to \psi \, U^{\dagger}_R \ ,  \chi \to U_L \, \chi \]
leaving the Lagrangian invariant where $U_{L,R}$ are independent 3 by 3 unitary matrices. Since the resulting $SU(3)_L \times SU(3)_R$ global symmetry is physical it should be realized on the spectrum of QCD (ie. the composites should form multiplets of the full  $SU(3)_L \times SU(3)_R$ symmetry). However, only one $SU(3)$ is actually realized on the spectrum, which is Gell-Mann's $SU(3)_V$ leading to the eightfold way \cite{GellMann:1961ky}. Hence we conclude that the dynamics of QCD must be breaking the $SU(3)_L\times SU(3)_R \to SU(3)_V$ by forming a quark condensate as a result of the strong dynamics:
\[ \langle \bar{q} q \rangle = \langle \bar{q}_{L i} q_{R j} + h.c. \rangle \propto \delta_{ij} \Lambda_{QCD}^3\ . \]
This structure of the condensate will ensure that $SU(3)_V$ remains unbroken, while $SU(3)_A$ is broken, resulting in 8 GB's, forming the pseudo-scalar octet $\pi^{\pm\ ,0}, K^{\pm}, K^0, \bar{K}^0$ and $\eta$. However they are not true Goldstone bosons: the $SU(3)_A$ axial symmetry is exact only in the $m_q\to 0$ limit. For finite quark masses there will be small explicit breaking terms which will render the octet to be pseudo-Nambu-Goldstone bosons (pNGBs) rather than true Goldstone bosons, lifting the masses of the pions and the other members of the octet. 

How do the electroweak gauge interactions fit into this picture? The $SU(2)_L\times U(1)_Y$ electroweak gauge symmetry can be embedded into the chiral global symmetries of the strong interactions: $SU(2)_L\times U(1)_Y \subset SU(3)_L\times SU(3)_R \times U(1)_B$. Clearly $SU(2)_L$ can just be identified with the upper left two by two corner of $SU(3)_L$, which will transform the $(u_L,d_L)$ quarks into each other. Incorporating the strange quark is slightly more complicated since as we know $(c_L,s_L)$ also form an $SU(2)_L$ doublet, but the charm mass is large $m_c> \Lambda_{QCD}$ and hence it is integrated out from the low-energy effective theory. The proper description would be to start with four quarks and an $SU(4)_L$ chiral symmetry, embed the $SU(2)_L$ weak interactions twice into this $SU(4)$, then turn on a large charm quark mass and integrate it out before confinement and chiral symmetry breaking is implemented. We will not follow through this procedure explicitly, but instead focus on the $SU(2)_L$ subgroup of $SU(3)_L$. Similarly, hypercharge can be embedded into the chiral symmetries as 
\[ Y = T_{R3} +\frac{B}{2} \ .\]
As the electroweak symmetries are weakly gauged, they also explicitly break the global symmetries of QCD (and hence split the charged and neutral pion masses). 

\subsection{The low-energy effective Lagrangian for QCD}

Let us now proceed and start explicitly constructing the effective Lagrangian for QCD. Since the condensate breaking the global symmetries $\langle \bar{q} q \rangle$ is a bidoublet under $SU(3)_L \times SU(3)_R$ we will take for $\Sigma_0$ a bifundamental VEV
\[ \Sigma_0 = f \left ( \begin{array}{ccc} 1 \\ & 1 \\ & & 1 \end{array} \right) \]
and act on this VEV with the broken global symmetries. A bifundamental transforms generically as 
$\Sigma_0 \to U_L \Sigma_0 U_R^\dagger$. Of course for us $\Sigma_0$ is a constant, and it will be left invariant for $U_L=U_R$, that is the vector-like $SU(3)_V$ transformations of Gell-Mann. On the other hand the axial elements can be identified by $U_R=U_L^\dagger$, leading to the non-linearly realized pion field 
\begin{equation}
\Sigma(x) = e^{i \Pi^a T^a/f} \Sigma_0 e^{i \Pi^a T^a/f} = e^{2 i \Pi^a T^a/f} ,
\end{equation}
where $\Pi^a (x)$ is now the Goldstone boson field identified with the members of the pseudo-scalar octet of QCD. How do the pions transform under the unbroken and the broken global symmetries? Let us first look at the case of the unbroken symmetries:
\[ \Sigma(x) \to U_V\, \Sigma(x)\, U_V^\dagger \ .\]
Linearizing $\Sigma(x)$ in the pion fields we find:
\begin{equation}
\Pi^a T^a \to U_V \, \Pi^a T^a \, U_V^\dagger
\end{equation}
yielding the usual linearly realized transformation of an adjoint under $SU(3)_V$. However, under the broken symmetries the transformation will turn out to be non-linear! For this case
\begin{equation}
\Sigma(x)\to U_A \,\Sigma(x)\, U_A = e^{2i \Pi_a' T^a/f} 
\end{equation}
The broken transformations can be themselves written as $U_A = e^{i c^a T^a}$ where the $c^a$ are the global transformation parameters. Expanding both sides in powers of the pion field (as well as powers of the gauge transformation parameter) we find
\begin{equation}
{\Pi^a}' T^a = \Pi^a T^a + f c^a T^a + {\cal O} (\Pi^a)^2
\end{equation}
This is a very important equation which tells us that:
\begin{itemize}
\item To leading order the pions have a shift symmetry $\Pi^a \to \Pi^a + f c^a$. This provides another simple proof of Goldstone's theorem, since the shift symmetry forbids any non-derivative terms, in particular mass terms or any potential. 

\item The pions transform non-linearly under the axial rotation.
\end{itemize}

We are now ready to construct the leading order Lagrangian for the interacting pion fields. We will simply write down all the terms in $\Sigma(x)$ that are symmetric under the entire $SU(3)_L\times SU(3)_R$ global symmetry, including both unbroken and broken ones. Since we are looking for a low-energy effective Lagrangian, we will organize them by the number of derivatives. The simplest term would contain no derivatives, and there is a unique invariant one can form: ${\rm Tr}\, \Sigma^\dagger \Sigma$, however this term is obviously just the trace of the unit matrix and independent of the Goldstone fields. The first non-trivial term contains 2 derivatives and is
\begin{equation}
\frac{f^2}{4} {\rm Tr} [ (\partial_\mu \Sigma )^\dagger \partial^\mu \Sigma ]
\end{equation}
where the overall coefficient has been fixed such that one obtains a canonical kinetic term for the pions. Every term will contain two derivatives and an arbitrary number of pions once the exponential is expanded. This will give the leading pion interaction terms in the $p/f \to 0$ limit. Besides the pion kinetic terms it will contain 4-pion interactions terms with two derivatives contribution to $\pi - \pi$ scattering and higher order terms with more pions:
\begin{equation}
{\cal L} = {\rm Tr} [ \partial_\mu \Pi \partial^\mu \Pi ]+ \frac{4}{f^2} {\rm Tr} [ \partial_\mu  \Pi \partial^\mu \Pi \Pi^2 ] + {\cal O} (\Pi^6) 
\end{equation}
where $\Pi = \Pi^a T^a$ and we assumed the normalization of the generators ${\rm Tr}~T^a T^b = \frac{1}{2} \delta^{ab}$. 

\subsection{Gauging EW symmetry and dynamical gauge boson masses}

\begin{figure}
\begin{center}
\begin{tikzpicture}[baseline=(A.west)]
\begin{feynman}[]
\vertex (a){};
\vertex[right=1cm of a,blob] (b){};
\vertex[right=1cm of b] (c){};

\vertex[right=0.5cm of c] (d){};
\vertex[right=1cm of d] (e){};

\vertex[right=0.3cm of e] (f){};
\vertex[right=1cm of f] (g){};
\vertex[right=1cm of g] (h){};

\vertex[right=0.2cm of h] (i){};
\vertex[right=1cm of i] (j){};
\vertex[right=1.3cm of j] (k){};
\vertex[right=1cm of k] (l){};

\diagram* {
 (a) --[photon] (b)--[photon] (c),
 (d) --[photon] (e),
 (f) --[photon] (g)--[photon] (h),
 (i) --[photon] (j)--[photon] (k)--[photon] (l),
};
\node[align=center,black,opacity=1] at (2.2,0) {$=$};
\node[align=center,black,opacity=1] at (3.6,0) {$+$};
\node[align=center,black,opacity=1] at (5.9,0) {$+$};
\node[align=center,black,opacity=1] at (9.6,0) {$+\cdots$};
\filldraw (g) [fill=white] circle (0.4);
\filldraw (j) [fill=white] circle (0.4);
\filldraw (k) [fill=white] circle (0.4);
\node[align=center,black,opacity=1] at (g) {$\Pi$};
\node[align=center,black,opacity=1] at (j) {$\Pi$};
\node[align=center,black,opacity=1] at (k) {$\Pi$};
\end{feynman}
\end{tikzpicture}
\end{center}
\caption{The full propagator is obtained by summing the 1PI contributions.\label{fig:1PI}}
\end{figure}
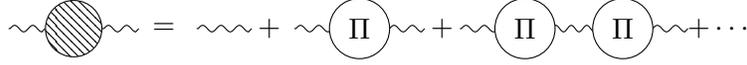

We can now weakly gauge the electroweak gauge group, by simply promoting the ordinary derivatives to covariant derivatives $\partial_\mu \to D_\mu$, defined as
\begin{equation}
\begin{tikzpicture}[baseline=(C.east)]

\matrix (A)[opacity=0,matrix of math nodes, nodes = {node style ge},row sep=0.38cm,column sep=0.19cm,
left delimiter=(,right delimiter={)}] at (0.05,-0.07)
{ & & & \\
   & &  &\\
  &  &  & &\\
};
\matrix (B) [right=2.1cm of A,opacity=0,matrix of math nodes, nodes = {node style ge},row sep=0.38cm,column sep=0.25cm,
left delimiter=(,right delimiter={)}]
{  & & & \\
   & &  & \\
  &  &  & \\
};
\node[align=center,black,opacity=1] (C) at (-2.5, 0) {$D_{\mu}\,=\, \plmu - ig\, W^a_\mu$};
\node[align=center,black,opacity=1] (CC) at (1.6, 0) {$- ig' \, B_\mu$};
\node[align=center,black,opacity=1] (A) at (-0.2, 0.2) {$\tau^{a}/2$};
\draw[-,black,opacity=0.9,line width=0.5 mm] (0.3,-0.55)  to (0.3,0.55) ;
\draw[-,black,opacity=0.9,line width=0.5 mm] (-0.55,-0.2)  to (0.63,-0.2) ;
\draw[-,black,opacity=0.9,line width=0.5 mm] (3.45,-0.55)  to (3.45,0.55) ;
\draw[-,black,opacity=0.9,line width=0.5 mm] (2.55,-0.2)  to (3.75,-0.2) ;
\node[align=center,black,opacity=1] (AA) at (2.95, 0.2) {$\frac{1}{6}\cdot \mathbb{1}$};
\node[align=center,black,opacity=1] (AAA) at (4.1, -0.05) {.};
\end{tikzpicture}
\end{equation}
In the above definition, the $\tau^a$ are the standard Pauli matrices.
An important side effect of the chiral symmetry breaking $SU(3)_L\times SU(3)_R \to SU(3)_V$ is that it also breaks electroweak symmetry! We can see this easily from the chiral Lagrangian: the $\Sigma$ field will contain a term independent of the pions $\Sigma=1+\ldots $, hence the covariant derivative will contain terms linear in the gauge fields:
\begin{equation} D_\mu \Sigma \subset -i \frac{g}{2} W_\mu^a \tau^a - i \frac{g'}{6} B_\mu \, ,
\end{equation}
hence the Lagrangian will contain the gauge boson mass terms 
\begin{equation}
\frac{f^2}{4} {\rm Tr} [ (D_\mu \Sigma )^\dagger D^\mu \Sigma ]\supset \frac{g^2f^2}{4} W_\mu^+ W^{\mu -}+ \frac{g^2+{g'}^2}{4} f^2 \frac{1}{2} Z_\mu Z^\mu \ .
\label{eq:covariant}
\end{equation}
The expressions obtained for the gauge bosons masses just like those from the ordinary SM Higgs mechanism, except v is replaced by the pion decay constant $f$. A simple way to convince yourself that the gauge boson indeed has become massive is to examine the fate of the Goldstone bosons. In addition to the gauge boson mass terms (\ref{eq:covariant}) also contains a derivative mixing term between the pions and the gauge bosons:
\begin{equation}
\frac{g}{2} f\, W_\mu^+ \partial^\mu \Pi^- + h.c.
\end{equation}
which is also the term that explains most easily the measured decay width of the charged pions to $\mu + \nu_\mu$, and can be used to fix the pion decay constant $f_\pi \sim 130$ MeV. 
This mixing will also contribute to the $W$ propagator and shift the location of the pole in the W propagator from $0$ to $g^2 f^2/4$. One way to think of the Lagrangian for the $W$ boson is to say that it is linearly coupled to weak currents made out of quarks:
\begin{equation}
{\cal L}_W = W_\mu^+  J^{\mu \ -} + h.c.,
\end{equation}
and the $W$ propagator $i \Pi_{\mu\nu} (q)$ is nothing but the current-current two-point function
\begin{equation}
i \Pi_{\mu\nu} (q) = \langle J_\mu^+ (q) J_\nu^-(-q) \rangle = i \left( \eta_{\mu\nu} -\frac{q_\mu q_\nu}{q^2} \right) \Pi (q^2)
\end{equation}
where the function $\Pi (q^2)$ encodes the effect of the strong dynamics. The full $W$ propagator can be obtained from summing up the 1PI contributions (see Fig.~\ref{fig:1PI}) 

\begin{equation}
\Delta_{\mu\nu} = \frac{-i}{q^2 - g^2 \Pi (q^2)} \left( \eta_{\mu\nu} -\frac{q_\mu q_\nu}{q^2} \right)
\end{equation}
hence shifting the mass of the $W$ boson to $g^2 \Pi (0)$. We know that the charged current generates the charged pions
\begin{equation}
\langle 0| J_\nu^+ | \Pi^- (p) \rangle = \frac{i f p_\mu}{\sqrt{2}}
\end{equation}
implying that $\Pi (q^2) = \frac{f^2}{2}$.

\subsection{Explicit breaking}

Next we discuss the effects of explicit breaking and how to incorporate them into the chiral Lagrangian. One source for explicit breaking are the charges of the quarks (that is the fact that the electromagnetic charges of the quarks are not uniform but different for up vs. down and strange). The charge matrix is given by 
\begin{equation}
Q= \left( \begin{array}{ccc} \frac{2}{3} \\ & -\frac{1}{3} \\ & & -\frac{1}{3} \end{array} \right) 
\end{equation}
in the $(u,d,s)$ basis for the quarks. We will use this quark charge matrix as a {\it spurion}: we imagine that there was a field $Q$ that transformed as a bifundamental under the $SU(3)_L\times SU(3)_R$ global symmetry, and try to write down invariants under the full symmetry including the spurion. For example we can write the term:\footnote{From the point of view of the symmetries one could also write the term ${\rm Tr} \, Q \Sigma$, but investigating the electromagnetic contributions we can quickly convince ourselves that every term must contain at least two $Q$ insertions, as a consequence of electromagnetic gauge invariance.}
\begin{equation}
\Delta {\cal L} = e^2 {\rm Tr} [Q \Sigma^\dagger Q \Sigma ] 
\end{equation}
where the presence of the overall $e^2$ factor follows from the observation that this term must vanish for $e\to 0$ (in other words $Q$ must always appear together with a factor of $e$). 
At this point we can freeze the spurion $Q$ to its VEV ${\rm diag} (\frac{2}{3},-\frac{1}{3},-\frac{1}{3})$ which will yield a mass contribution to the charged pions (but not the neutral ones) and hence explain the observed charged-neutral pion mass splittings. The exact same story can be repeated for the quark masses. The quark mass matrix is numerically given by 
\begin{equation}
M = \left( \begin{array}{ccc} m_u \\ & m_d \\ & & m_s \end{array} \right) 
\end{equation}
but we first promote it to a spurion transforming as a bi-fundamental under $SU(3)_L\times SU(3)_R$. The leading operator in this case is 
\begin{equation}
\Delta {\cal L} = \mu^3  {\rm Tr} \left[M \left(\Sigma+\Sigma^\dagger\right) \right] \, =\, \mu^3   {\rm Tr} \left[{\left(\frac{M}{f} \,\Pi^a T^a\right)}^2 \right] +\ldots
\end{equation}
where $\mu$ is a fixed dimensionful constant. This provides a shift to the pion mass squares of the form 
\begin{equation}
\Delta m_\pi^2 \propto \mu^3 \frac{m_q}{f^2} 
\end{equation}
yielding the famous Gell-Mann-Okubo mass formula
\begin{equation}
m_\eta^2 +m_\pi^2 = 4 m_K^2
\end{equation}
for $m_u \simeq m_d \ll m_s$.

\subsection{NDA and the cutoff scale}

An essential part of every EFT, including non-linearly realized Goldstone Lagrangians is its cutoff $\Lambda$. Since it is a non-renormalizable theory the best way to make sense of it is to assume that there is a region of validity for the theory characterized by the cutoff scale $\Lambda$. This is the scale where we definitely expect new particles to show up, and generically this is also the scale that cuts off the radiative corrections to the Higgs potential. The simplest method for estimating the size of the cutoff scale is called naive dimensional analysis (NDA). In this method we assume that all couplings are $\mathcal{O}(4\pi)$, and get an upper limit on how large the cutoff scale could be.

The cutoff is determined to be the energy scale in which the divergent loop corrections become as large as the tree level ones. For example in the chiral Lagrangian, the 4-point pion-pion interaction vertex is of the form 
\begin{equation}
\frac{\Pi^2 (\partial \Pi)^2}{f^2} \to \frac{p^2}{f^2} \ {\rm vertex}\, .
\end{equation}
The 4-point vertex allows us to write a loop term, and we are looking for $\Lambda$ for which
\begin{center}
\begin{tikzpicture}[baseline=(A.west)]
\begin{feynman}[]
\vertex (a){};
\vertex[above=1cm of a] (b){};
\vertex[below=1cm of a] (c){};
\vertex[right=1cm of a,dot] (d){};
\vertex[right=1cm of d] (f){};
\vertex[above=1cm of f] (g){};
\vertex[below=1cm of f] (h){};

\vertex[right=2cm of f] (i){};
\vertex[above=1cm of i] (j){};
\vertex[below=1cm of i] (k){};
\vertex[right=1cm of i,dot] (l) {};
\vertex[right=1.5cm of l,dot] (m) {};
\vertex[right=1cm of m] (n) {};
\vertex[above=1cm of n] (o){};
\vertex[below=1cm of n] (p){};
\diagram* {
 (b) --[scalar] (d),
 (c) --[scalar] (d),
 (d) --[scalar] (g),
 (d) --[scalar] (h),
 
(j) --[scalar] (l),
(k) --[scalar] (l),
(m) --[scalar] (o),
(m) --[scalar] (p),
(l) --[half left, scalar] (m),
(m)--[half left, scalar] (l),
};
\node[align=center,black,opacity=1,right=1cm of f] {$\approx$};
\node[align=center,black,opacity=1,left=0.5cm of d] {$\frac{p^2}{f^2}$};
\node[align=center,black,opacity=1,left=0.5cm of l] {$\frac{p^2}{f^2}$};
\node[align=center,black,opacity=1,right=0.5cm of m] {$\frac{p^2}{f^2}$};
\end{feynman}
\end{tikzpicture}
\end{center}
We have then
\begin{equation}
\frac{p^2}{f^2}\,\approx\, \frac{1}{f^4} \frac{1}{16 \pi^2} \int \frac{d^4k}{(k^2)^2} k^2 p^2  \sim \frac{p^2 \Lambda^2}{16\pi^2 f^4} = \left( \frac{\Lambda}{4\pi f} \right)^2 \frac{p^2}{f^2}
\end{equation}
We can see that the requirement that the one loop result be at most as large as the original tree-level vertex will limit the size of the cutoff to 
\begin{equation}
\Lambda \lsim 4 \pi f 
\label{eq:NDA}
\end{equation}
The scale $\Lambda$ is the physical scale where new particles have to appear (to be more precise this is an upper bound, new particles could also show up earlier). It is also the actual cutoff of the purely strongly coupled theory, where the interactions among the composites are also strongly coupled, $g_* \sim 4\pi$. This is called the NDA limit, the limit when (\ref{eq:NDA}) is saturated. However, new particles could also show up earlier than the maximal scale $\Lambda$. This happens, when the interaction strength of some of the composites is actually not so strong, $g_* <4\pi$. We expect these particles to show up at a lower scale 
\begin{equation}
m_\rho= g_* f \ .
\end{equation}
In particular, $g_* \sim 1$ means that some of the composites are actually weakly coupled, and 
the $m_\rho$ scale is actually $f$. This means that there will be new particles that can be used to cut off the quadratic divergences before reaching the full cutoff scale. As we will see these will be the so-called top partner and spin 1 partner particles. 

Similarly one can perform NDA for a generic term in the effective Goldstone Lagrangian. The rules are the following:
\begin{itemize}
\item Every Goldstone field will have a $1/f$ suppression (arising from expanding the exponential $e^{i \Pi^a T^a/f}$).

\item The remaining dimensions are made up by the $\rho$ mass scale $m_\rho = g_* f$. Thus the two dimensionless quantities are $x \equiv  \frac{\Pi}{f} = g_* \frac{\Pi}{m_\rho}$ and $y \equiv \frac{\partial}{m_\rho}$. 

\item We start with a dimensionless function of $x,y$ to give $\tilde{\cal L}(x,y)$, from which we get a dimension 4 Lagrangian $m_\rho^4 \tilde{\cal L}(x,y)$

\item The kinetic terms obtained using this rule are $m_\rho^4 y^2 x^2 = g_*^2 (\partial \Pi )^2$. Thus the entire Lagrangian needs to be rescaled by $1/g_*^2$. 
\end{itemize}

For example, a quartic two-derivative coupling would be estimated at $\frac{1}{g_*^2} m_\rho^4 \frac{\partial^2}{m_\rho^2} \frac{\Pi^4}{f^4} = \frac{(\partial \Pi )^2 \Pi^2}{f^2}$ as we saw from the explicit expansion for the chiral Lagrangian. A non-derivative tree-level quartic would be estimated at $\frac{1}{g_*^2} m_\rho^4  g_*^4 \frac{\Pi^4}{m_\rho^4} = g_*^2 \Pi^4$. If it is loop induced, NDA will give its size at $\frac{g_*^2}{16\pi^2} \Pi^4$. These will be the two basic magnitudes of quartics generically showing up in pNGB Higgs models.  

\subsection{Towards a composite Higgs model}

We are now ready to start constructing models with dynamical electroweak symmetry breaking. The simplest idea is to mimic the story already happening in QCD, where the strong dynamics breaks the global symmetries as $SU(3)_L\times SU(3)_R \to SU(3)_V$, except the symmetry breaking scale is too low. One can instead choose a group with strong dynamics that has chiral symmetry breaking patters $SU(2)_L\times SU(2)_R \to SU(2)_V$ and a much higher dynamical scale $\Lambda_{TC}$ giving rise to a condensate $\langle \bar{q} q \rangle = \Lambda_{TC}^3$. This is the main idea of {\it technicolor} models, where all formulae discussed above will be rescaled via $\Lambda_{QCD}\to \Lambda_{TC}$. The   $SU(2)_L\times SU(2)_R \to SU(2)_V$ breaking will produce 3 GBs which are the minimal number to provide for the longitudinal degrees of freedom of $W^\pm, Z$. In such minimal TC models there is no additional light particle, in particular no light Higgs boson would appear. The symmetry breaking pattern is similar in {\it higgsless models}, the main difference is that the higgsless models are weakly coupled and calculable. Since the physical Higgs boson has been discovered in 2012, TC and higgsless models are no longer viable options. The next simplest possibility is for the strong dynamics to not directly break the electroweak symmetry, but rather produce a light composite Higgs among the generic heavier composite states. In this case the hierarchy problem would be solved since there would be no true elementary scalars. However in generic (bona fide) strongly coupled theories the composite Higgs is expected to appear at $\Lambda_{strong}$, or at most a loop factor below, at $\Lambda_{strong}/(4\pi)$. If we could take $\Lambda_{strong} \sim 1$ TeV there would be no problem. However, in generic theories with a strong scale at 1 TeV one would expect a very diverse spectrum of new particles BSM showing up at 1 TeV. There are two problem with this: first the LHC has not (yet) observed any new particles, with several typical bounds well above 1 TeV. Second, generic new particles interacting under the SM gauge symmetries will give large corrections to electroweak precision observables (essentially higher dimensional operators suppressed by $1/\Lambda^2$). The LEP experiments at CERN have strongly constrained such corrections, with the conclusion that generic suppression scales should be more like $\Lambda_{strong} \gsim 5-10$ TeV. However in this case the expected size of the Higgs mass would be around $10\ {\rm TeV}/(4\pi) \sim 1$ TeV. This would still leave a tuning of about 
\begin{equation}
\left( \frac{125\ {\rm GeV}}{1\ {\rm TeV}}\right)^2 \sim 1 \%
\end{equation}
This percent level tuning is generically called the little hierarchy problem of generic composite Higgs models. In order to reduce the Higgs mass from 1 TeV to 125 GeV we will assume that it is Goldstone boson. The next sections will explore on how to implement the pNGB Higgs idea as part of the composite Higgs models and obtain realistic models of this sort.   

\section{Little Higgs models and Collective Symmetry Breaking\label{sec:Collective}}
 
In this section we will review the mechanism called collective symmetry breaking \cite{ArkaniHamed:2001nc, Littlest} , and show how it can lower the effective cutoff to the Higgs radiative potential. For other reviews of little Higgs models and collective symmetry breaking, see \cite{LHreviews}.
As in the last section, we consider a sector with some strong dynamics which confines at a scale $\Lambda$. The strongly coupled sector has some global symmetry $G$, which is spontaneously broken to a subgroup $H$ by the confining dynamics at the scale $\Lambda$. The Higgs is then among the Nambu-Goldstone bosons (NGBs) in the coset $G/H$ \cite{Kaplan:1983fs}. By Goldstone's theorem, NGBs have only derivative couplings and no mass or quartic. However, there is one more key element in the construction: the original $G$ is not exact, but rather an approximate global symmetry. Another way to say it is that $G$ is \textbf{explicitly broken}. The explicit breaking of $G$ makes the Higgs not an exact NGB but rather a pseudo-Nambu-Goldstone boson (pNGB) by generating a mass and a quartic term for it. The problem is that the loop-induced mass term is generically large since it is quadratically divergent. However, the quadratic divergence can be eliminated in scenarios which exhibit collective symmetry breaking.

To illustrate the idea of collective symmetry breaking, we will focus on a model called ``The Simplest Little Higgs" \cite{Schmaltz:2004de}.
In this model we consider a strongly coupled sector with a $G=SU(3)$ global symmetry, broken to the subgroup $H=SU(2)$ at the scale $\Lambda$. The first thing we can do is count the number of broken generators, which by Goldstone's theorem equals the number of NGBs:
\begin{eqnarray}
N_{\text{NGB}}=\left(3^2-1\right)\, -\, \left(2^2-1\right)\,=\,5 .
\end{eqnarray}
We can represent these NGBs graphically by looking at a generic $SU(3)$ matrix and splitting it to the unbroken $SU(2)$ part, and the NGB part:

\begin{equation}
\begin{tikzpicture}[baseline=(A.center)]
 \tikzset{BarreStyle/.style =   {opacity=.4,line width=5 mm,line cap=round,color=#1}}

\matrix (A) [opacity=0,matrix of math nodes, nodes = {node style ge},,column sep=0 mm,
left delimiter=(,right delimiter={)}]
{ a & a& \times  \\
 a  &a & \times  \\
\times & \times &  b \\
};
\fill [opacity=.8,black,rounded corners=10, draw]
   (A-2-1.south west) -- (A-2-2.south east) -- (A-1-2.north east) -- (A-1-1.north west) --
  cycle
  {};
 \draw [BarreStyle=black,opacity=.4] (A-1-3.center)  to (A-2-3.center) ;
 \draw [BarreStyle=black,opacity=.4] (A-3-1.center)  to (A-3-2.center) ;
 \draw [BarreStyle=black,opacity=.4] (A-3-3.center)  to (A-3-3.center) ;
\draw[->,opacity=.4,line width=0.5 mm] (1,0.4)  to (2,0.4) ;
\node[align=center,black,opacity=.6] (A) at (3.1, 0.21) {Broken \\ $SU(3)/SU(2)$};
\draw[->,black,opacity=.8,line width=0.5 mm] (-1.02,0.4)  to (-2,0.4) ;
\node[align=center,black,opacity=.8] (A) at (-3.1, 0.38) {Unbroken \\ $SU(2)$};
\end{tikzpicture}
\end{equation}
Every generator that has non-zero elements in one of the blue parts is a broken generator, while generators that only have nonzero elements in the red part are unbroken. The broken $SU(3)$ generators in this case are $\lambda_{4,\ldots,8}$, and we can represent the NGBs as:
\begin{equation}
\begin{tikzpicture}[baseline=(A.center)]

\node[align=center,black,opacity=1] (C) at (-5.5, 0) {$U_{\text{NGB}}\,\,\,\equiv \,\,\, \exp \left[\frac{\sqrt{2}i}{f} \pi^{\hat{a}} T^{\hat{a}} \right] \,\,\, \equiv~\exp \left[\frac{i}{f}\right.$};

\matrix (A) [matrix of math nodes, nodes = {node style ge},
left delimiter=(,right delimiter={)}]
{ \frac{\eta}{\sqrt{6}} & &  \\
   &\frac{\eta}{\sqrt{6}} &  \\
 &  &  -\frac{2\eta}{\sqrt{6}} \\
};
\node[align=center,black,opacity=1] (A) at (1.2, 0.6) {$H$};
\node[align=center,black,opacity=1] (B) at (-0.55, -1.1) {$H^\dagger$};
\draw[-,black,opacity=0.9,line width=0.7 mm] (0.5,-1.7)  to (0.5,1.7) ;
\draw[-,black,opacity=0.9,line width=0.52 mm] (-1.7,-0.5)  to (1.7,-0.5) ;
\node[align=center,black,opacity=1] (D) at (2.5, 0) {$\left.\right],$};
\end{tikzpicture}
\end{equation}
where the index $\hat{a}$ represents broken $SU(3)$ generators only. In the above equation,
\begin{eqnarray}
H \equiv \frac{1}{\sqrt{2}}\colvec{h_1 +i h_2\\h_3+ih_4}\, 
\end{eqnarray}
is the SM $SU(2)_L$ Higgs doublet.
As stated above, the $SU(3)$ symmetry must be explicitly broken in order to get a Higgs potential. This is achieved by the gauging of an $SU(2)$ subgroup. Gauging only a subgroup of a global symmetry explicitly breaks the symmetry---in particular the gauge bosons only transform under the subgroup. Note that the $SU(2)$ subgroup doesn't generically coincide with the subgroup $H=SU(2)$ that survives the spontaneous breaking at $\Lambda$. The explicit breaking generates a potential for the (p)NGBs arising from loops of $SU(2)$ gauge bosons. We can parameterize this potential by writing a $nl\sigma m$. The $nl\sigma m$ field is defined as:
\begin{equation}\label{eq:nlsmslh}
\begin{tikzpicture}[baseline=(A.center)]
\node[align=center,black,opacity=1] (A) at (0, 0) {$\Sigma\,=\, U_{NGB}\, \colvec{0\\0\\f}=\colvec{ \\ \\f-\frac{H^{\dagger}H}{2f}}\,+\, \eta \text{ dependent and higher order terms}\, ,$};
\node[align=center,black,opacity=1] (B) at (-1.6, 0.2) {$iH$};
\end{tikzpicture}
\end{equation}
where we remember that $H$ is a complex doublet.
The leading-order $nl\sigma m$ Lagrangian is simply
\begin{eqnarray}
\mathcal{L}_{nl\sigma m}={\left(D_{\mu}\, \Sigma\right)}^{\dagger}\,\left(D^{\mu}\, \Sigma\right)\, ,
\end{eqnarray}
where $D_{\mu}$ is the gauged $SU(2)$ covariant derivative 
\begin{equation}
\begin{tikzpicture}[baseline=(C.east)]

\node[align=center,black,opacity=1] (C) at (-2.5, 0) {$D_{\mu}\,=\, \plmu - ig W^a_\mu$};
\node[align=center,black,opacity=1]  at (1.2, -0.1) {$.$};
\matrix (A) [opacity=0,matrix of math nodes, nodes = {node style ge},row sep=0.38cm,column sep=0.44cm,
left delimiter=(,right delimiter={)}] at (0.03,-0.05)
{  & &  \\
   &  &  \\
 &  &  \\
};
\node[align=center,black,opacity=1] (A) at (-0.22, 0.2) {$\tau^{a}/2$};
\draw[-,black,opacity=0.9,line width=0.4 mm] (0.35,-0.5)  to (0.35,0.5) ;
\draw[-,black,opacity=0.9,line width=0.3 mm] (-0.6,-0.2)  to (0.69,-0.2) ;
\end{tikzpicture}
\end{equation}
Expanding this Lagrangian in terms of $H$, we get:
\begin{eqnarray}\label{eq:lnlsm}
\mathcal{L}_{nl\sigma m}\,=\,{\left|\dlmu H\right|}^2\,\left(1+\frac{H^{\dagger}H}{f^2}+\ldots\right),
\end{eqnarray}
where
\begin{eqnarray}
D_{\mu} H \, =\, \left(\plmu - i\frac{g}{2} W^a_{\mu} \tau^a\right)\, H \, .
\end{eqnarray}
The $SU(2)$ gauge bosons $W$ explicitly break the global $SU(3)$ invariance via interactions of the type ${\left|gW_{\mu} h\right|}^2$. At the one loop level, these generate a quadratic contribution to the Higgs potential:
\begin{equation}\label{eq:diags}
\begin{tikzpicture}[baseline=(E.west)]
\begin{feynman}[]
\vertex (a){H};
\vertex[right=1cm of a,dot] (b){};
\vertex[right=1cm of b] (c) {H};
\vertex[right=1cm of c] (a1) {H};
\vertex[right=1cm of a1,dot] (a2){};
\vertex[right=1cm of a2,dot] (a3){};
\vertex[right=1cm of a3] (a4){H};
\diagram* {
(a1) -- [scalar] (a2) -- [scalar] (a3) -- [scalar] (a4),
(a2) -- [half left, photon] (a3),
 (a) --[scalar] (b) [dot] --[scalar] (c),
};
\node[align=center,black,opacity=1] (B) at (2.5,0) {$+$};
\node[align=center,black,opacity=1] (C) at (1.6,0.7) {$W$};
\node[align=center,black,opacity=1] (D) at (5.3,0.35) {$W$};
\path (b)--++(90:0.4) coordinate (A);
\draw (A) [photon]circle(0.4);
\node[align=center,black,opacity=1] (E) at (8,0.2) {$\propto~~~~ \frac{3g^2}{64\pi^2}\,\Lambda^2\,\,H^{\dagger}H\,\,+\, \, \ldots$};
\end{feynman}
\end{tikzpicture}
\end{equation}
The quadratic divergence generated by the gauging of $SU(2)$ should not come as a surprise. After all, the interaction with the $SU(2)$ gauge bosons is similar to the SM, where the gauge quantum corrections contribute all the way up to the cutoff. We seem to have come full circle then: first the Higgs was an NGB in the coset $G/H$, with zero potential due to the shift symmetry protecting NGBs. Then we broke the original $G$ symmetry, but unfortunately got the quadratic divergences back. Have we achieved anything?

To understand better what is going on, let us calculate the Higgs potential in a more systematic way. The first step is to extract the term quadratic in $W$ from the Lagrangian Eq.~(\ref{eq:lnlsm}):
\begin{equation}
\begin{tikzpicture}[baseline=(C.east)]

\node[align=center,black,opacity=1] (C) at (-2.4, 0.1) {$\mathcal{L}_{nl\sigma m}~~ \ni ~~\left|\, g\right.$};
\node[align=center,black,opacity=1] (D) at (3.7, 0.1) {${\left. \Sigma~\right|}^2\,~~=~~M^2(H)_{ab}~\, W^{a}_{\mu}W^{b\mu}\, ,$};
\matrix (A) [opacity=0,matrix of math nodes, nodes = {node style ge},row sep=0.38cm,column sep=0.58cm,
left delimiter=(,right delimiter={)}] at (0,-0.05)
{  & &  \\
   &  &  \\
 &  &  \\
};
\node[align=center,black,opacity=1] (A) at (-0.2, 0.25) {$W_\mu$};
\draw[-,black,opacity=0.9,line width=0.3 mm] (0.35,-0.6)  to (0.35,0.6) ;
\draw[-,black,opacity=0.9,line width=0.4 mm] (-0.65,-0.23)  to (0.7,-0.23) ;
\end{tikzpicture}
\end{equation}
where 
\begin{equation}
\begin{tikzpicture}[baseline=(C.east)]
\node[align=center,black,opacity=1] (C) at (-2.75, 0.1) {$M^2(H)_{ab}~~ = ~~\frac{g^2}{4}\Sigma^\dagger~
$};
\node[align=center,black,opacity=1] (D) at (1.25, 0.1) {$\Sigma$};
\matrix (A) [opacity=0,matrix of math nodes, nodes = {node style ge},row sep=0.38cm,column sep=0.58cm,
left delimiter=(,right delimiter={)}] at (0,-0.05)
{  & &  \\
   & &  \\
 &  &  \\
};
\node[align=center,black,opacity=1] (A) at (-0.2, 0.25) {$\tau^a\tau^b$};
\draw[-,black,opacity=0.9,line width=0.3 mm] (0.35,-0.6)  to (0.35,0.6) ;
\draw[-,black,opacity=0.9,line width=0.4 mm] (-0.65,-0.22)  to (0.7,-0.22) ;
\end{tikzpicture}
\end{equation}
 is a Higgs dependent mass matrix for $W_\mu$, which we wrote in terms of an $SU(3)$-breaking spurion $P$. From this term we can compute the Coleman-Weinberg potential \cite{Coleman:1973jx} for the Higgs due to radiative corrections from the gauge bosons. This potential is given by the formula:
\begin{eqnarray}
V_{CW}(H)~=~\underbrace{\frac{\Lambda^2}{16\pi^2}\,Tr\left[M^2_{ab}\right]}_{\text{Quadratically div. term}}
+\,~
\underbrace{
\frac{3}{64\pi^2}\,Tr\left\{M^4_{ab}\,\log\left[\frac{M^2_{ab}}{\Lambda^2}\right]\right\}}_{\text{Log div. + finite term}}\, .
\end{eqnarray}
We see that the Higgs potential is quadratically divergent because $M^\dagger M$ is not $\, \propto \mathbb{1}$. In other words,
\begin{eqnarray}\label{eq:quad}
V_{CW}(H)~=~\frac{3g^2\Lambda^2}{64\pi^2}\,\Sigma^{\dagger}\,P\,\Sigma~=~\frac{3g^2\Lambda^2}{64\pi^2}\,\,H^\dagger H \, ,
\end{eqnarray}
where $P$ is an $SU(3)$ breaking spurion  
\begin{eqnarray}\label{eq:Pspur}
P~=~\colmatth{1&0&0\\0&1&0\\0&0&0}\,.
\end{eqnarray}
Note that if not for the spurion $P$, the quadratic divergence would be proportional to $\Sigma^{\dagger} \Sigma$ which is independent of $H$.

Our next idea to get rid of the quadratic divergence is to get rid of the spurion $P$ in Eq.~(\ref{eq:quad}) by simply gauging the entire $SU(3)$ instead of just an $SU(2)$ subgroup. In this case the symmetry is no longer explicitly broken because the gauge bosons come in a complete adjoint of $SU(3)$. Accordingly, the quadratic divergence becomes
\begin{eqnarray}
V_{CW}(H)~=~\frac{g^2\Lambda^2}{12\pi^2}\,Tr\left[\Sigma^{\dagger}\,\Sigma\right]\, ,
\end{eqnarray}
which is independent of $h$. However, there is one major problem in this scenario: there are now additional $SU(3)$ gauge bosons that become massive due to the $SU(3)/SU(2)$ spontaneous breaking. Unfortunately, that means that the would-have-been Higgs is now ``eaten" by these additional gauge bosons---there is no longer a physical scalar in the theory to break electroweak symmetry. We somehow need a better solution, one which preserves a global $SU(3)$ symmetry but also has an uneaten scalar. 

This reasoning leads us to the third, and final version of our story.
Consider a strongly interacting sector with a global symmetry which is not $SU(3)$, but a larger $SU(3)\times SU(3)$. The strong sector confines at $\Lambda$, breaking the global symmetry to $SU(2)\times SU(2)$. 

We can parameterize this breaking by not one, but two $nl\sigma m\,$ fields $\Sigma_1$ and $\Sigma_2$, in the first and second $SU(3)/SU(2)$ coset, respectively:
\begin{eqnarray}
\Sigma_{1}\,\,\,\equiv \,\,\, e^{i \,\pi_1 /\sqrt{2}f }\,\colvec{0\\0\\f}~~,~~\Sigma_{2}\,\,\,\equiv \,\,\, e^{i \,\pi_2 /\sqrt{2}f }\,\colvec{0\\0\\f}\, ,
\end{eqnarray}
where we have taken $f_1=f_2=f$ for simplicity. 
Additionally, we break the $SU(3)\times SU(3)$ explicitly by gauging only the diagonal subgroup ${SU(3)}_D$.
In terms of the $nl\sigma m\,$ fields the nonlinear Lagrangian is
\begin{eqnarray}
\mathcal{L}_{nl\sigma m}={\left|D_{\mu}\, \Sigma_1\right|}^2\,+\,{\left|D_{\mu}\, \Sigma_2\right|}^2\, ,
\end{eqnarray}
where $D_\mu$ is now the $SU(3)_D$ covariant derivative. Because we gauged the entire ${SU(3)}_D$, there is no $SU(3)$ breaking spurion in the Lagrangian.

What about the physical Higgs? This time we started with $SU(3)\times SU(3)$ spontaneously broken to $SU(2)\times SU(2)$, with 
 \begin{eqnarray}
N_{\text{NGB}}=2\left(3^2-1\right)\, -\, 2\left(2^2-1\right)\,=\,10 ,
\end{eqnarray}
out of which $\left(3^2-1\right)-\left(2^2-1\right)=5$ were eaten by heavy gauge bosons, leaving us with an uneaten complex doublet $H\,\propto \,\, \pi_1-\pi_2$ and a real scalar $\eta$. For its potential, we obtain the leading quadratically divergent piece
\begin{eqnarray}
V_{CW}(H)~=~\frac{\Lambda^2}{16\pi^2}\,\left\{Tr\left[\Sigma^{\dagger}_1\,\Sigma_1\right]\,\,+\,\, Tr\left[\Sigma^{\dagger}_2\,\Sigma_2\right] \right\}\, ,
\end{eqnarray}
which is again independent of the NGBs. 

We see that the problem has been solved, at least on a technical level. There is a physical Higgs and no quadratic divergence in the Higgs potential. The deeper reason for this cancellation is called \textbf{collective breaking}.

\subsection{Collective breaking}
To illustrate the concept of collective breaking, we will study the quadratic part of the $nl\sigma m\,$ Lagrangian 
\begin{eqnarray}\label{eq:coll}
\mathcal{L}_{nl\sigma m}\,\ni\,{\left|g\, W_{\mu}\, \Sigma_1\right|}^2\,+\,{\left|g\, W_{\mu}\, \Sigma_2\right|}^2\, . \label{eq:NLSM_collective}
\end{eqnarray}
Let us analyze the symmetries of this Lagrangian. Both terms are present because the diagonal group ${SU(3)}_D$ is gauged. Note that the existence of a \textit{gauge} ${SU(3)}_D$ symmetry also indicates the existence of a \textit{global} ${SU(3)}_D$, symmetry under which:
\begin{eqnarray}
\Sigma_1\rightarrow e^{i\alpha^aT^a}\Sigma_1~,~\Sigma_2 \rightarrow e^{i\alpha^aT^a}\Sigma_2~,~W_{\mu}\rightarrow e^{i\alpha^aT^a} W_{\mu} e^{-i\alpha^aT^a}\, .
\end{eqnarray}
Both $\Sigma_1, \Sigma_2$ must rotate with the same transformation angles $\alpha^a$ as $W_\mu$ such that Eq.~(\ref{eq:NLSM_collective}) is invariant.

Now we reach the key part of our analysis: what would the symmetry be if only the first term of Eq.~(\ref{eq:coll}) was present? That case would correspond to having a \textit{gauge} $SU(3)$ symmetry, and also an $SU(3)$ \textit{global} symmetry, i.e. there is an overall $SU(3)\times SU(3)$ global symmetry:
\begin{eqnarray}
\Sigma_1\rightarrow e^{i\alpha^a_1 T^a}\Sigma_1~,~\Sigma_2 \rightarrow e^{i\alpha^a_2T^a}\Sigma_2~,~W_{\mu}\rightarrow e^{i\alpha^a_1T^a} W_{\mu} e^{-i\alpha^a_1T^a}\, ,
\end{eqnarray}
where $\alpha_{1,2}$ are independent $SU(3)$ rotation parameters. Since the second term is absent, we are free to do rotations on $\Sigma_2$ that are independent of the rotations on $\Sigma_1, W_\mu$. The same is true if only the second term was present in  Eq.~(\ref{eq:coll}), this time with an $SU(3)\times SU(3)$ under which
\begin{eqnarray}
\Sigma_1\rightarrow e^{i\alpha^a_1 T^a}\Sigma_1~,~\Sigma_2 \rightarrow e^{i\alpha^a_2T^a}\Sigma_2~,~W_{\mu}\rightarrow e^{i\alpha^a_2T^a} W_{\mu} e^{-i\alpha^a_2T^a}\, .
\end{eqnarray}

We have seen that each term in Eq.~(\ref{eq:coll}) separately conserves an extended $SU(3)\times SU(3)$ symmetry, and only in the presence of both terms, the global symmetry is reduced to ${SU(3)}_D$. In other words, the symmetry is \textbf{collectively broken} in the presence of the two terms: if only one of the two terms were present, we would have a full $SU(3)\times SU(3)/SU(2)\times SU(2)$ of exact NGBs, out of which an $SU(3)/SU(2)$ worth is eaten by the gauged $SU(3)$, leaving us with exactly massless NGBs in the coset $SU(3)/SU(2)$ (the Higgs included). In particular, the $SU(3)$ gauge bosons cannot generate a potential to the NGBs by Goldstone's theorem---i.e. any radiative correction that only involves one of the terms in Eq.~(\ref{eq:coll}) has to be exactly zero. 

What happens then in the full picture, when both terms in Eq.~(\ref{eq:coll}) are included? In that case we still have an $SU(3)/SU(2)$ coset of uneaten bosons, but this time they are only pNGBS of the $SU(3)\times SU(3)$ with is explicitly broken to the diagonal ${SU(3)}_D$. However, we now have an important insight into the radiative corrections to the potential of the pNGBs: it only involves diagrams that combine both terms in Eq.~(\ref{eq:coll}), for example
\begin{equation}
\begin{tikzpicture}[baseline=(E.west)]
\begin{feynman}[]
\vertex (a){\(\Sigma^\dagger_1\)};
\vertex[below=0.7cm of a] (b){};
\vertex[below=0.7cm of b] (c){\(\Sigma_1\)};
\vertex[right=1cm of b,dot] (d) {};
\vertex[right=1cm of d,dot] (e) {};
\vertex[right=1cm of e] (f) {};
\vertex[above=0.7cm of f] (g){\(\Sigma^\dagger_2\)};
\vertex[below=0.7cm of f] (h){\(\Sigma_2\)};
\diagram* {
 (a) --[scalar] (d),
 (c) --[scalar] (d) ,
 (d) --[half left, photon] (e),
 (d) --[half right, photon] (e),
 (e) --[scalar] (g),
 (e) --[scalar] (h),
};
\node[align=center,black,opacity=1] (A) at (5,-0.75) {$\propto~~~~ {\left|\Sigma^\dagger_1 \Sigma_2\right|}^2\,\frac{g^4}{16\pi^2}\,\,\log\left[\frac{\Lambda^2}{\mu^2}\right]\, .$};
\end{feynman}
\end{tikzpicture}
\end{equation}
The way to understand this diagram is by expanding $\Sigma_{1,2}$ in terms of $H$, so that some of the terms in the expansion are quadratic in $H$. By simple power counting, diagrams involving both terms are only logarithmically divergent, because they contain one more propagator than the would be quadratically divergent diagrams Eq.~(\ref{eq:diags2}), which are independent of $H$ by the collective breaking argument. We see that the leading $SU(3)\times SU(3)$ breaking invariant is 
\begin{eqnarray}
{\left|\Sigma^\dagger_1 \Sigma_2\right|}^2\,\sim \, f^2 - 2H^\dagger H + \ldots 
\end{eqnarray}
and so the dominant contribution to the Higgs mass is $\sim \frac{g^4}{16\pi^2}\,\,\log\left[\frac{\Lambda^2}{\mu^2}\right]$. For $f\sim1\,\text{TeV}$, we get $m^2_h \sim {\left(100\,\text{GeV}\right)}^2$, which is in the right ballpark. 

The one question that remains is: what exactly cancels the SM gauge boson quadratic contributions? To understand that, note that our SM ${SU(2)}_L$ gauge symmetry is now embedded in a larger gauge symmetry, ${SU(3)}_D$. After the spontaneous breaking $SU(3)^2/SU(2)^2$, the 3 SM ${SU(2)}_L$ gauge bosons remain massless, while the extra 5 gauge bosons get masses proportional to $f$. Let us introduce some notation.
We denote the SM (isospin) ${SU(2)}_L$ gauge bosons $W^{\pm,3}_\mu$.
The extra gauge bosons are then denoted $X^{\pm}_\mu,\,Y^{1,2}_\mu,$ and $A^8_\mu$. These fields are embedded in the adjoint of $SU(3)$ as
 \begin{eqnarray}
\colmatth{&W^+&X^+\\W^-&&Y^1+iY^2\\X^-&Y^1-iY^2&}\,+\, W^3,\,A^8 \,\text{ on the diagonal}\, .
\end{eqnarray}
In terms of these fields, the $nl\sigma m\, $ Lagrangian is
 \begin{eqnarray}
&&\mathcal{L}_{nl\sigma m} \, =\, {\left|\left(\plmu - ig W^a_\mu T^a\right)\Sigma_{1,2}\right|}^2=\nonumber\\
&&= \frac{g^2}{4}\,H^\dagger H \left[2W^+_\mu W^{- \mu}+W^3_\mu W^{3 \mu}-X^+_\mu X^{- \mu}-\frac{1}{2}\left(Y^1_\mu Y^{1\mu}+Y^2_\mu Y^{2\mu}\right)-A^3_\mu A^{3 \mu}\right]\, .\nonumber\\
\end{eqnarray}
The would be quadratically divergent contributions from the gauge bosons are
\begin{equation}\label{eq:diags2}
\begin{tikzpicture}[baseline=(E.west)]
\begin{feynman}[]
\vertex (a){H};
\vertex[right=1cm of a,dot] (b){};
\vertex[right=1cm of b] (c) {H};
\vertex[right=1cm of c] (a1) {H};
\vertex[right=1cm of a1,dot] (a2){};
\vertex[right=1cm of a2,dot] (a3){};
\vertex[right=1cm of a3] (a4){H};
\diagram* {
(a1) -- [scalar] (a2) -- [scalar] (a3) -- [scalar] (a4),
(a2) -- [half left, photon] (a3),
 (a) --[scalar] (b) [dot] --[scalar] (c),
};
\node[align=center,black,opacity=1] (B) at (2.5,0) {$+$};
\node[align=center,black,opacity=1] (C) at (1.6,0.7) {$W$};
\node[align=center,black,opacity=1] (D) at (5.3,0.35) {$W$};
\path (b)--++(90:0.4) coordinate (A);
\draw (A) [photon]circle(0.4);
\end{feynman}
\end{tikzpicture}
\end{equation}
but the quaratic contributions from all of the gauge bosons cancel. In the first diagram, for example,
\begin{eqnarray}
 \frac{g^2}{64\pi^2}\,\Lambda^2\,\,H^{\dagger}H\, \left[\underbrace{2}_{W^{\pm}} + \underbrace{1}_{W^3} - \underbrace{1}_{X^{\pm}} - \underbrace{1}_{Y^{1,2}}- \underbrace{1}_{A^8}\right]=0\, .
\end{eqnarray}
A similar cancellation occurs for the second diagram in Eq.~(\ref{eq:diags2}), and so we are left with no quadratic divergences. This of course had to be true by the collective symmetry argument.
A notable feature of collective symmetry breaking that is evident here is that the cancellation happens between \textbf{same}-spin partners (in this case spin-1). This is in contrast with Supersymmetry, where the cancellation happens between \textbf{opposite}-spin partners. A similar cancellation happens in the fermion sector, as we will now see.

\subsection{The fermion sector}

We have seen how collective breaking can eliminate the quadratic correction to the Higgs potential arising from the gauge sector. However, the most dominant SM quadratic corrections to the Higgs potential come from the fermion sector, more specifically from the top quark due to its large Yukawa coupling. Carrying over the lesson we have learned form the gauge boson case, we have to introduce top partners in multiplets of the global symmetry. We will begin by introducing top partners in the $\left(\mathbf{3},\mathbf{1}\right)+\left(\mathbf{1},\mathbf{3}\right)$ of $SU(3)\times SU(3)$, from which we keep only the degrees of freedom in the diagonal ${SU(3)}_D$ part. Keeping only a part of the full multiplet constitutes an explicit breaking of $SU(3) \times SU(3) \rightarrow SU(3)_D$. This is in direct analogy to the gauge boson case, where only $SU(3)_D$ was gauged. The $\mathbf{3}$ of ${SU(3)}_D$ contains the SM $Q_L=\left(t,b\right)$ ${SU(2)}_L$ doublet plus an additional top partner $T$:
\begin{eqnarray}
\Psi = \colvec{t_L\\b_L\\T_L}\equiv \colvec{Q_L\\T_L}\, .
\end{eqnarray}
In addition, we have two right handed $t_{1,2}$ in the $\mathbf{1}$ of ${SU(3)}_D$.
The fermion sector of the $nl\sigma m$ Lagrangian is then
 \begin{eqnarray}\label{eq:ferm}
\mathcal{L}_{\text{top}} \, &=& \, \lambda_1 \bar{\Psi}\Sigma_1 t_1\,+\, \lambda_2 \bar{\Psi}\Sigma_2 t_2\, .
\end{eqnarray}
Generically $\lambda_1\neq\lambda_2$ but we will set them equal for simplicity, and the symmetries allow for mixing between $t_1, t_2$, $t_1 \Sigma_1$, \textit{etc}. but we can rotate these away with unitary transformations. We can check that this Lagrangian exhibits collective breaking in a very similar manner to Eq.~(\ref{eq:coll}). In the presence of both terms, the overall global symmetry is ${SU(3)}_D$. However, if one of the terms is turned off, the symmetry of the above Lagrangian is enhanced to $SU(3)\times SU(3)$ where $\Sigma_{1,2}$ can rotate differently. The radiative corrections contributing to the Higgs potential must involve both terms in Eq.~(\ref{eq:ferm}), and the quadratic divergences cancel. Let us verify this cancellation explicitly. Expanding the Lagrangian in $H$, we find:
 \begin{eqnarray}
\mathcal{L}_{\text{top}} \, &=& \, \frac{\lambda}{\sqrt{2}} \, \left[
\left(\bar{Q}_L \, ,\, \bar{T}_L\right)\colvec{iH\\ f-\frac{H^\dagger H}{2f}} t_1\,+\,\left(\bar{Q}_L\, ,\, \bar{T}_L\right)\colvec{-iH\\ f-\frac{H^\dagger H}{2f}} t_2\right]\, ,
\end{eqnarray}
or in the mass basis
 \begin{eqnarray}\label{eq:massb}
\mathcal{L}_{\text{top}} \, &=& \lambda \bar{Q}_L H t_R \, + \, \lambda f \left(1-\frac{H^\dagger H}{2f^2}\right)\bar{T}_L T_R\, ,
\end{eqnarray}
with $t_R=\frac{i}{\sqrt{2}}\left(t_1-t_2\right)~,~T_R=\frac{1}{\sqrt{2}}\left(t_1+t_2\right)$.
We get the SM top Yukawa and a heavy top partner $T$ of mass $\lambda f$. It is this top partner that cancels the top quadratic divergences to the Higgs potential.
This time, the cancellation is between two different diagrams:
\begin{equation}\label{eq:toploops}
\begin{tikzpicture}[baseline=(E.west)]
\begin{feynman}[]

\vertex (a1) {h};
\vertex[right=1cm of a1,dot] (a2){};
\vertex[right=1cm of a2,dot] (a3){};
\vertex[right=1cm of a3] (a4){h};
\vertex[right=1cm of a4] (a){h};
\vertex[right=1cm of a,dot] (b){};
\vertex[right=1cm of b] (c) {h};
\diagram* {
(a1) -- [scalar] (a2),
(a3) -- [scalar] (a4),
(a2) -- [half left, fermion] (a3),
(a3) -- [half left, fermion] (a2),
 (a) --[scalar] (b) [dot] --[scalar] (c),
};
\node[align=center,black,opacity=1] (B) at (3.5,0) {$+$};
\node[align=center,black,opacity=1] (C) at (1.85,0.7) {$t$};
\node[align=center,black,opacity=1] (E) at (0.8,0.2) {$\lambda$};
\node[align=center,black,opacity=1] (E) at (2.2,0.2) {$\lambda$};
\node[align=center,black,opacity=1] (D) at (5.6,0.9) {$T$};
\node[align=center,black,opacity=1] (E) at (5.0,1) {\Large $\times$};
\node[align=center,black,opacity=1] (E) at (5.0,1.3) {$\lambda f$};
\node[align=center,black,opacity=1] (E) at (4.9,-0.3) {$-\lambda/f$};
\path (b)--++(90:0.5) coordinate (A);
\draw (A) [fermion] circle(0.5);
\end{feynman}
\end{tikzpicture}
\end{equation}
We see that the quadratic divergences cancel out, leaving a finite piece of order $\frac{3\lambda}{16\pi^2}{\left(\lambda f\right)}^2$. By collective breaking, the leading divergence has to come from a diagram that involves both terms in Eq.~(\ref{eq:ferm}). As usual it is convenient to write a diagram for the $nl\sigma m$ fields which, when expanded, becomes a contribution to the Higgs mass. The result is
\begin{equation}\label{eq:toploopdiv}
\begin{tikzpicture}[baseline=(E.west)]
\begin{feynman}[]

\vertex (a) {$\Sigma_1$};
\vertex[right=1cm of a,dot] (b){};
\vertex[right=1cm of b] (c){};
\vertex[right=1cm of c,dot] (d){};
\vertex[right=1cm of d] (e){$\Sigma_2$};
\vertex[above=1cm of c,dot] (f){};
\vertex[above=1cm of f] (g){$\Sigma^\dagger_1$};
\vertex[below=1cm of c,dot] (h){};
\vertex[below=1cm of h] (i){$\Sigma^\dagger_2$};

\diagram* {
(a) -- [scalar] (b),
(d) -- [scalar] (e),
(f) -- [scalar] (g),
(h) -- [scalar] (i),
(b) -- [quarter left] (f)
(f) -- [quarter left, fermion] (d)
(d) -- [quarter left] (h)
(h) -- [quarter left, fermion] (b)
};
\node[align=center,black,opacity=1] (B) at (3.1,0.9) {$\Psi$};
\node[align=center,black,opacity=1] (B) at (0.9,-0.9) {$\Psi$};
\node[align=center,black,opacity=1] (B) at (0.9,0.9) {$t_1$};
\node[align=center,black,opacity=1] (B) at (3.1,-0.9) {$t_2$};
\node[align=center,black,opacity=1] (A) at (7,0) {$\propto~~~~ {\left|\Sigma^\dagger_1 \Sigma_2\right|}^2\,\frac{\lambda^4}{16\pi^2}\,\,\log\left[\frac{\Lambda^2}{\mu^2}\right]\, ,$};
\end{feynman}
\end{tikzpicture}
\end{equation}
which is only a log divergent contribution to the Higgs potential
 \begin{eqnarray}\label{eq:topdiv}
\frac{\lambda^4}{16\pi^2}\,\,\log\left[\frac{\Lambda^2}{\mu^2}\right]\, H^\dagger H\, .
\end{eqnarray}
This concludes our survey of the collective breaking and cancellation of quadratic divergences in the ``Simplest Little Higgs" \cite{Schmaltz:2004de}.

\subsection{Other versions of little Higgs models}
\subsubsection{The littlest Higgs}
The littlest Higgs \cite{Littlest} is an example of a model with collective symmetry breaking which is not based on a product group (such as $SU(3)\times SU(3)$ in the simplest little Higgs). In this model the Higgs is a pNGB in the coset $SU(5)/SO(5)$. A quick counting of generators gives
\begin{eqnarray}
N_{\text{NGB}}=\left(5^2-1\right)\, -\, \frac{5\left(5-1\right)}{2}\,=\,14 .
\end{eqnarray}
In contrast with the simplest little Higgs, where the $SU(3)\rightarrow SU(2)$ breaking was triggered by a fundamental, the $SU(5)\rightarrow SO(5)$ breaking is triggered by a VEV of the form
\begin{equation}
\Sigma_0 \, =\, f\, \colmatth{0 & 0 & \mathbb{1} \\
   0 &1&0  \\
 \mathbb{1} & 0 &  0
}\, ,
\label{eq:nlsmlh}
\end{equation}
where $\mathbb{1}$ are $2\times 2$ unit matrices ($\Sigma_0$ is a $5\times 5$ matrix).
To parameterize our pNGBs in an $SU(5)$ matrix, we note that as usual
\begin{equation}\label{eq:pionm}
\begin{tikzpicture}[baseline=(A.center)]

\node[align=center,black,opacity=1] (C) at (-6.1, 0) {$U_{\text{NGB}}\,\,\,\equiv \,\,\, \exp \left[\frac{2i}{f} \pi^{\hat{a}} T^{\hat{a}} \right] ~=~\exp \left[\frac{i}{f}\right.$};
\matrix (A) [matrix of math nodes, nodes = {node style ge},,column sep=0 mm,
left delimiter=(,right delimiter={)}]
{ &  &  & \phi_{++} & \phi_+ \\
 &  &  & \phi_+ & \phi_0 \\
 &  & ~~~ ~&  &  \\
\phi^*_{++} & \phi^*_{+} &  &  &  \\
\phi^*_{+} & \phi^*_{0} &  &  &  \\
};
\draw[-,black,opacity=0.9,line width=0.52 mm] (-2.3,-0.4)  to (2.3,-0.4) ;
\draw[-,black,opacity=0.9,line width=0.52 mm] (-2.3,0.4)  to (2.3,0.35) ;
\draw[-,black,opacity=0.9,line width=0.52 mm] (-0.4,-2.3)  to (-0.4,2.3) ;
\draw[-,black,opacity=0.9,line width=0.52 mm] (0.4,-2.3)  to (0.4,2.3) ;
\node[align=center,black,opacity=1] (B) at (0,1.3) {\Large $H$};
\node[align=center,black,opacity=1] at (0.05,-1.33) {\Large $H^*$};
\node[align=center,black,opacity=1] at (1.37,0) {\Large $H^T$};
\node[align=center,black,opacity=1] at (-1.35,0) {\Large $H^\dagger$};
\node[align=center,black,opacity=1] at (3.4,0) {$\left.\right]\,,$};
\end{tikzpicture}
\end{equation}
where for simplicity we only include the Higgs complex doublet and a complex triplet $\phi$. We will soon see that the other 4 NGBs are eaten by gauge bosons, so we omit them here. 
The $SU(5)/SO(5)$ coset is then parameterized by
\begin{eqnarray}
\Sigma\,=\, U_{\text{NGB}}\, \Sigma_0 \, U^\dagger_{\text{NGB}}\, .
\end{eqnarray}
In addition to the spontaneous $SU(5)/SO(5)$ breaking, we also break the $SU(5)$ explicitly by gauging an ${\left[SU(2)\times U(1)\right]}^2$ subgroup, comprising of the generators:
\begin{eqnarray}
Q^a_1 \, &=&\, \colmatth{\sigma^a/2 & 0 & 0 \\
   0 &0&0  \\
0 & 0 &  0
},~~~~~~~~~~~~~~
Q^a_2 \, =\, \colmatth{0 & 0 & 0 \\
   0 &0&0  \\
0 & 0 &  -\sigma^{a*}/2
}\, \nonumber\\
Y_1&=&\frac{1}{10}\, \text{diag}\left(3,3,-2,-2,-2\right)~,~
Y_2=\frac{1}{10}\, \text{diag}\left(2,2,2,-3,-3\right)\, .
\end{eqnarray}
This explicit breaking results in the NGBs becoming pNGBs. The $nl\sigma m\,$ Lagrangian is in this case
\begin{eqnarray}\label{eq:nlsmllh}
\mathcal{L}_{nl\sigma m}=\frac{1}{4}{\left|D_{\mu}\, \Sigma\right|}^2\, ,
\end{eqnarray}
where the covariant derivative for $\Sigma$ in the adjoint of the ${\left[SU(2)\times U(1)\right]}^2$ is given by
\begin{eqnarray}
D_{\mu}\,\Sigma\,=\, \plmu - ig_{1,2}\, W^{1,2;a}_{\mu}\, \left\{Q^a_{1,2},\Sigma\right\}\, -\, i g'_{1,2}\, B^{1,2}_{\mu}\, \left\{Y_{1,2},\Sigma\right\}\, ,
\end{eqnarray}
where $g_{1,2}, g'_{1,2}$ are the ${\left[SU(2)\times U(1)\right]}^2$ gauge couplings. 
Unsurprisingly, the spontaneous $SU(5)/SO(5)$ breaking results in the breaking of the ${\left[SU(2)\times U(1)\right]}^2$ to $SU(2)\times U(1)$, as can be seen by expanding Eq.~(\ref{eq:nlsmllh}). One combination of the ${\left[SU(2)\times U(1)\right]}^2$ gauge bosons ``eats" 4 of the 14 pNGBs and becomes massive, while the other remains massless as well as the SM $SU(2)\times U(1)$ gauge bosons. These combinations are given by
\begin{eqnarray}
\colvec{W_\mu \\ W'_\mu}=\colmatt{-\cos \alpha & \sin \alpha\\ -\sin \alpha & -\cos \alpha}\colvec{W^1_\mu \\W^2_\mu}~~,~~
\colvec{B_\mu \\ B'_\mu}=\colmatt{-\cos \alpha' & \sin \alpha'\\ -\sin \alpha' & -\cos \alpha'}\colvec{B^1_\mu \\ B^2_\mu}\, ,\nonumber\\
\end{eqnarray}
with $\tan \alpha = g_2/g_1$ and $\tan \alpha' = g'_2/g'_1$. This pattern of symmetry breaking leads to collective breaking, which can be seen as follows. When $g_2,g'_2\rightarrow 0$ in Eq.~(\ref{eq:nlsmllh}), the unbroken $SU(5)$ generators are the ones commuting with $Q^a_{1},Y_1$. These live in the lower-right corner of
\begin{equation}\label{eq:adj}
\begin{tikzpicture}[baseline=(A.center)]

\node[align=center,black,opacity=1] (C) at (-3, 0) {${\text{Adj}}_{SU(5)}~=$};
\matrix (A) [white,matrix of math nodes, nodes = {node style ge},row sep=0.63cm,column sep=0.65cm,
left delimiter=(,right delimiter={)}]
{ &  &  &  &  \\
 &  &  & & \\
 &  & &  &  \\
& &  &  &  \\
 &&  &  &  \\
};
\draw[-,black,opacity=0.9,line width=0.52 mm] (-1.4,0.1)  to (1.4,0.1) ;
\draw[-,black,opacity=0.9,line width=0.52 mm] (-0.24,-1.3)  to (-0.24,1.3) ;
\node[align=center,black,opacity=1] (C) at (-0.9, 1.1) {$SU(2)$};
\node[align=center,black,opacity=1] (C) at (-0.88, 0.8) {$\times$};
\node[align=center,black,opacity=1] (C) at (-0.85, 0.5) {$U(1)$};
\node[align=center,black,opacity=1] (C) at (0.6, -0.6) {$SU(3)$};
\node[align=center,black,opacity=1] (B) at (2,0) {,};
\end{tikzpicture}
\end{equation}
\vspace*{0.1cm}
and constitute an $SU(2)\times U(1)\times SU(3)$ global symmetry. The $SU(3)$ part of this symmetry protects the Higgs from corrections just like in the simplest little Higgs. When all $g_{1,2},g'_{1,2}$ are present, the unbroken $SU(5)$ generators are the ones commuting with $Q^a_{1,2},Y_{1,2}$, i.e. , ${\left[SU(2)\times U(1)\right]}^2$. There is no $SU(3)$ global symmetry, and the Higgs isn't protected. Consequently, we expect all the gauge boson contributions to the Higgs potential to depend both on $g_{1,2}$ and on ${g'}_{1,2}$. Indeed, expanding Eq.~(\ref{eq:nlsmllh}) in $H$ we get
 \begin{eqnarray}
\mathcal{L}_{nl\sigma m}=\frac{1}{4}\, H^\dagger H \, \left[ g_1g_2 W^1_\mu W^{2;\mu} + g'_1g'_2 B^1_\mu B^{2;\mu}\right] \, + \ldots
\end{eqnarray}
The important thing to notice is that there are only off diagonal couplings between $W^{1}_\mu$ and $W^{2}_\mu$ and similarly for $B^{1,2}_\mu$. This leads to the softening of the gauge contribution to the Higgs potential, since there's no way to close a loop with just a single gauge boson. In the mass eigenbasis this becomes:
 \begin{eqnarray}
\mathcal{L}_{nl\sigma m}=\frac{1}{4}\, H^\dagger H \, \left[ g^2\left( W_\mu W^{\mu} - W'_\mu W^{'\mu}-2\cot 2\alpha\, W'_\mu W^{\mu}\right)\,+\,\text{term for}\, B\right] \, + \ldots\nonumber\\
\end{eqnarray}
It is then easy to see that the quadratic divergences cancel
\begin{equation}\label{eq:LHc}
\begin{tikzpicture}[baseline=(AA.center)]
\begin{feynman}[]
\vertex (a){H};
\vertex[right=1cm of a,dot] (b){};
\vertex[right=1cm of b] (c) {H};
\vertex[right=2cm of c] (d){H};
\vertex[right=1cm of d,dot] (e){};
\vertex[right=1cm of e] (f) {H};
\diagram* {
 (a) --[scalar] (b) [dot] --[scalar] (c),
 (d) --[scalar] (e) [dot] --[scalar] (f),
};
\path (b)--++(90:0.5) coordinate (A);
\draw (A) [photon]circle(0.5);
\path (e)--++(90:0.5) coordinate (B);
\draw (B) [photon]circle(0.5);
\node[align=center,black,opacity=1] (AA) at (3,0.45) {+};
\node[align=center,black,opacity=1] at (1.6,1) {$W$};
\node[align=center,black,opacity=1] at (5.6,1) {$W'$};
\node[align=center,black,opacity=1] at (1,-0.3) {$g^2$};
\node[align=center,black,opacity=1] at (5,-0.3) {$-g^2$};
\node[align=center,black,opacity=1] at (7.2,0.45) {$=~~~0\, .$};
\end{feynman}
\end{tikzpicture}
\end{equation}
A similar cancellation happens for $B$ and $B'$. The leading divergence involves the mixed $W'_\mu W^\mu$ and $B'_\mu B^\mu$ terms in Eq.~(\ref{eq:LHc}), but these are only give rise to log divergences.

The fermion sector of the model is not very different from the simplest little Higgs. We embed the left handed doublet $Q_L$ in an $SU(3)$ triplet
\begin{eqnarray}
\bar{\Psi} = \left(\bar{t}_L\, ,\,\bar{b}_L\, , \,\bar{T}_L\right)\equiv \left(\bar{Q}_L\, , \,\bar{T}_L\right)\, .
\end{eqnarray}
In terms of this multiplet, the effective $SU(3)$ invariant Largrangian is:
 \begin{eqnarray}
\mathcal{L}_{\text{top}} \,= \, -\frac{\lambda_1 f}{2} \bar{\Psi}_i\, \epsilon_{ijk}\epsilon_{mn}\Sigma_{jm} \, \Sigma_{kn} \, t_1\,-\, \lambda_2 f\, \bar{T}_L\, t_2\,  ,
\end{eqnarray}
where $i,j,k\in\left[1,2,3\right]$ and $m,n\in\left[4,5\right]$. The resulting Higgs couplings and collective breaking is similar to the simplest little Higgs case.

The novel part in the littlest Higgs is the emergence of an $\mathcal{O}(1)$ quartic self coupling for the Higgs. This sounds impossible at first, since we know that the Higgs is a pNGB and its tree level potential should vanish, and the radiative corrections are one-loop suppressed. The solution to this conundrum is that the quartic is both radiatively generated but still $\mathcal{O}(1)$, which is known as a \textbf{collective quartic}. 
To see this, consider the quadratically divergent gauge boson contribution to the Coleman-Weinberg potential for $H$ and $\phi$. We know from collective breaking that this should not depend on $H$.
 \begin{eqnarray}
&&V^{\text{quad}}_{\text{CW}}\left(H,\phi\right) \,=\nonumber\\
&&\, a\, \frac{\Lambda^2}{16\pi^2}\,f^2\, \sum_{j=1,2} \,\left\{g^2_j\sum_a Tr\,\left[ Q^a_j\Sigma Q^{b*}_j\Sigma^* \right]+{g'}^2_j Tr\,\left[ Y_j\Sigma Y^{*}_j\Sigma^* \right]\right\}\, ,
\end{eqnarray}
with $a$ a model dependent $\mathcal{O}(1)$ coefficient. Expanding in $H$ and $\Phi$, we get:
 \begin{eqnarray}\label{eq:colCW}
V^{\text{quad}}_{\text{CW}}\left(H,\phi\right) \,&=&a f^2\, \left\{\left(g^2_1+{g'}^2_1\right){\left|\phi_{ij}+\frac{i}{4f}\left(H_iH_j+H_jH_i\right)\right|}^2+\right.\nonumber\\
 &&~~~~~\left.+\left({g}^2_2+{g'}^2_2\right){\left|\phi_{ij}-\frac{i}{4f}\left(H_iH_j+H_jH_i\right)\right|}^2\right\}\, ,
\end{eqnarray}
where we have used $\Lambda \sim 4\pi f$. As expected, the quadratic divergence cancels for $H$. However, there's still a quadratically divergent mass term for $\phi$ and also a quadratically divergent $H\phi\phi$ coupling. Below $M_{\phi}\sim af$, we can integrate $\phi$ out and get a quartic term for the Higgs:
  \begin{eqnarray}
\lambda \, = \, a \frac{\left(g^2_1+{g'}^2_1\right)\left(g^2_2+{g'}^2_2\right)}{g^2_1+{g'}^2_1+g^2_2+{g'}^2_2}\, .
\end{eqnarray}
The alert reader might notice two main qualities of the above quartic: 1) it is $\mathcal{O}(1)$ and 2) it is collective, in the sense that it is nonzero only when both terms exist in Eq.~(\ref{eq:colCW}). The fact that the quartic is $\mathcal{O}(g^2)$ might come as a surprise, after all we are used to quartic couplings arising at loop level. But notice that the quartic is due to a tree level $\phi$ exchange
\begin{equation}\label{eq:tphi}
\begin{tikzpicture}[baseline=(AA.center)]
\begin{feynman}[]
\vertex (a){H};
\vertex[below=1cm of a] (b){};
\vertex[below=1cm of b] (c) {H};
\vertex[right=1cm of b,dot] (d){};
\vertex[right=1.5cm of d,dot] (e){};
\vertex[right=1cm of e] (f) {};
\vertex[above=1cm of f] (g) {H};
\vertex[below=1cm of f] (h) {H};
\diagram* {
 (a) --[scalar] (d), 
 (c) --[scalar] (d),
 (d) --[scalar] (e),
 (e) --[scalar] (g),
 (e) --[scalar] (h), 
};

\node[align=center,black,opacity=1] (AA) at (1.75,-0.7) {$\Phi$};
\node[align=center,black,opacity=1] (BB) at (4,-1.3) {$,$};
\end{feynman}
\end{tikzpicture}
\end{equation}
where $M_{\phi}\sim f$ and $c_{\phi HH}\sim f$ is the coupling. The fermion sector in this model contributes a similar, $\mathcal{O}(g^2)$ collective quartic.

Summing up, the Higgs potential in the model is of the form
  \begin{eqnarray}
V(H) \, &=& \, \left[\,\underbrace{-\frac{3 y^2_t M^2_T}{8\pi^2}\log \left(\frac{\Lambda^2}{M^2_T}\right)}_{top}+
\underbrace{\frac{3}{64\pi^2}\left(3g^2 M^2_{W'}\log \left(\frac{\Lambda^2}{M^2_{W'}}\right)+{g'}^2 M^2_{B'}\log \left(\frac{\Lambda^2}{M^2_{B'}}\right)\right)}_{gauge}+\right.\nonumber\\
&&\left.\underbrace{\frac{\lambda}{16\pi^2}M^2_{\phi}\log \left(\frac{\Lambda^2}{M^2_{\phi}}\right)}_{scalar}
\,\right]\,{\left|H\right|}^2\,+\,\lambda {\left|H\right|}^4\, ,
\end{eqnarray}
where $M_T,~M_{W',B'}$ and $M_{\phi}$ are the masses of the top partners, gauge partners, and heavy pNGB, respectively. The scalar contribution comes from Higgs loops, and is cut at the $M_{\phi}$. For a rough estimate of the naturalness in this model, we can keep only the dominant contribution due to top loops. The potential is then roughly
  \begin{eqnarray}
V(H) \, &=& \, -\frac{g^2_{SM} M^2_T}{16\pi^2}\log \left(\frac{\Lambda^2}{M^2_T}\right)\,{\left|H\right|}^2\,+\,\mathcal{O}(1)\,g^2_{SM}\, {\left|H\right|}^4\, ,
\end{eqnarray}
where $g_{SM}$ represents a generic SM weak coupling. Minimizing the potential provides us with a natural VEV $v\sim\frac{M_T}{4\pi}=\mathcal{O}\left(100\,\text{GeV}\right)$. This is a beautiful example of a fully natural electroweak symmetry breaking model, where the separation between the scales $v$ and $f$ is automatic. Ironically this same mechanism leads also to a prediction for a rather heavy Higgs in this model: 
  \begin{eqnarray}
m_h =\sqrt{2\lambda}v\sim\sqrt{2}g_{SM}v\sim 200-300\, \text{GeV}.
\end{eqnarray}
The origin of the heaviness of the Higgs is the large tree-level Higgs quartic (which is exactly also the reason behind the fully natural EWSB potential). To obtain the experimentally measured $m_h=125\,\text{GeV}$, the parameters of the model have to be slightly tuned to reduce the quartic (but then one also has to further reduce the quadratic term to maintain the separation between $v$ and $f$).

\subsubsection{The Minimal Composite Higgs Model}

In the previous sections we've seen how little Higgs models predict top and gauge partners around the compositeness scale times their interaction strength $g_* f$. This is generically in tension with electroweak precision observables, e.g. S \& T parameters. The more constraining of the two, the T-parameter, is the experimental fact that:
  \begin{eqnarray}\label{eq:Tpar}
\rho\equiv\frac{M^2_W}{M^2_Z \cos^2 \theta_W}\approx 1\, ,
\end{eqnarray}
to within $1\%$.
In little Higgs models, the top and gauge partners generate radiative corrections to the gauge boson masses which violate this relation. In the absence of any protective symmetry, these unwanted corrections push the compositeness scale $f$ to the multi-TeV regime, making the model unnatural. The minimal Composite Higgs (MCH) model~\cite{MCH} (which was inspired by~\cite{CNP}) and related models~\cite{SILH} greatly reduce the tension with electroweak constraints. They do this by incorporating a global symmetry known as \textbf{custodial symmetry}.

Custodial symmetry is a way to protect the T-parameter from correction involving the top and gauge partners. The S-parameter is not protected, but it is also far less constraining than the T-parameter. To illustrate how custodial symmetry works, let us look at the SM Higgs sector:
  \begin{eqnarray}
\mathcal{L}_{H}\,=\,-\mu^2 {\left|H\right|}^2 + \lambda {\left|H\right|}^4\, .
\end{eqnarray}
Ignoring gauge symmetry and Yukawa couplings for the moment, we see that the Higgs potential is invariant under an $SU(2)_L\times SU(2)_R$ global symmetry, under which:
  \begin{eqnarray}
\left(i\tau_2 H^*,H\right)\,\rightarrow\, U_L\,\left(i\tau_2 H^*,H\right)\,U^{\dagger}_R\, ,
\end{eqnarray}
where $U_L \in SU(2)_L~,~U_R\in SU(2)_R$

This symmetry is unbroken even when the gauging of $SU(2)_L$ is taken into account. In fact, under the global $SU(2)_R$, the $W^{\pm}$ and $Z$ bosons transform as a triplet, which ensures that $M_W=M_Z$. This relation is modified at tree level due to the gauging of $U(1)_Y\in SU(2)_R$, which explicitly breaks $SU(2)_R$ and yields $\rho=1$ at tree level. The difference between the up-type and down-type Yukawa couplings also breaks $SU(2)_R$, but this breaking only leads to loop level corrections to Eq.~(\ref{eq:Tpar}).

This quick illustration of custodial symmetry in the standard model makes it clear how to protect the T-parameter in composite Higgs models: all we have to do is make sure that the new physics introduced respects $SU(2)_R$. This is exactly what happens in the MCH models. The global symmetry is these models is $SO(5)\times U(1)$, spontaneously broken to $SO(4)\times U(1)$, so that the Higgs is a pNGB in the coset $SO(5)/SO(4)$ (for other possible cosets, see \cite{Mrazek:2011iu,BCS}). The counting of broken generators is
\begin{eqnarray}
N_{\text{NGB}}=\frac{5\left(5-1\right)}{2}\, - \, \frac{4\left(4-1\right)}{2}\,=\, 4 \, ,
\end{eqnarray}
Exactly the right number of broken generators to make up the Higgs doublet $H$. The important thing to notice is that $SO(4) \cong SU(2)_L\times SU(2)_R$, so the Higgs sector in this type of model is custodially symmetric. Under $SU(2)_L\times SU(2)_R$, the Higgs transforms as a bi-doublet $\left(\mathbf{2},\mathbf{2}\right)$. The $nl\sigma m\,$ field is then
\begin{eqnarray}
\Sigma\,=\,e^{i\frac{\sqrt{2}}{f}\pi^{\hat{a}}T^{\hat{a}}}{\left(0,0,0,0,f\right)}^T\,=\,\frac{\sin \frac{h}{f}}{f}\,{\left(\,h_1,h_2,h_3,h_4,h\,\cot \frac{h}{f}\, \right)}^T\, ,
\end{eqnarray}
where $h\equiv\sqrt{h_a h^a}$.
The $nl\sigma m\,$ Lagrangian is as usual
\begin{eqnarray}
\mathcal{L}_{nl\sigma m}=\frac{1}{2}{\left(D_{\mu}\, \Sigma\right)}^{\dagger}\,\left(D^{\mu}\, \Sigma\right)\, ,
\end{eqnarray}
where $D_{\mu}$ is the gauged $SU(2)\times U(1)$ covariant derivative 
\begin{equation}
\begin{tikzpicture}[baseline=(C.east)]

\matrix (A) [opacity=0,matrix of math nodes, nodes = {node style ge},row sep=0.58cm,column sep=0.37cm,
left delimiter=(,right delimiter={)}] at (-0.06,0)
{ & & &  &\\
   &  &  & &\\
 &  &  & &\\
  &  &  & &\\
};
\matrix (B) [right=2.2cm of A,opacity=0,matrix of math nodes, nodes = {node style ge},row sep=0.58cm,column sep=0.38cm,
left delimiter=(,right delimiter={)}]
{  & & &  &\\
   &  &  & &\\
 &  &  & &\\
  &  &  & &\\
};
\node[align=center,black,opacity=1] (C) at (-3, 0) {$D_{\mu}\,=\, \plmu - ig\, W^a_\mu$};
\node[align=center,black,opacity=1] (C) at (1.95, 0) {$- ig' \, B_\mu$};
\node[align=center,black,opacity=1] (A) at (-0.65, 0.58) {$\tau^{a}/2$};
\node[align=center,black,opacity=1] (A) at (0.05, -0.1) {$\tau^{a}/2$};
\draw[-,black,opacity=0.9,line width=0.5 mm] (0.55,-0.9)  to (0.55,0.9) ;
\draw[-,black,opacity=0.9,line width=0.5 mm] (-0.95,-0.5)  to (0.85,-0.5) ;
\draw[-,black,opacity=0.9,line width=0.6 mm] (4.45,-0.9)  to (4.45,0.9) ;
\draw[-,black,opacity=0.9,line width=0.5 mm] (3,-0.5)  to (4.8,-0.5) ;
\node[align=center,black,opacity=1] (A) at (3.35, 0.58) {$\frac{1}{6}\cdot \mathbb{1}$};
\node[align=center,black,opacity=1] (A) at (3.95, -0.1) {$\frac{2}{3}\cdot \mathbb{1}$};
\node[align=center,black,opacity=1] (A) at (5.1, -0.05) {.};
\end{tikzpicture}
\end{equation}

Expanding the $nl\sigma m \,$ Lagrangian in the gauge fields, we get
\begin{eqnarray}\label{eq:gl}
&&\mathcal{L}_{\text{gauge}}~=\nonumber\\
&&\frac{f^2}{8}\sin^2 \frac{h}{f}\,\left(g'^2 B_\mu B_\nu + g^2 W^3_\mu W^2_\nu -2g g' W^3_\mu B_\nu +2g^2 W^+_\mu W^-_\nu \right)\,\left(\eta_{\mu \nu}-\frac{q^\mu q^\nu}{q^2}\right)\, .\nonumber\\
\end{eqnarray}
Setting $v\equiv f\sin \frac{\left<h\right>}{f}$ where $v = 246$ GeV and $\left<h\right>$ is the actual physical Higgs VEV, we get the right $W$ and $Z$ masses.

The appearance of the $\sin \frac{h}{f}$ might seem strange at first, but note that it is an essential part in the description of the Higgs as a pNGB in any coset, including the $SO(5)/SO(4)$ case of the MCH.\footnote{We could write the non-linear fields in the LH models using trigonometric functions as well, there we simply chose to expand those functions to lowest powers in $\frac{h}{f}$ to follow the literature.} We can think of the NGBs in a $G/H$ coset as just the set of rotation angles connecting different vacua that break $G$ but preserve $H$. In this way, pNGBs always enter the $nl\sigma m$ Lagrangian inside trigonometric functions. In fact, the $nl\sigma m\,$ fields in the little Higgs, Eq.~(\ref{eq:nlsmslh},\ref{eq:nlsmlh}) also depend on the pNGBS through trigonometric functions. All we did before was expand these functions to second order. 

A direct consequence of the fact the the Higgs only enters the Lagrangian through $\sin \frac{h}{f}$ is a modification of Higgs couplings with respect to their SM values. This is a general prediction in composite Higgs models. To see this, note the following expansion:
\begin{eqnarray}
f^2 \sin^2 \frac{h}{f} \,=\, f^2\left[\sin^2 \frac{\left<h\right>}{f}+\left(2\sin \frac{\left<h\right>}{f}\cos \frac{\left<h\right>}{f}\right)\, \frac{h}{f}+\left(1-2\sin^2 \frac{\left<h\right>}{f}\right)\,\frac{h^2}{f^2}+\ldots\right]\ ,\nonumber\\
\end{eqnarray}
 which by our definition of $v$ becomes
 \begin{eqnarray}
v^2 \,+\, 2v\sqrt{1-\xi}\, h \,+\, \left(1-2\xi\right)\,h^2\,+\,\ldots
\end{eqnarray}
with $\xi\equiv \frac{v^2}{f^2}$. Using this expansion in Eq.~(\ref{eq:gl}), we see that the Higgs-gauge boson couplings are modified:
 \begin{eqnarray}
g_{VVh}\,=\,g^{SM}_{VVh}\,\sqrt{1-\xi}~~,~~g_{VVhh}\,=\,g^{SM}_{VVhh}\,\left(1-2\xi\right)\, .
\end{eqnarray}

These couplings will be experimentally measured to within $10\%$ at the high-luminosity LHC and to within $1\%$ at the ILC, providing bounds on $v/f$. The current leading bound on $v/f$ comes from the $S$ parameter \cite{Agashe:2005dk,Barbieri:2007bh}: $f>3v$.

Another important thing to note, is the absence of gauge partners in the Lagrangian Eq.~(\ref{eq:gl}), which makes the Higgs potential quadratically divergent. However, the fact that the gauge partners our missing from Eq.~(\ref{eq:gl}) does not mean that they are absent in the model. In fact, Eq.~(\ref{eq:gl}) should be taken as the effective action \textit{below} the gauge partner mass. The gauge partners enter at a scale $\sim g_{SM} f$ in complete $SO(5)$ multiplets, and cut the quadratic divergences in a similar way to little Higgs models. One can write a collective symmetry breaking argument in this case based on a two- or three-site model \cite{Contino:2006nn,Panico:2011pw}.

The more general way of writing the effective Lagrangian Eq.~(\ref{eq:gl}) is to allow for generic momentum dependent form factors, to account for the effect of integrating out composite degrees of freedom below the compositeness scale $g^*f$. These form factors reflect the fact that the gauge bosons are partially composite---their nonlocal substructure is encoded in the form factors for momenta $\gtrsim f$. These are similar in spirit to the momentum dependent factors that arise in the chiral Lagrangian for pions below $\Lambda_{QCD}$. In terms of these the Lagrangian is
\begin{eqnarray}\label{eq:lagff}
&&\mathcal{L}_{\text{gauge}}~=~\frac{1}{2}P^{\mu \nu}_T\,\left[
\left(\Pi^X_0(q^2)+\Pi_0(q^2)+\frac{\sin^2 \frac{h}{f}}{4}\Pi_1(q^2)\right) B_\mu B_\nu+\right.\nonumber\\
&&\left. +\left(\Pi_0(q^2)+\frac{\sin^2 \frac{h}{f}}{4}\Pi_1(q^2)\right) W^a_\mu W^a_\nu +
2\sin^2 \frac{h}{f} \,\, \Pi_1(q^2)\,H^\dagger\, T^a_L \, Y\, H \, W^a_\mu B_\nu \right]\, .\nonumber\\
\end{eqnarray}
The formula for the Coleman-Weinberg potential in the presence of these momentum dependent form factors is given by
\begin{eqnarray}\label{eq:lagff}
V_{CW}(h)\,=\,\frac{9}{2}\int\,\frac{d^4q}{{\left(2\pi\right)}^4}\,\log \left[1+\frac{1}{4}\frac{\Pi_1(q^2)}{\Pi_0(q^2)}\sin^2 \frac{h}{f}\right]\, .
\end{eqnarray}
The composite substructure encoded in the form factors for $p\gtrsim f$ damps the integrand, making the integral UV finite. In the next section we will show how to calculate the form factors exactly in an equivalent five-dimensional setting. The UV finiteness will be clearer from that perspective.

\subsection{Partial Compositeness}
Previously, when discussing the fermion sector of the littlest Higgs, we contended to take the top to be in an $SU(3)$ triplet, even though the composite sector was invariant under a larger $SU(5)$ global symmetry. Clearly, our choice to include only a fraction of an $SU(5)$ multiplet is in need of some UV completion. In composite Higgs models, there is an easy way to do that, called \textbf{partial compositeness} \cite{Kaplan:1991dc,NimaMartin,YuvalMatthias,RSGIM}.

The idea behind partial compositeness is to separate the spontaneous $G/H$ breaking from the explicit breaking of $G$ due to partial gauging and incomplete fermion multiplets. In the partial compositeness picture there are two sectors: the composite sector and the elementary sector. 
In the \textbf{composite sector}, composite resonances come in complete $G$ multiplets. In the MCH, for example, the left handed tops can come in the $\mathcal{O}_L=\mathbf{5}$ of $SO(5)$, while the right handed top can be in the $\mathcal{O}_R=\mathbf{1}$ of $SO(5)$. These multiplets are split due to the spontaneous $SO(5)/SO(4)$ breaking.
In the \textbf{elementary sector}, states come in $SU(2)_L\times U(1)_Y$ representations, for example in $\Psi_L=\mathbf{2}_{\frac{1}{3}}$ and $\Psi_R=\mathbf{1}_{\frac{4}{3}}$, since the elementary sector does not know about the $SO(5)$ of the composite sector.

The Lagrangian of the model can then be written as \cite{Contino:2006nn}
\begin{eqnarray}\label{eq:partc}
\mathcal{L}_{\text{CH}}\,=\,\mathcal{L}_{\text{elementary}}\,+\,\mathcal{L}_{\text{composite}}\,+\,\mathcal{L}_{\text{mix}}\, .
\end{eqnarray}
In the equation above $\mathcal{L}_{\text{elementary}}$ is a Lagrangian involving the elementary fields $\Psi_L, \Psi_R$. These fields are charged under an $\left(SU(3)\right)\times SU(2)_L\times U(1)$ gauge symmetry. On the other hand, $\mathcal{L}_{\text{composite}}$ is a $nl\sigma m\,$ Lagrangian involving complete $SO(5)$ multiplets $\mathcal{O}_L~,~\mathcal{O}_{R}$ and the $nl\sigma m\,$ field $\Sigma$.
The third term in Eq.~(\ref{eq:partc}) is a linear mixing term
\begin{eqnarray}
\mathcal{L}_{\text{mix}}\,=\,f\,\bar{\Psi}_L\lambda_L\,\mathcal{O}_L\,+\,f\,\bar{\Psi}_R\lambda_R\,\mathcal{O}_R\, ,
\end{eqnarray}
where $\lambda_{LR}$ are spurions that break $SO(5)\times \left[SU(2)\times U(1)\right]$ to the diagonal $SU(2)\times U(1)$.
For every elementary state that couples to a composite state, there are two mass eigenstates. The heavy of the two is at the compositeness scale $f$, while the light one is simply the corresponding SM fermion. In this way the SM fermions are \textit{partially composite}. Heuristically, we can write the mass and Yukawa terms in Eq.~(\ref{eq:partc}) as follows:
\begin{eqnarray}\label{eq:mx}
\underbrace{f\,\bar{\Psi}_L\lambda_L\,\mathcal{O}_L\,+\,f\,\bar{\mathcal{O}}_R\lambda_R \Psi_R\,}_{\mathcal{L}_{\text{mix}}}+\underbrace{M_L\,\bar{\mathcal{O}}_L\,\mathcal{O}_L\,+\,M_R\,\bar{\mathcal{O}}_R\,\mathcal{O}_R+Y\,\bar{\mathcal{O}}_L\,H\,\mathcal{O}_R}_{\mathcal{L}_{\text{composite}}}\, .
\end{eqnarray}
Rotating to the mass basis, we have
\begin{eqnarray}
\colvec{\bar{\Psi}^{SM}_L\\ \bar{\Psi}^{H}_L}=\colmatt{1 & -f_L \\ f_L& 1 }\colvec{\bar{\Psi}_L\\ \bar{\mathcal{O}}_L}~,~
\colvec{\Psi^{SM}_R\\ \Psi^{H}_R}=\colmatt{1 & -f_R \\ f_R& 1 }\colvec{\Psi_R\\ \mathcal{O}_R}\, ,
\end{eqnarray}
with $f_L\sim\frac{\lambda_L f}{M_L}$ and $f_R\sim\frac{\lambda_R f}{M_R}$, and we assume $M_{L,R}\gg \lambda_{L,R}f$. Substituting back in Eq.~(\ref{eq:mx}), we get massless SM fermions $\bar{\Psi}^{SM}_L,\,\Psi^{SM}_R$ with a Yukawa coupling
\begin{eqnarray}
y \, \bar{\Psi}^{SM}_L \, H \,  \Psi_R\, ,
\end{eqnarray}
with $y=f_L \, Y \, f_R$. Generalizing this to include down-type fermions and three generations, we get:
\begin{eqnarray}
y^u_{ij} \,&=&\, f^q_{im}\,Y^u_{mn}\,f^{u}_{nj}\nonumber\\
y^d_{ij} \,&=&\, f^q_{im}\,Y^d_{mn}\,f^{d}_{nj}\, ,
\end{eqnarray}
where $f^q=\text{diag}\left(f^q_1\, ,\, f^q_2\, ,\, f^q_3\right)$ and similarly for $f^{u,d}$.
The attractive feature of partial compositeness is that it allows for flavor structure in the SM Yukawa matrices $y^u~,~y^d$ even when the original composite couplings $Y^u~,~Y^d$ are $\mathcal{O}(1)$. This is called \textbf{anarchic flavor}. To accommodate this possibility, the mixing matrices $f^{q,u,d}$ have to be hierarchical. This can happen, for example due to a large anomalous dimension:
\begin{eqnarray}
f^{q,u,d}_i\left(\Lambda_C\right)\sim f^{q,u,d}_i\, \, \left(\Lambda_F\right) {\left(\frac{\Lambda_C}{\Lambda_F}\right)}^{d^{q,u,d}_i-\frac{5}{2}}\, .
\end{eqnarray}

By approximately diagonalizing the SM Yukawa matrices $y^{u,d}$, we can infer the CKM structure for anarchic flavor from the $f^{q,u,d}$ hierarchy. Up to $\mathcal{O}(1)$ numbers, we have
\begin{eqnarray}
y^{u,d}\,=\,f^q\,Y^{u,d}\,f^{u,d}\,\equiv\, L_{u,d}\, y^{u,d}_{\text{diag}}\,R^\dagger_{u,d}\, ,
\end{eqnarray}
with
\begin{eqnarray}
y^{u,d}_{\text{diag}}&\sim& \text{diag}\left(f^q_1\,f^{u,d}_1\,,\,f^{q}_2\,f^{u,d}_2\,,\,f^q_3\,f^{u,d}_3\right)\nonumber\\
L^u_{ij}\sim L^d_{ij}&\sim& \text{min}\left(\frac{f^q_i}{f^q_j}\,,\,\frac{f^q_j}{f^q_i}\right)\nonumber\\
R^{u,d}_{ij}&\sim& \text{min}\left(\frac{f^{u,d}_i}{f^{u,d}_j}\,,\,\frac{f^{u,d}_j}{f^{u,d}_i}\right)\, .
\end{eqnarray}
If we set
\begin{eqnarray}
\frac{f^q_1}{f^q_2}\,\sim \lambda~,~\frac{f^q_2}{f^q_3}\,\sim \lambda^2~,~\frac{f^q_1}{f^q_3}\,\sim \lambda^3\, ,
\end{eqnarray}
where $\lambda\sim0.22$ is the Cabbibo angle, we get the phenomenologically viable structure for the CKM matrix and mass hierarchy
\begin{eqnarray}
V_{\text{CKM}}\,=\,L_u\,L^{\dagger}_d~,~m^{u,d}_i\,\sim\,f^q_i\,f^{u,d}_i\,v\, .
\end{eqnarray}
Note that to get an $\mathcal{O}(1)$ Yukawa coupling for the top, we need large mixing
\begin{eqnarray}
f^{q,u,d}_3\,=\,\frac{\lambda^{q,u,d}_3\, f}{\sqrt{{\left(\lambda^{q,u,d}_3\right)}^2+M^2_T}}\sim\, \mathcal{O}(1)\, ,
\end{eqnarray}
where this time we did not take the limit $M_T \gg \lambda f$. In composite models we have $M_T=g_\Psi\,f$, where $g_\Psi<4\pi$ is the (dimensionless) interaction strength among composite fermions. We see that to get a large mixing, we need $g_\Psi\ll 4\pi$, i.e. top partners that are much lighter than the cutoff of our $nl\sigma m$:
\begin{eqnarray}
M_T \ll \Lambda=4\pi f\, .
\end{eqnarray}
If we take the cutoff to be $10\,\text{TeV}$, we could have top partners as light as $1.5-2\,\text{TeV}$.

\subsection{RS-GIM Mechanism}
All composite and little Higgs models reviewed here predict composite gauge partners which cut the quadratic divergences to the Higgs potential. But the existence of these heavy gauge partners is severely constrained by the experimental bounds on flavor violation. For example, a composite $Z'$ could mediate tree level $\Delta F=2$ \textbf{flavor changing neutral currents} (FCNC) through the following s-channel exchange:
\begin{equation}
\begin{tikzpicture}[baseline=(AA.center)]
\begin{feynman}[]
\vertex (a){d};
\vertex[below=1cm of a] (b){};
\vertex[below=1cm of b] (c) {\(\bar{s}\)};
\vertex[right=1cm of b,dot] (d){};
\vertex[right=1.5cm of d,dot] (e){};
\vertex[right=1cm of e] (f) {};
\vertex[above=1cm of f] (g) {s};
\vertex[below=1cm of f] (h) {\(\bar{d}\)};
\vertex[right=2.8cm of f] (i){};
\vertex[above=0.5cm of i,dot] (j){};
\vertex[above=1cm of j] (k){};
\vertex[left=1cm of k] (l){d};
\vertex[right=1cm of k] (m){s};
\vertex[below=0.5cm of i,dot] (o){};
\vertex[below=1cm of o] (p){};
\vertex[left=1cm of p] (q){\(\bar{s}\)};
\vertex[right=1cm of p] (r){\(\bar{d}\)};
\diagram* {
 (a) --[fermion] (d), 
 (d) --[fermion] (c),
 (d) --[photon] (e),
 (e) --[fermion] (g),
 (h) --[fermion] (e), 
 (l) --[fermion] (j), 
 (j) --[fermion] (m),
 (j) --[photon] (o),
 (o) --[fermion] (q),
 (r) --[fermion] (o), 
};

\node[align=center,black,opacity=1] (AA) at (1.75,-0.7) {$Z'$};
\node[align=center,black,opacity=1] (CC) at (6.75,-0.9) {$Z'$};
\node[align=center,black,opacity=1] (BB) at (4.5,-1.0) {$+$};
\end{feynman}
\end{tikzpicture}
\end{equation}
At energies lower than the mass of the gauge partners, we can express the flavor violation through dimension 6 operators such as:
\begin{eqnarray}
C_{4K}~\,\left(\bar{s} \, \gamma^\mu\,d\right)\,\left(\bar{d}\,\gamma_\mu\,s\right)\, ,
\end{eqnarray}
where 
\begin{eqnarray}
C_{4K} \sim A_{4K}\,\frac{g^2_\rho}{M^2_\rho}\, ,
\end{eqnarray}
where $A_{4K}$ is a dimensionless coefficient and $m_\rho$ and $g_\rho\sim \mathcal{O}(1)$ are the mass and dimensionless coupling strength among the composite vectors.
The stringent constraints from $K-\bar{K}$ mixing and other flavor violating process severely constrain the coefficients $C_{4K}$ to be hierarchically small, for example $\left|\text{Re} \,C_{4K}\right|<3.6\cdot10^{-9}\,\text{TeV}$.
In generic models with heavy vectors and an anarchic flavor structure, we would expect $A_{4K}$ to be $\mathcal{O}(1)$, and so $m_\rho\gtrsim 3\cdot10^4\,\text{TeV}$. This is clearly a disaster for any LHC phenomenology, and also means $\mathcal{O}(10^5)$ tuning in the Higgs potential.
In composite Higgs models the situation is different, as the coupling to the composite vector resonances also involve the mixing parameters $f^{q,u,d}$. In fact, this creates an approximate alignment between the Yukawa matrices and the couplings to the composite vectors. Consequently:
\begin{eqnarray}
A_{4K}\sim f^{d\dagger}_1f^{q}_2f^{q\dagger}_1f^{d}_2\sim\mathcal{O}(10^{-4})\, ,
\end{eqnarray}
and so composite Higgs models with anarchic flavor can accommodate composite vectors at $\sim20\,\text{TeV}$ \cite{Csaki:2008zd}. The same alignment comes into play with all other $\Delta F=1$ and $\Delta F=2$ flavor bounds. This mechanism is called RS-GIM \cite{RSGIM}, as it was first discovered in the context of Randall-Sundrum models \cite{RS} and it suppresses FCNCs similar to the GIM mechanism in the SM.

\subsection{More About the MCH Model}
The minimal composite Higgs model \cite{MCH,Contino:2006qr}
 is a specific implementation of the composite Higgs idea with a global symmetry $SO(5)\times U(1)_X$, broken to $SO(4)\times U(1)_X$ at a scale $f$. The pattern of symmetry breaking uniquely determines the gauge partner sector of the model.
In the fermion sector, however, there are many different choices of $SO(5)$ representations to use for partial compositeness. On possibility is:
\begin{eqnarray}
\mathcal{L}_{\text{top}}\,=\,\lambda_q\,\bar{q}_L\,\mathcal{O}_q\,+\,\lambda_u\,\bar{u}_R\,\mathcal{O}_u\,+\,\lambda_d\,\bar{d}_R\,\mathcal{O}_d\, ,
\end{eqnarray}
with $\mathcal{O}_q\,,\,\mathcal{O}_u$ and $\mathcal{O}_d$ in the $\mathbf{5}_{-\frac{2}{3}}\,,\,\mathbf{5}_{-\frac{2}{3}}$ and $\mathbf{10}_{-\frac{2}{3}}$ of $SO(5)\times U(1)_X$. Under the low energy $SU(2)_L\times SU(2)_R$ these multiplets decompose as:
\begin{eqnarray}
\mathbf{5}\,\rightarrow \, \left(\mathbf{2},\mathbf{2}\right)\,+\, \left(\mathbf{1},\mathbf{1}\right)~,~
\mathbf{10}\,\rightarrow \, \left(\mathbf{2},\mathbf{2}\right)\,+\, \left(\mathbf{1},\mathbf{3}\right)\,+\, \left(\mathbf{3},\mathbf{1}\right)\, .
\end{eqnarray}
These include the SM $q_L\,,\,t_R\,,b_R$, as well as other top and bottom partners with masses $\sim g_\Psi f$.
Other options for the composite multiplets are discussed in \cite{MCH,Contino:2006qr,Csaki:2008zd,Panico:2012uw,Pappadopulo:2013vca}. We know from collective symmetry breaking that the combined contribution of all of these to the Higgs potential is free from quadratic divergences. In fact, we will soon encounter an extra-dimensional realization of composite Higgs in which even the log divergences cancel out, leaving a Higgs potential which is strictly \textit{finite}.

An additional thing to note is that the $SO(5)/SO(4)$ coset contains only the Higgs and no additional pNGBs. Consequently, there cannot be a collective quartic in the MCH. The Higgs quartic only arises at loop level due to interactions with the SM top and gauge bosons. This has important phenomenological implications that we will see momentarily. 

\subsection{The Higgs Potential in the Littlest Higgs vs. the MCH}
The typical Higgs potential has the form
 \begin{eqnarray}
V(H)\,\sim\,-\mu^2\,{\left|H\right|}^2\,+\,\lambda\,{\left|H\right|}^4\, .
\end{eqnarray}
In both the littlest Higgs and the MCH the quadratic and quartic couplings are loop-induced. The quadratic term in both models scales like
 \begin{eqnarray}
\mu^2 \, \sim\, g^2_{SM}f^2\,\frac{g^2_\psi}{16\pi^2}\, .
\end{eqnarray}
The quartics the two models are different:
 \begin{eqnarray}
\lambda_{LH}\sim g^2_{SM}~~,~~\lambda_{MCH}\sim g^2_{SM}\,\frac{g^2_\psi}{16\pi^2}\, .
\end{eqnarray}
This is a major difference between these two models. 
In the Littlest Higgs the VEV is suppressed with respect to the compositeness scale $v\sim\frac{f}{4\pi}$ and the Higgs mass is naturally heavy $m_h\sim\sqrt{2}g_{SM}v$.
In the MCH, the Higgs mass is naturally light $m_h\sim\frac{g_\psi}{4\pi}\sqrt{2}g_{SM}v$, but the VEV is naturally $v=f$, which is unacceptable since electroweak precision constraints demand $f>3v$. By playing with the dimensionless parameters of the model we can always get $f>3v$, but this comes at the cost of an $\frac{v^2}{f^2}$ tuning. 

In the next section we will study a concrete, calculable realization of composite Higgs as a warped 5D model\cite{MCH,CNP} . This realization draws inspiration from AdS/CFT, but is not necessarily based on it.

\section{Extra dimensions\label{extradimensions}}

Geometries with extra compact dimensions provide a calculable framework for studying BSM physics, including many of the ideas we have already discussed in a 4D context. As we will see, the geometry of the extra dimension can be responsible for solving the hierarchy problem. Alternatively, we could take another point of view in which the extra dimension is a tool which allows us to perform weakly coupled calculations that are dual to a 4D strongly coupled field theory as a consequence of holography and AdS/CFT duality. In this section, we will develop the necessary machinery to do calculations in extra dimensions and review some of the most interesting results. For complementary introductions to extra dimensions see~\cite{Csaki:2004ay,Csaki:2005vy,Sundrum:2005jf,Gherghetta:2006ha,Rattazzi:2003ea,Serone:2009kf,Cheng:2010pt,TonyTASI,Ponton:2012bi,CT}

Throughout the rest of this paper, we will consider $d$ extra compact spatial dimensions such that the total number of spacetime dimensions is $D = 4 + d$ with the $(+,-,\ldots,-)$ signature. We will use Roman letters, $\textit{e.g.}$ $M$, $N$, to enumerate the full $D$-dimensional spacetime indices. Greek letters will be used to denote ordinary $4D$ spacetime coordinates. The spacetime interval is given by
\begin{equation}
	ds^2 = g_{MN} dx^M dx^N 
\end{equation}
where in a flat spacetime background the metric can be written as \vspace*{0.05cm}
\begin{equation}
g_{MN} =
\begin{pmatrix}
	1 & & & & &\\
	& -1 & & & & \\
	& & -1 & & & \\
	& & & -1 & & \\
	& & & & \ddots & \\
	& & & & & -1
\end{pmatrix}.
\end{equation}
\vspace*{0.3cm}
For now we will focus on flat backgrounds, although later warped gravitation backgrounds will play an important role in addressing the hierarchy problem.

\subsection{KK Decomposition}

As a first step, we must review how to construct a 4D effective theory from a fundamental Lagrangian with extra dimensions. \textbf{Kaluza-Klein (KK) decomposition}~\cite{KK}, which is essentially a normal mode expansion, converts a $(4+d)$-dimensional Lagrangian into a 4D Lagrangian with an infinite spectrum of 4D particles. To perform the dimensional reduction, we must integrate out the extra dimensions by putting the bulk part of the fields on their equation of motion (EOM) and then integrate over the $d$ extra dimensions.  

As a concrete example, let's focus on the case of a free real scalar with one extra dimension ($d=1$) compactified on a circle of radius $R$. For simplicity, we take the scalar potential to be absent. The action takes the form
\begin{align}
	S &= \int d^4x dy ~\frac{1}{2} \partial_M \phi(x,y) \partial^M \phi(x,y) \\
	&=  \int d^4x dy ~\frac{1}{2}\left[ (\partial_\mu \phi)^2  - (\partial_y \phi)^2\right]
	\label{eq:freescalar}
\end{align}
with $M = 0,1,2,3,5$ and $x_5 = y$. Variation of this action leads to the bulk EOM for $\phi$
\begin{equation}
	\partial_\mu^2 \phi - \partial_y^2 \phi = 0
\end{equation}
which, given the factorizable geometry and the periodic boundary conditions for $\phi$, is separable and admits a periodic solution of the form $\phi(x,y) = \frac{1}{\sqrt{2\pi R}}\sum_{n=-\infty}^\infty \phi^{(n)}(x) e^{i \frac{n}{R} y}$ with $\phi^{(n)*} = \phi^{(-n)}$ in order to guarantee $\phi$ is real. Substitution of this ansatz back into Eq.~\ref{eq:freescalar} and use of the orthogonality relations for Fourier modes yields the effective action 
\begin{equation}
	S = \int d^4x \sum_{n > 0} ~\partial_\mu \phi^{(n)\dagger} \partial^\mu \phi^{(n)} - \frac{n^2}{R^2}|\phi^{(n)}|^2.
\end{equation}

The main point is that in the 4D effective theory each 5D field corresponds to a KK tower of particles with masses $m_n = n/R$. The momentum along the compact direction is quantized by the boundary conditions, and its spectrum appears as a 4D tower of particles. The generalization to more dimensions ($\textit{e.g.}$ a torus) is simple:
\begin{equation}
	m^2_{n_5,n_6\ldots} = m_0^2 + \frac{n_5^2}{R_5^2} + \frac{n_6^2}{R_6^2} + \ldots
\end{equation}
where $m_0^2$ would arise if we had included a 5D mass term for the scalar. 

\subsection{Gauge Fields in Extra Dimensions}

Now we wish to study theories with gauge fields which propagate in the extra dimension. We will focus on an abelian gauge theory, although the generalization to non-abelian theories is straightforward. Gauge fields must still be periodic in $y$, so we can apply KK decomposition to the gauge sector
\begin{equation}
A_M(x,y) = \frac{1}{\sqrt{2\pi R}} \sum_{n= -\infty}^{\infty} A_M^n e^{i \frac{n}{R} y }
\end{equation}
with the one complication being that the 5D vector decomposes as a 4D vector $A_\mu$ and a 4D scalar $A_5$ under 4D Poincare transformations:
\begin{equation}
\begin{tikzpicture}[baseline=(C.east)]

\node[align=center,black,opacity=1] (C) at (-1.5, 0.1) {$A_M  \ = $};
\matrix (A) [opacity=0,matrix of math nodes, nodes = {node style ge},,column sep=0 mm,
left delimiter=(,right delimiter={)}]
{ a \\
  a \\
};
\node[align=center,black,opacity=1] (A) at (0., 0.3) {$A_\mu$};
\draw[-,black,opacity=0.9,line width=0.52 mm] (-0.3,-0.1)  to (0.3,-0.1) ;
\node[align=center,black,opacity=1] (A) at (0., -0.5) {$A_5$};
\end{tikzpicture}.
\end{equation}

After KK expansion, the action contains the following quadratic part
\begin{align}
S_{\text{gauge}} &= \int d^4x dy ~-\frac{1}{4} F_{MN}F^{MN} \\
&=  \int d^4x ~\sum_n \left[ -\frac{1}{4}F_{\mu\nu}^{(-n)} F^{(n)\mu\nu} 
+ \frac{1}{2} \left( \partial_\mu A_5^{(-n)} -\partial_5 A_\mu^{(-n)} \right)
\left( \partial^\mu A_5^{(n)} - \partial_5 A^{(n)\mu} \right)
\right]
\end{align}
where $F_{MN} = \partial_M A_N - \partial_N A_M$. In the last line, there is mixing between $A_\mu^{(n)}$ and $A_5^{(n)}$ (absent for $n=0$ since $A^{(0)}_\mu$ is flat) which suggests that $A_5^{(n)}$ is eaten by $A_\mu^{(n)}$ in order for the KK vectors to become massive. In fact, the mixing can be removed by moving to the 5D axial gauge (defined by the gauge transformation parameter $\alpha(z) = - \int A_5 dy$):
\begin{equation}
	A_\mu^{(n)} \rightarrow A_\mu^{(n)} - \frac{i}{n/R} \partial_\mu A_5^{(n)} \hspace{1cm}
	A_5^{(n)} \rightarrow 0
\end{equation}
which for $n \neq 0$ removes $A_5^{(n \neq 0)}$ from the action. 

The 4D effective action becomes
\begin{align}
S = \int d^4 x \left[ -\frac{1}{4} (F_{\mu\nu}^{(0)})^2 + \frac{1}{2} (\partial_\mu A_5^{(0)})^2
+ \sum_{n \geq 1} 2\left( -\frac{1}{4} F_{\mu\nu}^{(-n)} F^{(n)\mu\nu} + \frac{1}{2} \frac{n^2}{R^2} A_\mu^{(-n)} A^{(n)\mu}
 \right)
 \right]
\end{align}
which contains one massless zero-mode gauge boson, one zero mode $A_5^{(0)}$ scalar, and a tower of massive $A_\mu^{(n\neq 0)}$. The KK excitations of the $A_5$ scalar are unphysical: they were eaten so that $A_\mu^{(n\neq 0)}$ could become massive. 

For a $(4+d)$-dimensional theory, the 5D vector decomposes as again a 4D vector $A_\mu$ with additional scalars $A_5,~\ldots,~A_{4+d}$:
\begin{equation}
\begin{tikzpicture}[baseline=(C.east)]

\node[align=center,black,opacity=1] (C) at (-1.5, 0.1) {$A_M \ = $};
\matrix (A) [opacity=0,matrix of math nodes, nodes = {node style ge},,column sep=0 mm,
left delimiter=(,right delimiter={)}]
{ a \\
  a \\
  a \\
};
\node[align=center,black,opacity=1] (A) at (0., 0.65) {$A_\mu$};
\draw[-,black,opacity=0.9,line width=0.52 mm] (-0.4,0.3)  to (0.4,0.3) ;
\node[align=center,black,opacity=1] (A) at (0., -0.1) {$A_5$};
\node[align=center,black,opacity=1] (A) at (0., -0.45) {$\vdots$};
\node[align=center,black,opacity=1] (A) at (0., -0.8) {$A_{4+d}$};
\end{tikzpicture}
\end{equation}
In this case, one linear combination of the $A_5,~\ldots,~A_{4+d}$ towers is eaten by the KK tower of $A_\mu$, while the remaining combinations lead to $(d-1)$ scalar towers.

We can also consider the degrees of freedom arising from the higher-dimensional graviton. The metric is a $D\times D$ symmetric tensor with $D(D+1)/2$ independent components. However general relativity has $D$ dimensional general coordinate invariance, requiring $2D$ conditions to fix the the gauge. Thus the $(4+d)$-dimensional graviton has $D(D-3)/2 = (d+4)(d+1)/2$ physical degrees of freedom. Therefore, the graviton will contain additional DOFs in addition to the ordinary 4D graviton with 2 helicity states. 

The massless 4D graviton and its KK modes live in the upper 4 by 4 block of the metric tensor. 
\begin{equation}
\begin{tikzpicture}[baseline=(C.east)]

\node[align=center,black,opacity=1] (C) at (-2.2, 0.1) {$G_{MN} = $};
\matrix (A) [opacity=0,matrix of math nodes, nodes = {node style ge},,column sep=0 mm,
left delimiter=(,right delimiter={)}]
{ a & a & a \\
  a & a & a \\
};
\node[align=center,black,opacity=1] (A) at (-0.4, 0.4) {$g_{\mu\nu}$};
\draw[-,black,opacity=0.9,line width=0.52 mm] (-1.,0.)  to (1.,0.) ;
\draw[-,black,opacity=0.9,line width=0.52 mm] (.3,-.75)  to (.3,.75) ;
\node[align=center,black,opacity=1] (A) at (0.75, 0.4) {$A_{\mu j}$};
\node[align=center,black,opacity=1] (A) at (0.75, -0.5) {$\phi_{i j}$};
\end{tikzpicture}
\end{equation}
The tensor additionally contains $d$ vectors $A_{\mu j}$ and $d(d+1)/2$ scalars $\phi_{ij}$. However, in order for the KK tower of gravitons to become massive (5 helicity states), they must eat one gauge boson tower and one scalar tower ($5 = 2 + 2 + 1$).  This leaves us with $d-1$ gauge boson towers which each must eat a scalar KK tower to become massive. Finally, we are left over with $d(d-1)/2$ uneaten scalar towers.

\subsection{Matching of 5D and 4D Couplings}

The dimensions of fields and couplings depend on the number of spacetime dimensions. The simplest example is a scalar in $(4+d)$-dimensions. Focusing on the kinetic term, we can learn the classical dimension of the field from the requirement that the action be dimensionless.
\begin{equation}
	\int d^{4+d}x  (\partial \phi)^2 \ \  \Rightarrow \ \ [\phi] = 1 + \frac{d}{2}
\end{equation}
The brackets are used to express the energy dimension of the argument in natural units. Similarly from the fermion kinetic term one finds
\begin{equation}
	[\psi] = \frac{3}{2} + \frac{d}{2}
\end{equation}

How does one recover the canonical 4D dimension (1 for scalars, 3/2 for fermions), which the KK modes must have as 4D particles? The KK modes generally have additional dimensionful prefactors since the profile along the extra dimensions is dimensionful. More generally the energy dimension of higher dimensional couplings and their 4D effective couplings are mismatched, and the integration of the profiles along the extra directions in any given interaction vertex will make up for the mismatch and provide the relation between the $(4+d)$-dimensional couplings and the effective 4D ones. 

As an example, let's consider the bulk gauge interactions which arise from the covariant derivative
\begin{equation}
	D_M = \partial_M - i g_{(d)} A_M
\end{equation}
Since $[\partial_M] = 1$ and $[A_M] = 1 +\frac{d}{2}$, the bulk gauge coupling $g_{(d)}$ must be dimensionful, $[g_{(d)}] = - d /2$. However, the 4D effective gauge coupling must be dimensionless, so the zero mode profile must absorb the dimensionality. Specifying to $d=1$ compactified on a circle or radius $R$ and plugging in the zero mode expression whose profile in the extra dimension is flat
\begin{equation}
	A_\mu(x,y) = \frac{1}{\sqrt{2 \pi R}} A_\mu^0(x) + \ldots,
\end{equation}
we find that the effective coupling of the zero mode gauge boson is given by
\begin{equation}
	g_4 = \frac{g_5}{\sqrt{2\pi R}}
\end{equation}
where we identify the length of the extra dimension to be $L = 2 \pi R$. Here we see that the 4D coupling $g_4$ does indeed come out dimensionless. This relationship generalizes to geometries with more (flat) dimensions to 
\begin{equation}
	g_4^2 = \frac{g_{(4+d)}^2}{Vol_{(d)}}
\end{equation}
where $Vol_{(d)}$ is the volume of the $d$ compact extra dimensions.

Likewise, we can perform the matching of the gravitational coupling. Let the fundamental (higher dimensional) Planck scale be $M_{(4+d)}$. This is the energy scale where gravity becomes strongly interacting and requires UV completion. The Ricci tensor carries dimension two ($[R_{MN}] = 2$) in any dimension, so the action must have a prefactor of $M^{2+d}_{(4+d)}$ in order to be dimensionless. The higher dimension Einstein-Hilbert action takes the form 
\begin{align}
	S_{(4+d)} &= - M^{2+d}_{(4+d)} \int d^{4+d}x~ \sqrt{g_{(4+d)}} ~R_{(4+d)} \\
	&= - M^{2+d}_{(4+d)} Vol_{(d)} \int d^4 x ~\sqrt{g_{(4)}} ~R_{(4)} + \ldots
\end{align}
where in the second line we integrated over the compact dimensions and used the relation 
\begin{equation}\label{eq:rel}
	R_{(4+d)} = R_{(4)}
\end{equation}
for flat extra dimensions, valid to linear order  (see \cite{Csaki:2004ay} for more details). 

Matching onto the 4D action we obtain
\begin{equation}
	M_{\text{pl}}^2 = M^{2+d}_{(4+d)} Vol_{(d)}
\end{equation}
where $M_{\text{pl}}$ is the effective 4D Planck scale. This result holds only in scenarios with a flat gravitational background. In a warped background, the relation Eq.~(\ref{eq:rel}) no longer holds and the integral over the extra dimensions no longer has the interpretation of a volume.  

We can now describe the traditional (pre-branes) flat extra dimension scenario. If we assume one extra dimension and that all SM fields propagate in the bulk, then all gauge couplings and the gravitational coupling are set by a single scale, the radius of the extra dimension, in a natural theory. If we take $M_* \equiv M_{(4+d)}$ as the fundamental scale, then the higher dimensional gauge coupling is related by $g_{(4+d)} \sim M_{*}^{-d/2}$. Matching the 4D gauge couplings predicts for the radius of the extra dimension
\begin{equation}
	R \sim \frac{1}{M_{\text{Pl}}} g_{(4)}^{1 + \frac{2}{d}}.
\end{equation}
The size of the extra dimension is forced to be roughly Planck-length in order to simultaneously match the correct gravity coupling and gauge couplings of the SM. All of the KK modes would be near the Planck scale in mass, which is much too high to observe their effects at a collider. Moreover, the fundamental scale which acts as a cutoff of the SM effective field theory would be large since $M_* \sim 1 / R \sim M_{\text{Pl}}$ in a natural theory. Therefore, this framework can not solve the hierarchy problem by lowering the SM cutoff to the TeV scale.  

\subsection{Branes and Large Extra Dimensions}

A major breakthrough in the development of modern extra dimensional theories was the introduction of \textbf{branes}. Branes are hypersurfaces, $(3+1)$-dimensional in our case, with localized energy-momentum which can trap fields on their surfaces\footnote{We want to draw a distinction between our strictly phenomenological definition of ``branes'' and more formal definitions like D-branes. We will not care about the microphysics underlying the existence of branes, which may or may not be String Theory.}. The existence of $(3+1)$-dimensional branes in an extra dimensional theory means that some fields can propagate only on the brane, while others are free to propagate in the bulk of the extra-dimension. The brane-localized degrees of freedom are inherently four-dimensional, and their gauge couplings for example are decoupled from the size of the extra dimension. The introduction of branes allows one to separate gravity from SM gauge interactions allowing the size of the extra dimension to be much larger than we previously estimated.  
\begin{figure}
\begin{center}
\includegraphics[width=0.5\textwidth]{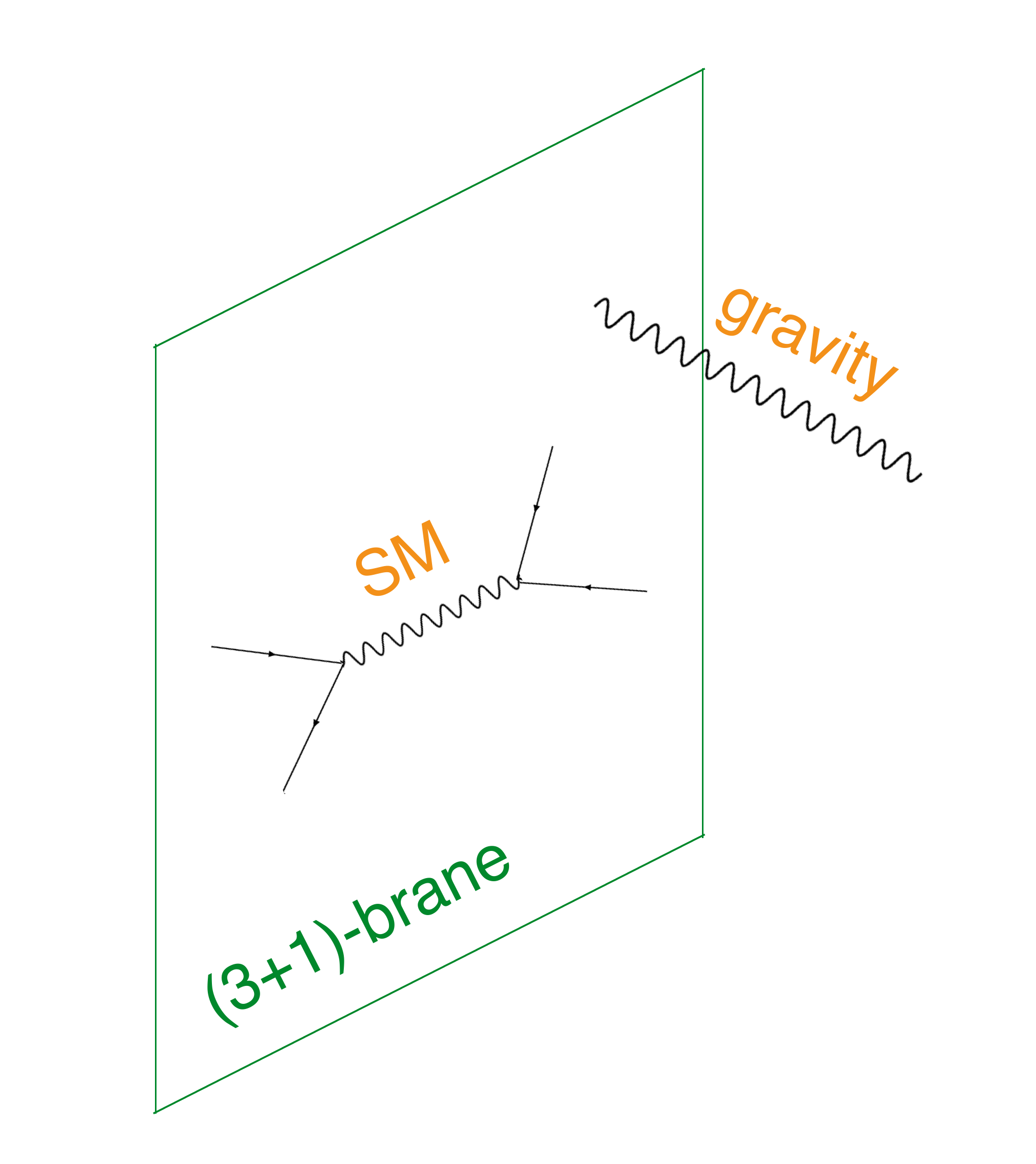}
\end{center}
\caption{The large extra dimensions scenario: SM fields are trapped and only propagate on a lower-dimensional $(3+1)$-brane, while the graviton propagates into the bulk. \label{fig:ADD}}
\end{figure}

A simple implementation of branes in an extra dimensional scenario is \textbf{Large Extra Dimensions}~\cite{ADD}, proposed by Arkani-Hamed, Dimopoulos, and Dvali (ADD). Imagine that the SM fields are trapped on a (3+1)-brane in a larger dimension bulk, but the graviton freely propagates in all the extra dimensions as shown in Fig.~\ref{fig:ADD}. The fundamental scale of the higher dimensional gravity $M_*$ is related to the 4D Planck scale by a dilution factor arising from the volume of the extra dimensions. As a consequence of the dilution, 4D gravity appears much weaker than one would have naively expected given that $M_* \ll M_{\text{Pl}}$. The fundamental scale $M_*$ acts as the cutoff for SM calculations\footnote{In String Theory, the Planck scale serves as a cutoff for its low energy EFTs, including the SM. There are daring ideas in which this is not necessarily true, like \cite{Salvio:2014soa}.}, so lowering $M_* \sim~\text{TeV}$ has the potential to eliminate the hierarchy problem. 

Can the cutoff be pushed this low? Unlike the previous scenario where all fields propagated in the bulk, here we only have to perform the matching for the graviton since the gauge fields which are brane-localized. Assuming that all of the extra dimension have similar radii, we get
\begin{equation}
	R_i~\sim~R~ =~ \frac{1}{M_*} \left( \frac{M_{\text{Pl}}}{M_*} \right)^{\frac{2}{d}}\, .
\end{equation}
Since we haven't seen strongly coupled gravity at colliders, we require $M_* \gtrsim ~\text{TeV}$ which leads to
\begin{equation}
R \lesssim \left( \frac{1}{1~\text{TeV}} \right) 10^{\frac{32}{d}}\sim 10^{\frac{32}{d}-17}\,\text{cm}\,	.
\end{equation}
Thus we see the size of the extra dimensions can actually be large, even macroscopic. 

However, this leads to $\mathcal{O}(1)$ deviations from $1/r^2$ Newtonian gravity on length scales smaller than $R$, due to the fact that gravity is actually propagating in more dimensions. Experiments which test Einstein/Newtonian gravity thus provide a bound on these models. 

How many large extra dimensions do we need?
\begin{itemize}
	\item $d=1$: One extra dimension leads to a radius $R \sim 10^{15}$ cm about the size of the solar system, which is very much ruled out.
	\item $d=2$: Two extra dimensions predict $R\sim 0.1$ cm, already ruled about by Cavendish-type experiments for $M_* = 1$ TeV.
	\item $d\geq 3$: Three extra dimensions bring us to $R<10^{-6}$, which is sufficient to evade experimental constraints. 
\end{itemize}
Some of the most sensitive experimental results come from the E\"ot-Wash experiment\footnote{A pun on the name of Lor\'and E\"otv\"os, who pioneered the experimental validation of the equivalence principle, and the University of Washington where the experiment is conducted.} which probes $\mathcal{O}(10^{-3})$ deviations from Newtonian gravity down to distances of $10^{-3}$ cm. For $d=2$, the bound is $R < 37$ $\mu$m which requires $M_* > 1.4$ TeV. A different bound on these models comes from the prediction of KK gravitons with a mass gap set by $m_{KK} \sim 1 / R$, which can result in stringent cosmological bounds. For example, the bound on KK gravitons from their effect on supernova cooling implies $M_* > 10-100$ TeV for $d=2$.

In ADD, hierarchy between the weak scale and the Planck scale is just a mirage because gravity appears weak on large length scales, but it is only weak from having to propagate in very large extra dimensions. However, why are the radii so large? One would expect in a natural model that there is only one fundamental scale $M_*$ which is related to the radius by $M_* \sim 1 / R$. The ADD idea however requires
\begin{equation}
	R = \frac{1}{M_*} \left( \frac{M_{\text{Pl}}}{M_*} \right)^{\frac{2}{n}} \gg \frac{1}{M_*}
\end{equation}
which is unnatural from naive dimensional analysis arguments. One needs to explain where this large number is coming from. Large extra dimensions only translates the hierarchy of $m_W / M_{\text{Pl}}$ to the hierarchy between the large radius $R$ and $1/M_*$. Stabilizing this hierarchy dynamically in a natural model turns out to be very difficult. 

\subsection{Warped Extra Dimensions}

In this section we explore extra dimensions that are warped, i.e. their metric is non-factorizable. In 5 dimensions, this can be written generically as:
\begin{equation}\label{eq:warp}
	ds^2 ~=~a(z)^2 \left( \eta_{\mu\nu} dx^\mu dx^\nu -dz^2\right)
\end{equation}
where $z$ is the conformal coordinate along the extra dimension and $a(z)$ is called the scale factor or warp factor.
Warped extra dimensions were first proposed by Randall and Sundrum (RS). In a seminal paper \cite{RS}, they showed how a metric of the form Eq.~(\ref{eq:warp}) can arise as a solution to Einstein's equations on a 5D interval with a negative cosmological constant $\Lambda$, sandwiched between two branes of tensions $\pm\Lambda$. The resulting metric is called 5 dimensional Anti de-Sitter (AdS$_5$), in which the warp factor assumes the form: 
\begin{equation}
	a(z)~=~\frac{R}{z}\, .
\end{equation}
For more details on how to get AdS$_5$ gravity solutions see~\cite{Csaki:2004ay,CEHS}. 
As we will see in detail, the AdS$_5$ form of the metric has far reaching implications for the Hierarchy problem, making the cutoff to the SM warped down with respect to the Planck scale. In fact, we can now get the weak-Planck Hierarchy from a Planck size extra dimension. This was indeed a revolutionary step towards a solution to the Hierarchy problem. 

Shortly after the proposal of this solution, it became clear that the RS model has a 4D CFT dual: it corresponds to a 4D strongly coupled theory which confines and dynamically generates an IR scale. In essence, this is just another formulation of the familiar dimensional transmutation that happens in QCD, which yields a confinement scale far below the Planck scale. The advantage of the RS construction is that it constitutes a calculable, weakly coupled description of a confining theory that generates an IR scale---in this case the weak scale.
 
There are many variants of 4D solutions to Hierarchy problem that are based on dimensional transmutation from confinement, of which we have already named a few: Technicolor, in which the condensate is directly responsible for EWSB, "old" composite Higgs, in which the Higgs is some composite of the confining dynamics, and modern composite Higgs, in which the Higgs is a pNGB of a global symmetry broken by the confinement. All of the above models have weakly coupled duals set in RS space. The duals to Technicolor, "old" composite Higgs and modern composite Higgs are called Higgsless models, bulk Higgs models, and models with Gauge-Higgs unification (GHU), respectively.
Towards the end of this section we will mainly explore GHU, and show how it provide a calculable, weakly coupled framework for modern composite Higgs models with partial compositeness.  But for now, let's focus on the generic features of the RS construction which will be useful for model building.

Much of what follows can be generalized to more general gravitational backgrounds, parameterized by a general warp factor $a(z)$. We choose to work in AdS$_5$ since it is in this background that the correspondence to a 4D CFT is best understood. We take two branes at $z=R$ and $z=R' > R$ which truncate the space in the $z$-direction. The $z=R$ brane is usually called the ``UV" brane since one usually has $1/R \sim M_{\text{Pl}}$, while the other is referred to as the ``IR" brane as typically $1/R' \sim 1 / {\text{TeV}}$ for models which address the hierarchy problem. One could consider a more general background which truncates space without the need for branes, sometimes called soft-walls, but this will only affect the details of the KK spectrum. 

To see how RS resolves the hierarchy between the weak-scale and gravity, we first perform the gravity coupling matching for RS.
\begin{align}
S_g &= M_*^3 \int_R^{R'} \left( \frac{R}{z} \right)^3 \int d^4x \sqrt{g_{(4)}} R_{(4)} \\
&= M_*^3 \frac{1}{2}\left(1 - \frac{R^2}{R'^2} \right)
\end{align}
From this result we can read off the effective Planck scale
\begin{equation}\label{eq:sup}
	M_{\text{Pl}}^2 = M_*^3 R \left(1-\frac{R^2}{R'^2} \right) \sim M_*^2
\end{equation}
where last equality follows from the fact that the natural size for $R$ is $1/M_*$, the fundamental scale of the 5D theory. 

This result Eq~(\ref{eq:sup}) is very different from that of ADD, and at first glance does not seem like a solution to the hierarchy at all. After all, if $M_{\text{Pl}} \sim M_*$ then there is no apparent hierarchy between 4D gravity and 5D gravity! There must be some other mechanism at hand. Indeed, in warped extra dimensions the fundamental scale of gravity \textit{is} the 4D Planck scale, and the reason for the hierarchy is that the weak scale itself is \textit{warped down}. Below we will show this by examining the 4D effective action for the Higgs. For now we will only state heuristically that the 4D Higgs mass and VEV end up being related to $M_{\text{Pl}}$ through a warp factor evaluated at the position where the Higgs is localized (or peaked if we allow the Higgs to propagate in the bulk)
\begin{equation}
	v \sim M_{\text{Pl}} \frac{R}{R'} \ll M_{\text{Pl}}\, .
\end{equation}
In short, the weak scale is small because the Higgs is IR-localized and the warp factor in the IR provides a huge suppression. In contrast, we saw that the Planck scale itself is not suppressed at all. This is because the graviton is UV-localized, and the warp factor evaluated at $z=R$ is one. It is the combination of the a UV-localized graviton and an IR-localized Higgs that makes RS a successful solution to the Hierarchy problem.

To demonstrate the warping down of the Higgs potential, let's look at a concrete example: a simplified RS model with a Higgs on the IR brane and only gravity in the bulk (which is essentially the original RS proposal). The 5D action in this model is
\begin{align}
	S_5 &= \int d^5x \, \sqrt{-g}\,\left[R_{(5)}\,+\,\frac{\sqrt{-g_{\text{ind}}}}{\sqrt{-g}}\delta(z-R')\mathcal{L}_{H}\right]\, ,
\end{align}
where $g_{\text{ind}}$ is the induced metric on the IR brane and
\begin{align}
\mathcal{L}_{H}\,=\,g^{\mu\nu}_{\text{ind}}\partial_\mu H^* \partial_\nu H + \lambda \left( |H|^2 -\frac{v^2}{2} \right)^2\, ,
\end{align}
is the Higgs potential. At energies below $1/R'\sim\,\text{TeV}$, we can't resolve the extra dimension, and so the physics should be adequately described by a 4D EFT. To get this EFT, all we have to do is integrate the action over the extra dimension (for models with bulk fields we have to perform a KK decomposition). Plugging in $\sqrt{-g}=\frac{R^5}{z^5}~,~\sqrt{-g_{\text{ind}}}=\frac{R^4}{R'^4}$, we get
\begin{align}
	S_4 &= \int d^4x \,\left[R_{(4)}\,+\,\mathcal{L}^{4D}_{H}\right]\, ,
\end{align}
with
\begin{align}
\mathcal{L}^{4D}_{H}\,=\,{\left(\frac{R}{R'}\right)}^2\partial_\mu H^* \partial^\mu H +{\left(\frac{R}{R'}\right)}^4 \lambda \left( |H|^2 -\frac{v^2}{2} \right)^2\, .
\end{align}
Notice that the Higgs kinetic term is not canonically normalized. Rescaling the Higgs field, we obtain:
\begin{align}
\mathcal{L}^{4D}_{H}\,=\,\partial_\mu H^* \partial^\mu H +\lambda \left( |H|^2 -\frac{\tilde{v}^2}{2} \right)^2\, ,
\end{align}
where $\tilde{v}=v\,\frac{R}{R'}$ is the 4D Higgs VEV, which is warped down with respect to the 5D one. If we find a way to naturally set $\frac{R}{R'}\sim 10^{-18}$ (we will soon explain how this is possible), we get a weak scale 4D Higgs mass and VEV. 

We see that a warped 5D theory with an IR-localized Higgs corresponds to a 4D EFT with weak scale mass and VEV. In addition, 4D gravity is not warped down, so we explain the 4D weak-Planck hierarchy. However, in realistic theories, the 4D EFT contains the SM top and gauge fields, with the usual quadratically divergent corrections to the Higgs potential. How are these cut-off in an RS model? The answer is subtle. We note that the 4D EFT has a cutoff set by $\Lambda\sim 1/R'$. This is where we are starting to probe the fifth dimension and the 4D EFT is no longer adequate. From a bottom up point of view, the scale $\Lambda$ is where KK gravitons appear and become strongly coupled. In other words, the 4D itself does not have a hierarchy problem because its cutoff is close to the weak scale. The problem of radiative corrections thus goes over to the full 5D theory. But in the 5D theory, the bare Higgs mass and VEV can naturally be the Planck scale, in which case we do not expect any significant difference between the `bare' theory and the renormalized one.

One still needs to stabilize the extra dimension in order to provide an explanation for the hierarchy $R$ and $R'$. Unlike for ADD, such a natural explanation has been provided by Goldberger and Wise\cite{Goldberger:1999uk} who dynamically stabilized the distance between the two branes by the addition of a bulk scalar which obtains a VEV. The VEV generates a potential with a minimum due to two competing forces, one from from the scalar kinetic term which wants derivatives to be small and hence a large extra dimension and one from the potential which prefers a small radius.

As a side remark, note that $M_{\text{Pl}}$ remains fixed as $R'\rightarrow \infty$, so the large extra dimension can have infinite proper distance while still preserving the 4D Planck scale. In fact, one can localize SM fields on the UV brane and take the IR brane to $z\rightarrow \infty$ which is known as \textbf{RS2}. Although the effective Planck scale is finite as $R' \rightarrow \infty$ and 4D gravity is preserved, cosmology would be altered due to the emerging gapless continuum of KK gravitons. We will not consider this option further in this review.

\subsection{KK Decomposition in Warped Space}
In realistic RS and composite Higgs models, fields are generically not localized on the IR brane, but rather exist in the entire bulk. To get the 4D EFT for these fields, we need to perform a KK expansion.

For example, in the case of a complex bulk scalar, the 5D action is
\begin{equation}
	S_5 = \int_R^{R'} d^4x dz \sqrt{g} \left[ \partial_M \phi \partial_N \phi g^{MN} - m^2 |\phi|^2 \right] .
\end{equation}
We neglect localized boundary terms proportional to $\delta(z-R)$ or $\delta(z-R')$ which could be included. Their effect is to modify the boundary conditions on $\phi$. Variation of the action yields the bulk equation of motion
\begin{equation}
	\partial_M(\sqrt{g} g^{MN} \partial_N \phi) + \sqrt{g} m^2 \phi = 0.
\end{equation}
In deriving this equation, we integrated by parts picking up a boundary term. In order for the field to be on-shell, it is also necessary for the variation on the boundary to be vanishing\begin{equation}
	\phi^* \partial_z \phi \Big \rvert_{R,R'} = 0.
\end{equation}
We see that we can choose either Neumann or Dirichlet at both $z =R,R'$. This is our choice, and it will affect the spectrum of KK modes and, importantly, whether or not a zero mode is allowed in the spectrum.

We look for a solution in terms of KK eigenstates 
\begin{equation}
	\phi(x,z) = \frac{1}{\sqrt{R}} \sum_n \phi^{(n)} (x) f^{(n)}(z)
\end{equation}
Substitution of this ansatz into the EOM, we find that the profiles must satisfy
\begin{align}
	\left[ \partial_z^2 - \frac{3}{z}\partial_z + m_{n}^2 - \left(\frac{R}{z}\right)^2 m^2 \right] f^{(n)}(z) = 0
\end{align}
which is a Schrodinger-type problem with the appropriate field redefinition of $f(z)$.
The solutions are related to Bessel functions
\begin{equation}
	f^{(n)}(z) = z^2 \left[A_{n} J_\alpha(m_{n} z) + B_n Y_\alpha (m_{n} z) \right] \label{eq:scalarSol}
\end{equation}
where $\alpha = \sqrt{4+m^2 R^2}$, and the solutions satisfy orthogonality relations
\begin{equation}
	\int_R^{R'} \frac{1}{R} \left(\frac{R}{z}\right)^3 f^{(n)*}(z) f^{(m)}(z) = \delta_{m,n}.
\end{equation}

To determine which $m_{n}$'s are allowed, we must apply the chosen boundary conditions on the solution in Eq.~(\ref{eq:scalarSol}). The solution has two free coefficients. One is fixed by normalization (required to ensure the KK mode kinetic terms are canonically normalized), and the other by one of the two boundary conditions.
The other boundary condition provides a quantization condition, picking out discrete allowed values for the 4D masses. Excluding the zero mode, the first KK mode appears generically above a mass gap set by $m_{KK} \sim 1/R'$. 

For large $z$ and $m_{n} \neq 0$, the solutions are oscillatory $\sim z^{3/2} \sin( m_{n} z)$ and grow towards the IR. This means that KK modes are generally peaked at the IR brane. They interact most strongly with other IR-localized DOF. The only exception is for a possible zero mode with $m_{0} = 0$, in which case the solution take the form
\begin{equation}
	f^{(0)}(z) = A z^{2 + \sqrt{4 + m^2 R^2}} + B z^{2 - \sqrt{4 + m^2 R^2}}.
\end{equation}
The zero mode localization is controlled by the bulk mass parameter $m$ and is not necessarily IR localized. In most model building scenarios, SM degrees of freedom are usually associated with the zero modes of 5D fields. 

\subsection{AdS/CFT Correspondence}

One of the most important results related to extra dimensions is the AdS/CFT correspondence proposed by Maldacena~\cite{Maldacena:1997re} (see~\cite{Gherghetta:2010cj} for a complete review). This is a major avenue of research in formal theoretical physics, and here we will only give a quick heuristic sketch of it. Generally speaking, AdS/CFT is a duality between a weakly coupled gravitational theory in the bulk of AdS$_5$, and a strongly coupled 4D conformal field theory. We would say that the 4D CFT `lives' on the boundary of AdS$_5$. In its original formulation given by Maldacena, AdS/CFT is the duality:  
\begin{align*}
&\text{type IIB string theory}~~~~~~\Longleftrightarrow~~~~~~\text{$\mathcal{N}=4$ supersymmetric $SU(N)$ gauge theory }\\
& \hspace{.04\textwidth} \text{on AdS$_5 \times S^5$} \hspace{.32\textwidth} \text{on 4D Minkowski space}
\end{align*}
Correlation functions calculated in the theory on either side of the duality match given a dictionary for relating observables on both sides. The theory parameters on both sides of the duality are related by
\begin{equation}
	\frac{R^4}{l_s^4} = 4 \pi g_{YM}^2 N
\end{equation}
where $l_s$ is the string scale and $g_{YM}$ is the $SU(N)$ Yang-Mills gauge coupling. 

\begin{figure}
\begin{center}
\includegraphics[width=0.7\textwidth]{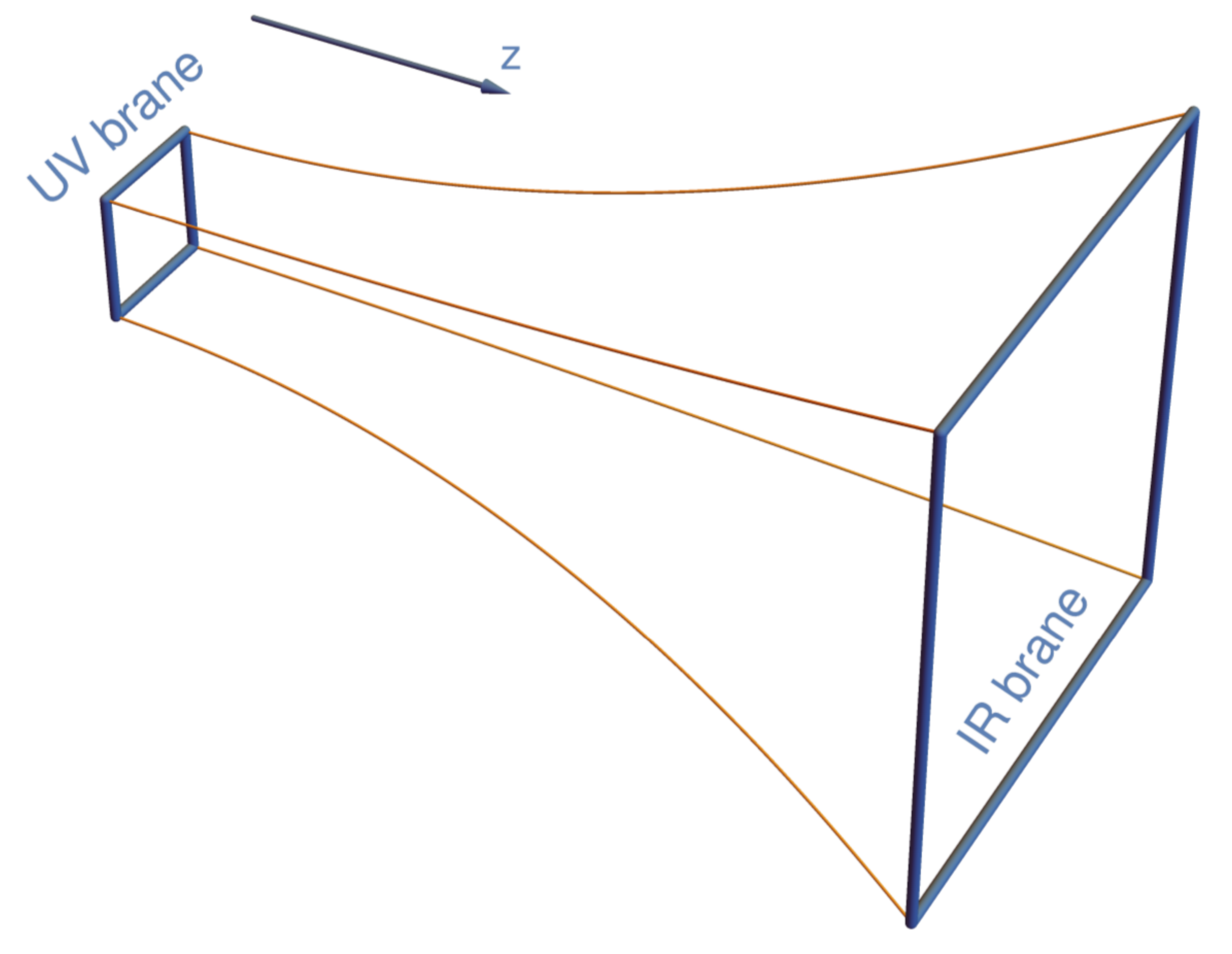}
\end{center}
\caption{Motion along $z$ scales the 4D coordinates and energy scale. \label{fig:RG}}
\end{figure}

In order for the bulk to be described by classical gravity, we should have $R \gg l_s$ so that we can neglect string corrections. This implies that $g_{YM}^2 N \gg 1$, but this is the requirement that the 4D dual CFT is strongly interacting. Now we can immediately see why this duality is useful: we can perform weakly coupled, classical gravity calculations on the 5D side which are dual to a strongly-coupled 4D CFT. This is not so surprising, we already know the 5D theory with an IR brane has a tower of states, which is something we would expect from a strongly coupled 4D theory (\textit{e.g.} resonances in QCD). 

We will not delve into specifics, but we now wish to heuristically explain why one might expect there to be such a correspondence. 
Consider the effect of the $z$-coordinate transformation 
\begin{align}
z \rightarrow e^\alpha z	
\end{align}
on a 4D slice of AdS
\begin{equation}
	ds^2 = \left( \frac{R}{z} \right)^2 \eta_{\mu\nu} dx^\mu dx^\nu.
\end{equation}
The 4D effective metric is rescaled by $e^{-2 \alpha}$, which
can be undone by the 4D coordinate transformation
\begin{align}
x \rightarrow e^\alpha x.
\end{align}
Therefore, we see that
\begin{equation*}
\text{motion along $z$} \Longleftrightarrow \text{rescaling 4D coordinates} .
\end{equation*}
This means that increasing $z$ is equivalent to increasing 4D length scales as in Fig~\ref{fig:RG} and thus decreasing the 4D energy scale. This is exactly what we found for the IR-localized Higgs VEV. We could have guessed this behavior from the form of the metric. 

This naturally leads us to the holographic interpretation of the extra dimension in which the $z$-coordinate corresponds to RG flow in the 4D CFT. A bulk profile which grows with $z$ corresponds to a CFT operator whose coefficient flows to larger values in the IR. How can we see this? Let's first just check the plausibility on the warped two brane RS scenario before describing the entire dictionary of the correspondence. Imagine localizing some 4D fields on a slice of the extra dimension at $z_0$. We will use these 4D fields to `probe' the CFT at different length scales by adjusting $z_0$. As we move $z_0$ deeper into the bulk towards the IR brane, we have seen that the overlap of the IR-localized KK modes with the $\delta(z-z_0)$ localized fields becomes large once $z_0$ approaches $R'$. Since $z_0$ sets the effective 4D length scale on the slice of AdS we are probing with our 4D fields, this would imply the 4D state dual to the RS KK mode is strongly interacting at energy corresponding to $1/R'$. This is exactly what we would expect if the CFT dual is a confining gauge theory with confinement scale $\Lambda \sim 1/R'$, and the KK modes are dual to the composite states. 

\begin{figure}
\begin{center}
\includegraphics[width=\textwidth]{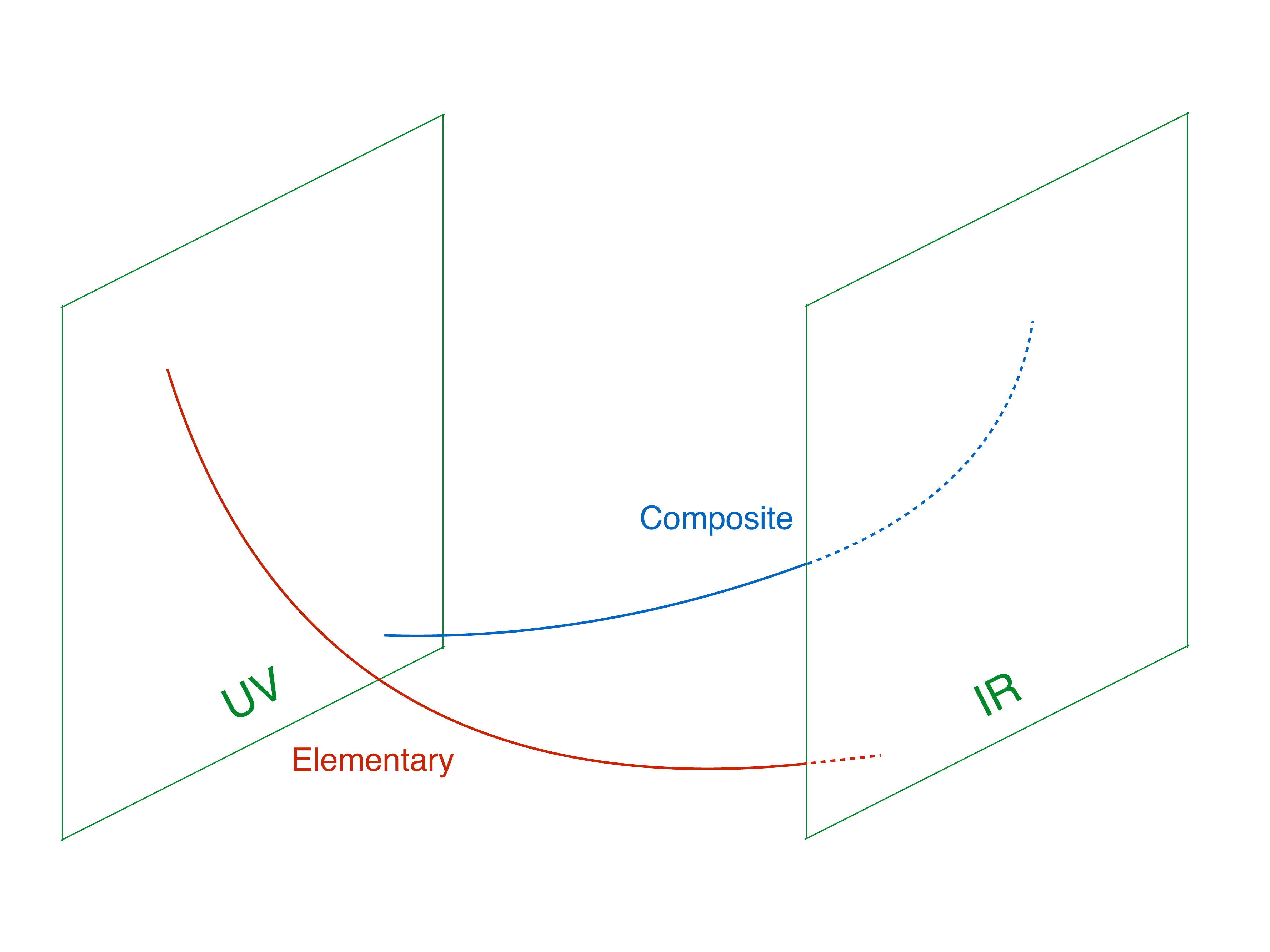}
\end{center}
\caption{AdS/CFT dictionary for localized fields. Elementary DOF are peaked on the UV brane, while composites are localized toward the IR brane. \label{fig:AdSCFT}}
\end{figure}

Maldacena's proof involved the entire AdS$_5$ space, not truncated by two branes as in RS, and the dual 4D theory was a true CFT (no confinement). What is the interpretation of the two branes in the RS scenario?
\begin{itemize}
	\item \underline{UV brane}: The 4D CFT is simply cutoff at a high energy scale $\Lambda \sim 1/R$. The cutoff introduces a mass scale into the CFT and is therefore a source of breaking.
\end{itemize}
Moving away from the UV brane, which corresponds to running down in energy in the 4D CFT, the bulk immediately becomes AdS implying the CFT should quickly become conformal below the UV cutoff scale. Any source of conformal breaking introduced by the cutoff must therefore be an irrelevant deformation of the CFT.

\begin{itemize}
	\item \underline{IR brane}: The IR brane sharply shuts off AdS space at $z=R'$ and corresponds to a relevant deformation of the CFT which ultimately leads to confinement occurring at scale $\Lambda \sim 1/R'$. 
 \end{itemize}
Indeed, we have seen that the IR brane forces a quantization condition on the allowed masses leading to KK modes with the lowest lying states near $1/R'$. The IR brane should be interpreted then as a simplified model of confinement. At some point in RG flow, a relevant deformation is introduced such that the beta function $\beta(g)$ is not completely vanishing.  The theory begins to flow away from its conformal fixed point, and eventually the theory becomes strongly interacting and confines, producing bound states. A more realistic 5D model of confinement would gradually shut off space, but the most important features of confinement are the mass gap and the discrete tower of states which the IR brane does capture. 

Now we have a beautiful picture starting to emerge. The profile of a particle's 5D wavefunction corresponds to RG flow of its couplings in the 4D CFT. Composite states should exist near the 4D confinement energy scale and therefore near the IR brane in 5D, so we can conclude that
\begin{itemize}
	\item  IR-localized fields (\textit{e.g.} KK modes, Higgs) are composite.
\end{itemize}
Fields which are UV-localized interact weakly with the composite states, and therefore should not be part of the strong dynamics. We can conclude that
\begin{itemize}
	\item UV-localized fields (\textit{e.g.} graviton) are elementary.
\end{itemize}
Now we can see that RS solves the hierarchy problem because the Higgs is composite! RS is the 5D dual of composite Higgs models of the sort we have discussed at the beginning of the lecture. We will be able to present explicit calculable constructions for the MCH model soon. However, we need one more ingredient, bulk gauge fields and their connection to global symmetries in the CFT.

We should add one more line to our AdS/CFT dictionary:
\begin{equation*}
	\text{5D gauge symmetry} \Longleftrightarrow \text{4D global symmetry}.
\end{equation*}
A bulk gauge symmetry in the full AdS$_5$ ($R' \rightarrow \infty$) corresponds to a global symmetry in the 4D CFT. In order to see this, we need to understand under which situations is there a massless gauge boson in the low-energy 4D effective theory. In a RS-type scenario with two branes, we can also get 4D gauge symmetries from 5D gauge symmetries since in RS we can get a massless gauge boson in the spectrum. 

Given a bulk gauge symmetry, the zero mode $A_\mu^{(0)}$ couples to the global current $J^\mu$ at each position in the bulk proportional to the 4D effective gauge coupling. The EOM for the zero mode has the form 
\begin{equation}
	\partial_M \left( \sqrt{g} g^{MN} F_{NP} \right) = 0,
\end{equation}
and one can check that the profile is exactly flat, $A_\mu(x,z) = N A_\mu(x)$ where $N$ is a normalization constant. In fact, it has to be flat because this is what ensures that the gauge boson couples diagonally to charged states. In order to get\ the effective 4D coupling, we normalize the zero mode such that its 4D kinetic term is canonically normalized:
\begin{equation}
	N = \left[ \int_R^{R'} \frac{R}{z} dz \right]^{-\frac{1}{2}}
\end{equation}

If we had taken the full AdS$_5$, \textit{i.e.} by sending $R'\rightarrow \infty$, then $N \rightarrow 0$ which shows that $A_\mu^{(0)}$ is not normalizable (it is absorbed into the $A_\mu$ KK mode continuum). The effective 4D gauge coupling, which is proportional to $N$, is zero, and therefore it decouples from the theory. In this case, we have a conserved current $ \partial_\mu J^\mu = 0$, which signals a true global symmetry in the limit $R' \rightarrow \infty$. 

However, for RS with a brane placed at finite $R'$, $N$ is finite, and we can have a normalizable zero mode. If the boundary conditions on the UV and IR brane admit a zero mode solution (Neumann)
\begin{equation}
	\partial_5 A_\mu \big \rvert_{R,R'} = 0 
\end{equation}
then the zero mode is allowed in the spectrum and the global symmetry in the CFT is weakly gauged since there is a massless gauge boson coupling to the current. 

What if only one of the boundary conditions is switched to Dirichlet (which is not compatible with the zero mode solution)?
\begin{itemize}
	\item $A_\mu (R) = 0$: The gauge symmetry in the 4D effective theory is broken by UV boundary conditions. The zero mode is removed from the spectrum leaving a residual global symmetry. The would-be zero mode gauge boson acquires a mass set by $1/R$ and decouples from the low-energy theory. 
	\item $A_\mu (R') = 0$: The gauge symmetry is broken by IR boundary conditions. Since the IR brane is dual to confinement, this corresponds to dynamical gauge symmetry breaking like technicolor. The would-be zero mode gauge boson acquires a mass set by the confinement scale or $1/R'$. 
\end{itemize}

Now imagine if we take Dirichlet boundary conditions on both branes: $A_\mu \big \rvert_{z=R,R'} = 0$. The first boundary condition we apply removes the zero mode gauge boson from the spectrum, converting the gauge theory in the 4D effective theory to a global symmetry. However, the second boundary condition provides a source of global breaking which corresponds to the global symmetry being spontaneously broken by confinement. This choice of boundary conditions should produce a Goldstone mode! We will show that it actually arises in the $A_5$ component.

This fact is easy to see once we realize the $A_5$ component should have opposite boundary conditions to that of $A_\mu$. In deriving the EOM for $A_5$ by varying the action, an integration by parts is required which generates boundary terms. By requiring the boundary terms vanish, one can show that if $A_\mu \big \rvert_{R,R'} = 0$, then $A_5$ should satisfy the boundary condition
\begin{align}
	\partial_5 \left( \frac{A_5}{z} \right)\Big \rvert_{z=R,R'} = 0.
	\label{eq:A5BC}
\end{align}
We will show this boundary condition does allow a zero mode in the $A_5$ component.

The gauge action in warped space is
\begin{equation}
S_{\text{gauge}} =  \int d^5 x ~ \frac{R}{z} \left[-\frac{1}{4} F_{\mu \nu} F^{\mu \nu} - \frac{1}{2} F_{\mu 5} F^{\mu 5}\right].
\end{equation}
The action contains the following mixing term between $A_5$ and $A_\mu$
\begin{align}
\int d^5x ~ \frac{R}{z} \partial_\mu A^5 \partial_5 A^\mu = \int d^5x ~ \partial_5 \left( \frac{R}{z} A^5 \right) \partial_\mu A^\mu 
+\text{boundary terms}.
\end{align}
The boundary terms will affect the boundary conditions but not the bulk equation of motion. The mixing can be removed by adding the gauge fixing term
\begin{align}
\int d^5x \frac{1}{2 \xi} \frac{R}{z} \left[ \partial_\mu A^\mu -\xi \partial_5 \left( \frac{R}{z} A^5 \right) \right]^2.
\end{align}

After gauge fixing, the quadratic $A_5$ part of the action contains the following terms
\begin{align}
\int d^5x ~ \frac{R}{z} \left[ \frac{1}{2} \partial_\mu A^5 \partial^\mu A^5  + \frac{1}{2} \xi \left( \partial_5 \left( \frac{R}{z} A^5 \right) \right)^2 \right]
\end{align}	
leading to the bulk EOM for $A_5$  
\begin{equation}
\partial^2 A_5 + \frac{R}{z} \xi \left[ \partial_5^2 \left(\frac{R}{z}\right) A_5 + 2 \partial_5 \left(\frac{R}{z}\right) \partial_5 A_5 +  \left(\frac{R}{z}\right) \partial_5^2 A_5 \right] = 0.
\end{equation}
If we replace $\partial^2 \rightarrow - m^2$, we see that for $m^2 \neq 0$, the $A_5$ KK mode masses are proportional to $\xi$ and are therefore unphysical. Remember $A_5^{(n\neq 0)}$ is eaten by $A_\mu ^{(n\neq 0)}$. However for the zero mode case ($m^2 = 0$) we have the EOM
\begin{equation}
	\partial_5^2 \left(\frac{R}{z}\right) A_5 + 2 \partial_5 \left(\frac{R}{z}\right) \partial_5 A_5 +  \left(\frac{R}{z}\right) \partial_5^2 A_5 = 0
\end{equation}
which has solutions
\begin{equation}
	A_5(x,z) = \left( a z + b z \log{z} \right) A_5(x).
\end{equation}
The boundary conditions in Eq.~(\ref{eq:A5BC}) pick out the solution proportional to $z$. In more general backgrounds, the EOM is obtained by the replacement $R/z \rightarrow a(z)$, and the $A_5^{(0)}$ profile  always scales as the inverse of the warp factor. 

Going through the full KK decomposition for $A_\mu$
\begin{equation}
	A_\mu(x,z) = \frac{1}{\sqrt{R}} \sum_n h^{(n)}(z) A_\mu^{(n)}(x)
\end{equation}
One can check that the solutions to the $A_\mu^{(n)}$ EOM are again Bessel functions
\begin{equation}
	h^{(n)}(z) = z\left( A_n J_1(m_n z) + B_n Y_1(m_n z) \right).
\end{equation}

\subsection{Fermions in RS}

We now will describe how to include bulk fermions in RS. The smallest irreducible representation of the 5D Lorentz group is the 4-component Dirac spinor. This implies that every bulk fermion field contains both left-handed (LH) and right-handed (RH) components, \textit{i.e.}
\begin{equation}
\Psi = \begin{pmatrix}
 	\chi \\
 	\bar{\psi}
 \end{pmatrix},	
\end{equation}
and the 5D theory is non-chiral. There is a way to get chiral SM matter content however since the boundary conditions in a RS-type model pick out chiral zero modes. The boundary conditions which allow a LH zero mode will not allow a zero mode in RH component of the same Dirac spinor and vice versa. The KK modes of the fermions are, however, vector-like. 

In this section, we will use Dirac matrices in the chiral representation:

\begin{align}
\gamma^\mu = 
	\begin{pmatrix}
		0 & \sigma^\mu \\
		\bar{\sigma}^\mu & 0 
	\end{pmatrix}, \ \ \
\gamma^5 =
	\begin{pmatrix}
		i & 0 \\
		0 & -i
	\end{pmatrix}
\end{align}
where $\sigma^0 = \bar{\sigma}^0 = -\mathbf{1}$ and $\sigma^i = -\bar{\sigma}^i$ are the usual Pauli spin matrices. The gamma matrices in warped space are related to the ordinary flat space ones~\cite{Sundrum:1998sj,Csaki:2005vy} by a factor known as the vielbein $e_a^M$, where $a$ indices denotes flat space indices, which satisfies 
\begin{equation}
e^M_a \eta^{ab} e_b^N = g^{MN},
\end{equation}
\begin{equation}
e^a_M = \frac{R}{z} \delta^a_M,
\end{equation}
\begin{equation}
\Gamma^M = e^M_a \gamma^a.
\end{equation}
Furthermore, the covariant derivatives require an additional piece called the spin-connection, which in AdS5 is 
\begin{align}
D_\mu \Psi &= (\partial_\mu + \frac{1}{4z} \gamma_\mu \gamma_5) \Psi \\
D_5 \Psi &= \partial_5 \Psi.
\end{align}

Working in terms of vielbeins and flat space gamma matrices, the 5D AdS fermion action can be written as
\begin{equation}
	S_{\text{fermion}} = \int d^5x \sqrt{g} \left( \frac{i}{2} \bar{\Psi} e^M_a \gamma^a D_M \Psi - \frac{i}{2} D_M \bar{\Psi} e_a^M \gamma^a \Psi - M \bar{\Psi}\Psi \right)
	\label{eq:fermionRS}
	\end{equation}
The spin-connection part of the covariant derivative cancels out leaving us (after integration by parts of the left-acting z derivatives) with
\begin{equation}
\int d^5x \left(\frac{R}{z} \right)^4 \bar{\Psi} \left( i \slashed{\partial}
	+ i \gamma^5 \partial_5 - i \frac{2}{z}\gamma^5 - \frac{c}{z} \right) \Psi
\end{equation}
where we have chosen to write the bulk mass in terms of a dimensionless bulk mass $c = M R$.

In terms of 2-component Weyl spinors, Eq~(\ref{eq:fermionRS}) becomes
\begin{align}
 &\int d^5x \ \left( \frac{R}{z} \right)^4 [ -i\bar{\chi}\bar{\sigma}^\mu \partial_\mu \chi - i \psi \sigma^\mu  \partial_\mu \bar{\psi} 
+ \frac{1}{2}(\psi \overleftrightarrow{\partial_5} \chi - \bar{\chi}\overleftrightarrow{\partial_5} \bar{\psi}) + \frac{ c }{z} (\psi \chi + \bar{\chi}\bar{\psi})]
\end{align}
where $\psi \overleftrightarrow{\partial_5} \chi = \psi \partial_5 \chi - \partial_5 \psi \chi$. Variation gives the 1st order coupled EOMs
\begin{align}
	- i \bar{\sigma}^\mu \partial_\mu \chi - \partial_5 \bar{\psi} + \frac{c+2}{z} \bar{\psi} &= 0 \nonumber \\
	-i \sigma^\mu \partial_\mu \bar{\psi} + \partial_5 \chi ~+~ \frac{c-2}{z} \chi & = 0
\end{align}

Now we can proceed with KK decomposition. As usual we expand the 5D fields as a sum of 4D eigenmodes
\begin{align}
	\chi = \sum g_n(z) \chi_n(x) \nonumber \\
	\bar{\psi} = \sum f_n(z) \bar{\psi}_n(x)
\end{align} 
where $\chi_n, \psi_n$ satisfy the ordinary 4D Dirac equation
\begin{align}
	-i \bar{\sigma}^\mu \partial_\mu \chi_n + m_n \bar{\psi}_n &= 0 \nonumber \\
	-i \sigma^\mu \partial_\mu \bar{\psi}_n + m_n \chi_n &= 0.
\end{align}
Substitution of the KK sum yields EOMs for the profiles
\begin{align}
&	f_n' + m_n g_n - \frac{c+2}{z} f_n = 0 \nonumber \\
&	g_n' - m_n f_n + \frac{c-2}{z} g_n = 0
\end{align}
These equations can be decoupled at the cost of turning them into two second order decoupled equations with relations among the coefficients of their solutions. The result is again Bessel functions:
\begin{align}
	& g_n(z) = z^{\frac{5}{2}} \left( A_n J_{c+1/2}(m_n z) + B_n Y_{c+1/2}(m_n z) \right) \nonumber \\
	& f_n(z) = z^{\frac{5}{2}} \left( A_n J_{c-1/2}(m_n z) + B_n Y_{c-1/2}(m_n z) \right).
\end{align}
Focusing on the zero mode solutions, we have
\begin{align}
	g_0 = A_0 \left( \frac{z}{R} \right)^{2-c}, \ \ \ \ \
	f_0 = B_0 \left( \frac{z}{R} \right)^{c+2} . 
\end{align} 

We will not go into much detail about how to derive the fermion boundary conditions\cite{Csaki:2003sh}. One can study the 1st order EOMs and show that if one chirality satisfies Dirichlet boundary conditions, then the other chirality must satisfy Neumann-type conditions in order for the EOM to be satisfied on the boundary. The main point is that one of the chiralities must have Dirichlet boundary conditions, which will eliminate the zero mode solution in that chirality. Thus, either $A_0$ or $B_0$ must be zero. This generates a chiral zero mode spectrum allowing us to get the SM fermion field content.

Moreover, just as we found for the bulk scalar, the bulk mass controls the localization of the zero mode. The fermion zero mode can be mostly elementary or mostly composite depending on our choice for the bulk mass parameter $c$. For a LH zero mode $\chi$,
\begin{itemize}
	\item $\chi$ is UV-localized (IR-localized) for $c > 1/2$ ($c < 1/2$)
\end{itemize}
and for a RH zero mode $\psi$,
\begin{itemize}
	\item $\psi$ is UV localized (IR localized) for $c < -1/2$ ($c > -1/2$) .
\end{itemize}	
The properly normalized fermion zero mode is
\begin{equation}
	\psi^0_{L,R}(x,z) = \frac{1}{\sqrt{R'}} \left( \frac{z}{R} \right)^2 \left( \frac{z}{R'} \right)^{\mp c} f_{\pm c} P_{L,R} \psi^0(x)
\end{equation}
where $f_c$ is known as the RS flavor function
\begin{equation}
	f_c = \sqrt{\frac{1-2c}{1 - \left( \frac{R}{R'} \right)^{1-2c} }}.
\end{equation}

\subsection{Construction of a Realistic RS Model}

In this section, we describe the process towards achieving a realistic RS model consistent with electroweak precision constraints which was worked out by Agashe, Delgado, May, and Sundrum\cite{Agashe:2003zs}. The first hurdle we face is to protect the $T$-parameter. Without a custodial symmetry incorporated, the strong dynamics from the composite sector generate large $T$- and $\rho$-parameter corrections. The custodial symmetry can also be useful to protect the $Z b \bar{b}$ coupling\cite{Agashe:2006at} which is highly constrained by LEP. 

We incorporate the custodial symmetry in the composite sector by enlarging the bulk gauge symmetry $G$  to contain the SM gauge symmetries plus a custodial $SU(2)_R$. For simplicity we can take $G = SU(2)_L \times SU(2)_R \times U(1)_X$. The SM hypercharge is embedded in $SU(2)_R \times U(1)_X$. 

However, we do not want additional massless $SU(2)_R$ gauge bosons in the 4D effective theory, so we break $SU(2)_R \times U(1)_X$ down to $U(1)_Y$ on the UV brane by applying Dirichlet boundary conditions on the gauge bosons corresponding to the generators we wish to break. The 4D CFT description of this scenario is a CFT with a $SU(2)_L \times SU(2)_R \times U(1)_X$ global symmetry whose $SU(2)_L \times U(1)_Y$ subgroup is gauged. If we take the Higgs to be a bidoublet under $SU(2)_L \times SU(2)_R$, the Higgs sector will have an approximate custodial symmetry thus reducing the bulk $T$-parameter contributions. The other option would be to break the unwanted $SU(2)_R$ generators on the IR brane, but this scenario would have larger custodial symmetry violation since this corresponds to the global $SU(2)_R$ symmetry being gauged  (and spontaneously broken by confinement). Usually in RS model building, the bulk gauge symmetry is broken down to SM gauge symmetries on the UV brane.

\begin{figure}
\begin{center}
\includegraphics[width=\textwidth]{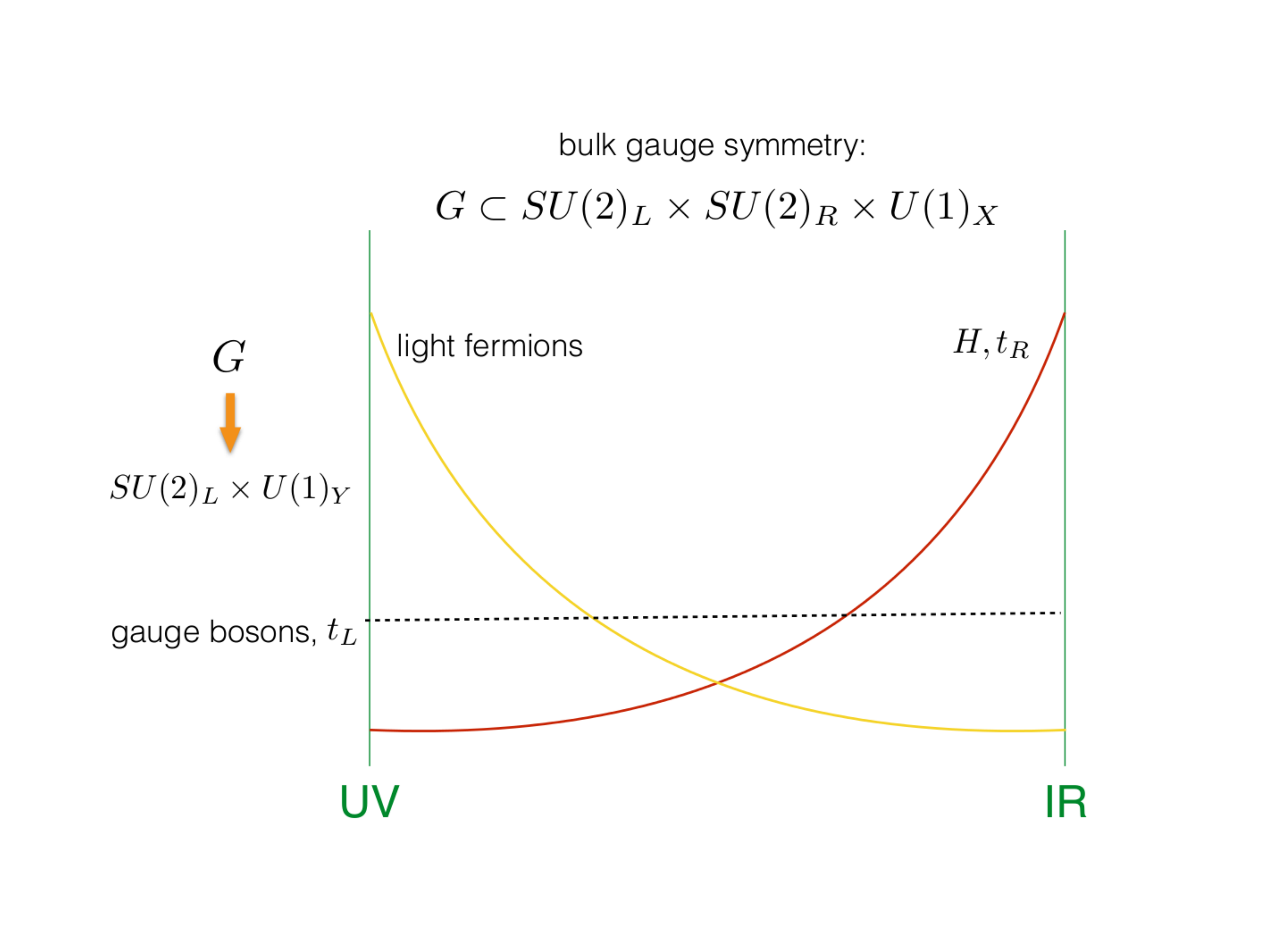}
\end{center}
\caption{The ``realistic" RS scenario. The bulk gauge symmetry contains the SM gauge symmetries with an additional custodial $SU(2)_R$ which is broken by boundary conditions on the UV brane. The Higgs and $t_R$ are IR-localized, the light quarks and leptons are UV-localized, and the gauge bosons and $t_L$ are approximately flat. \label{fig:realisticRS}}
\end{figure}

To address the hierarchy problem, the Higgs should be composite and thus IR-localized. In order to achieve a realistic top mass, the Higgs should have significant overlap with the top. However, $t_L$ cannot be IR localized since it is in the same doublet as $b_L$, and this would give large corrections to the $Z b \bar{b}$ coupling. Therefore $t_R$ should be significantly IR-localized, and it turns out we can get away with having $t_L$ approximately flat, $c_{t_L} \sim 1/2$. Light fermions should have small mass and therefore are UV-localized. This scenario is summarized in Fig.~\ref{fig:realisticRS}. 

There is a rich set of signatures from the realistic RS scenario\cite{Agashe:2006hk}. The most striking signal is the production of the KK gluon is produced via Drell-Yan with a large rate: $\sigma (q\bar{q} \rightarrow G^{(1)}) \sim 0.1$ pb for $m_{KK} \sim 3$ TeV. Gluon production is not important since the KK gluon profile is orthogonal to the zero mode gluon, and existing constraints already rule out KK gluon masses which are light enough to be pair-produced. Since $t_R$ is peaked on the IR brane, the KK gluons decay almost exclusively to $t \bar{t}$. Very heavy KK gluons decay to highly boosted tops, requiring the use of jet substructure to tag the tops. The current bound is roughly $m_{G^{(1)}} > 3$ TeV. In addition, one can also produce the other KK excitations: $Z^{(1)}$, $\gamma^{(1)}$, \textit{etc}. The KK modes tend to have largest overlap with the top, Higgs, and longitudinal gauge bosons. The most likely decays include $Z^{(1)} \rightarrow t\bar{t}, W^+ W^-, \ldots$, $\gamma^{(1)} \rightarrow Z h, t\bar{t},\ldots$ for example. KK decays to leptons are strongly suppressed since they are elementary.

Realistic RS is a natural implementation for partial composite anarchic flavor models. All we have to do is take the realistic RS scenario and which requires different $c$'s for the various SM fermions. The $c$'s control the localization of the zero modes, which generates exponential hierarchies in the zero mode overlap integrals.  Then just as in the general case we have
\begin{align}
	m_u &= \frac{v}{\sqrt{2}} f_q Y_u f_{-u} \\
	m_d &= \frac{v}{\sqrt{2}} f_q Y_d f_{-d}.
\end{align} 
where $f$ is now given by the RS flavor function, and we have adopted the short hand $q \equiv c_q$, $u \equiv c_u$, \textit{etc}. All of the generic partial compositeness discussion applies here. In the 4D CFT description, the bulk mass parameters control the anomalous dimension of the fermion mass term operators.

\begin{figure}
\begin{center}
\includegraphics[width=0.75\textwidth]{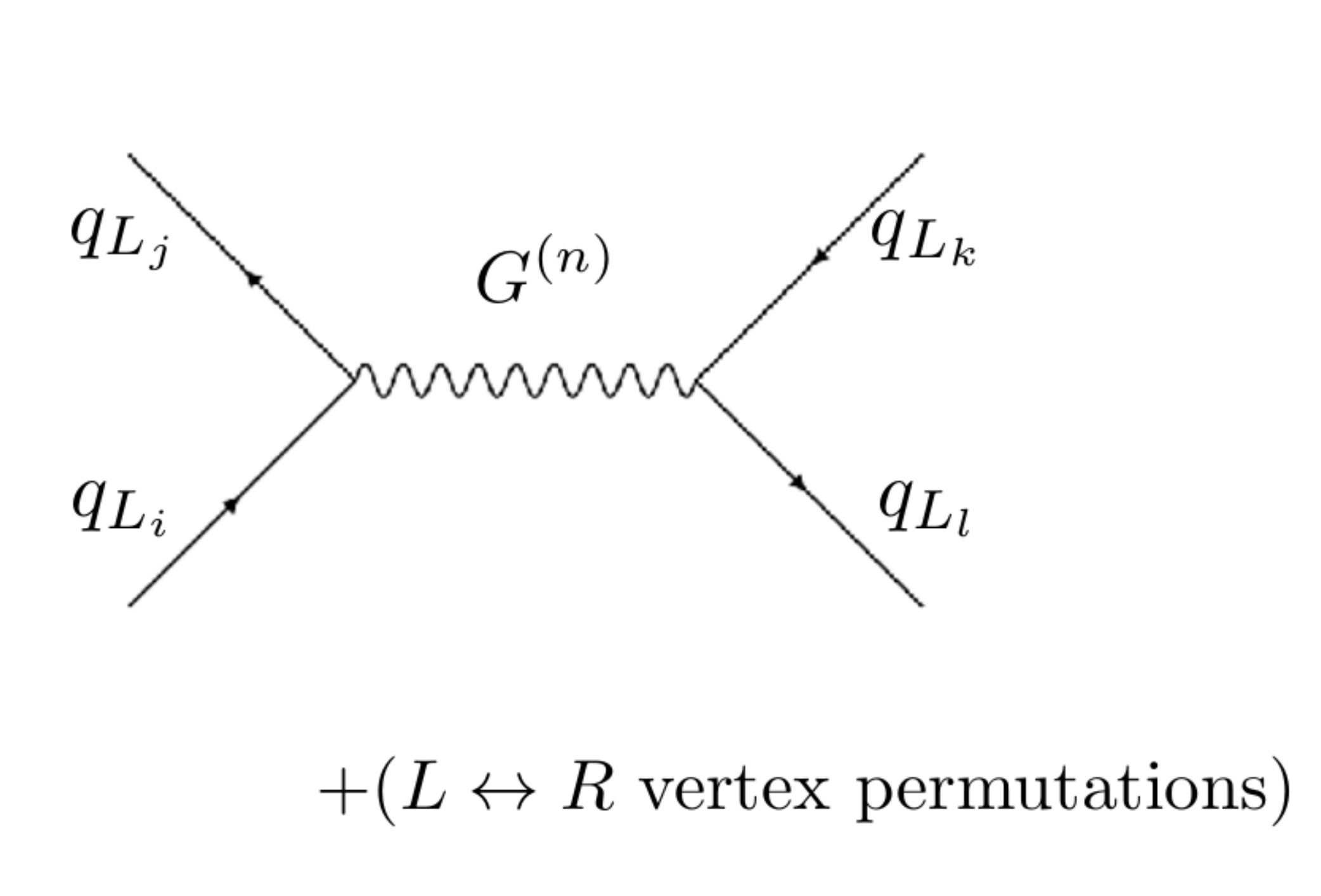}
\end{center}
\caption{Diagrams contributing to KK gluon induced 4-fermi operators. \label{fig:4fermi}}
\end{figure}

We can also explore the specific source of 4-fermi operators arising from KK gluon exchange. In the flavor basis (before mass diagonalization), the KK gluon's couplings to quark species $X$ are diagonal but not exactly universal\cite{Csaki:2008zd}
\begin{equation}
	g_X \simeq g_* \left( -\frac{1}{\log{\frac{R'}{R}}} + f_X^2 \Gamma(c_X) \right) .
\end{equation}
The first piece is universal and arises from the elementary part of the KK gluon coupling to the elementary fermions. The second term is the contribution from the composite sector via mixing with the elementary part of the KK gluon.  The coupling would be universal if the $f_X$'s were degenerate, but this is not the case if we wish to explain flavor with RS. 

After rotation to the mass basis, the off-diagonal couplings are of order
\begin{align}
	(g_{q_L})_{ij} &\sim g_* f_{q_i} f_{q_j} \\
	(g_{u_R})_{ij} &\sim g_* f_{-u_i} f_{-u_j} \\
	(g_{d_R})_{ij} &\sim g_* f_{-d_i} f_{-d_j} .
\end{align}
\begin{figure}
\vspace*{-0.4cm}
\begin{center}
\includegraphics[width=.75\textwidth]{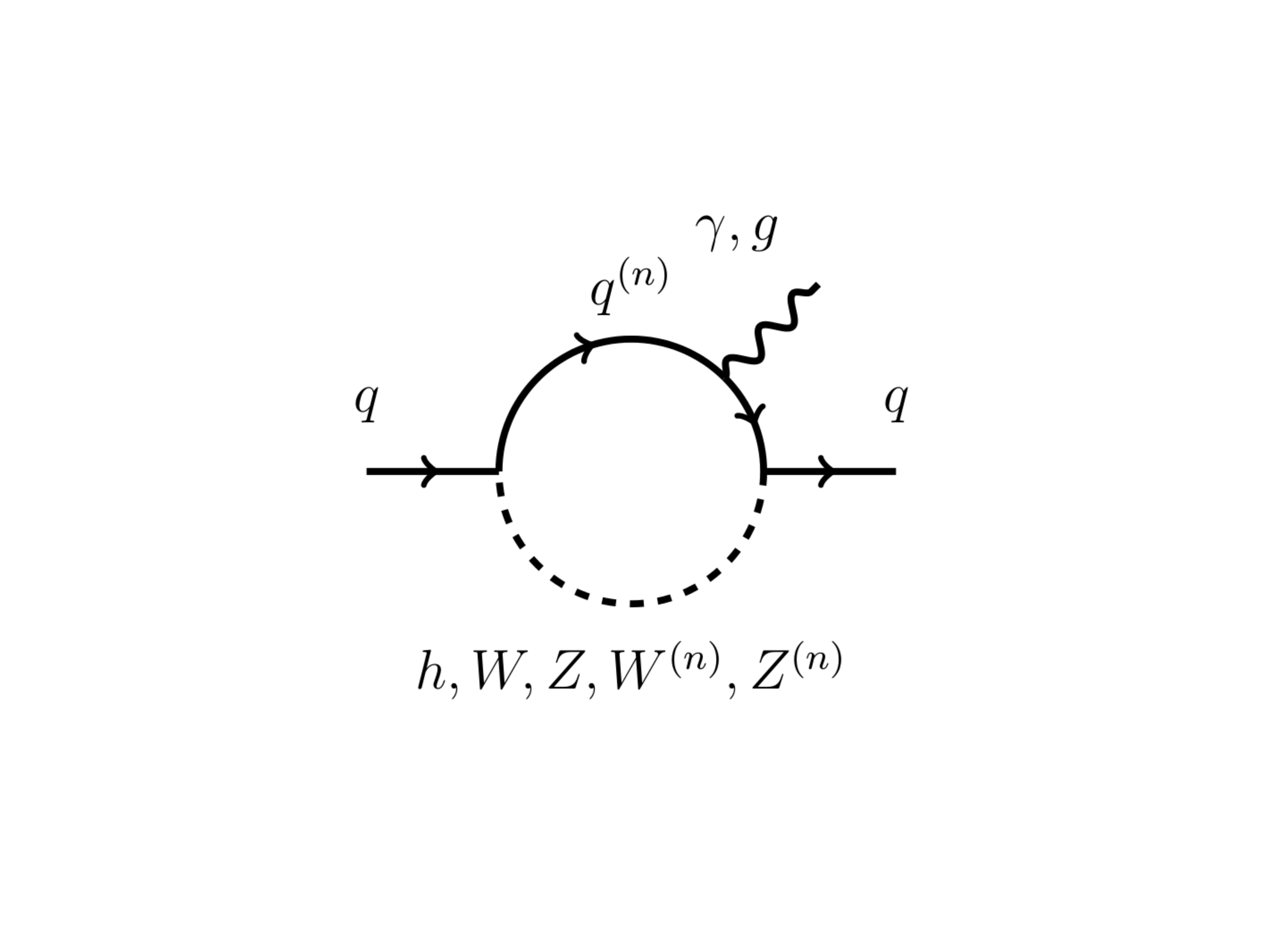}
\end{center}
\vspace*{-1.5cm}
\caption{Example of one contribution to the quark electric dipole moment arising from KK quarks. \label{fig:EDM}}
\end{figure}
Notice that the off-diagonal couplings of the quarks are suppressed by hierarchically small entries. This suppression is the appearance of the RS-GIM mechanism, which is a result since the off-diagonal couplings to the KK gluon are proportional to the fermion Yukawas.  The 4-fermi operators are generated from the diagrams in Fig.~\ref{fig:4fermi}. After integrating out the KK gluon and application of Fierz identities, we obtain the following operators parameterized in terms of Wilson coefficients $C^1$, $C^4$, $C^5$.
\begin{align}
	&  C^1 \left( \bar{q}^{i\alpha}_L \gamma^\mu q_{L\alpha}^j \right) \left( \bar{q}_L^{k\beta} \gamma^\mu q_{L\beta}^l \right) +
	C^4 \left( \bar{q}_R^{i\alpha} q_{L\alpha}^k \right) \left( \bar{q}_{L}^{l\beta}  q_{R\beta}^j \right)\nonumber \\
	&+ C^5 \left( \bar{q}_R^{i\alpha} q_{L\beta}^l \right) \left( \bar{q}_{L}^{k\beta}q_{R\alpha}^j\right)
\end{align}
where $\alpha, \beta$ are color indices. The most strongly constrained quantity is $C_{4K}$ which we estimate to be 
\begin{equation}
	C_{4K}^{RS} \sim \frac{g_*^2}{m_G^2} f_{q_1} f_{q_2} f_{-d_1} f_{-d_2} \sim \frac{1}{m_G^2} \frac{g_*^2}{Y_*^2} \frac{2 m_d m_s}{v^2}.
\end{equation}
Even with the RS-GIM mechanism the bound is still somewhat large, $m_G \gtrsim 20$ TeV. There is additionally another type of bound arising from electric dipole moments induced by KK fermion exchange as shown in Fig.~\ref{fig:EDM}. 

\subsection{Holographic Composite Higgs \& Higgs Potential}

We have seen how to obtain a realistic RS scenario. However, we wish to go one step further and incorporate a pNGB Higgs. The pNGB Higgs allows its mass to naturally be a loop factor below the strong dynamics, much as the pion is lighter than $\Lambda_{QCD}$. Moreover, the Higgs potential is finite and calculable. We will describe the RS setup for the MCH model. We know from AdS/CFT that to obtain a Goldstone boson we can break a bulk gauge symmetry on both the UV and IR branes, and we should get Goldstone bosons in the $A_5$ component of the gauge fields corresponding to the the broken generators. This kind of scenario is known as \textbf{Gauge Higgs Unification} since the Higgs boson is actually part of a higher-dimensional gauge field \cite{GHUnif, CNP}. 
\begin{figure}
\begin{center}
\vspace*{-1cm}
\includegraphics[width=\textwidth]{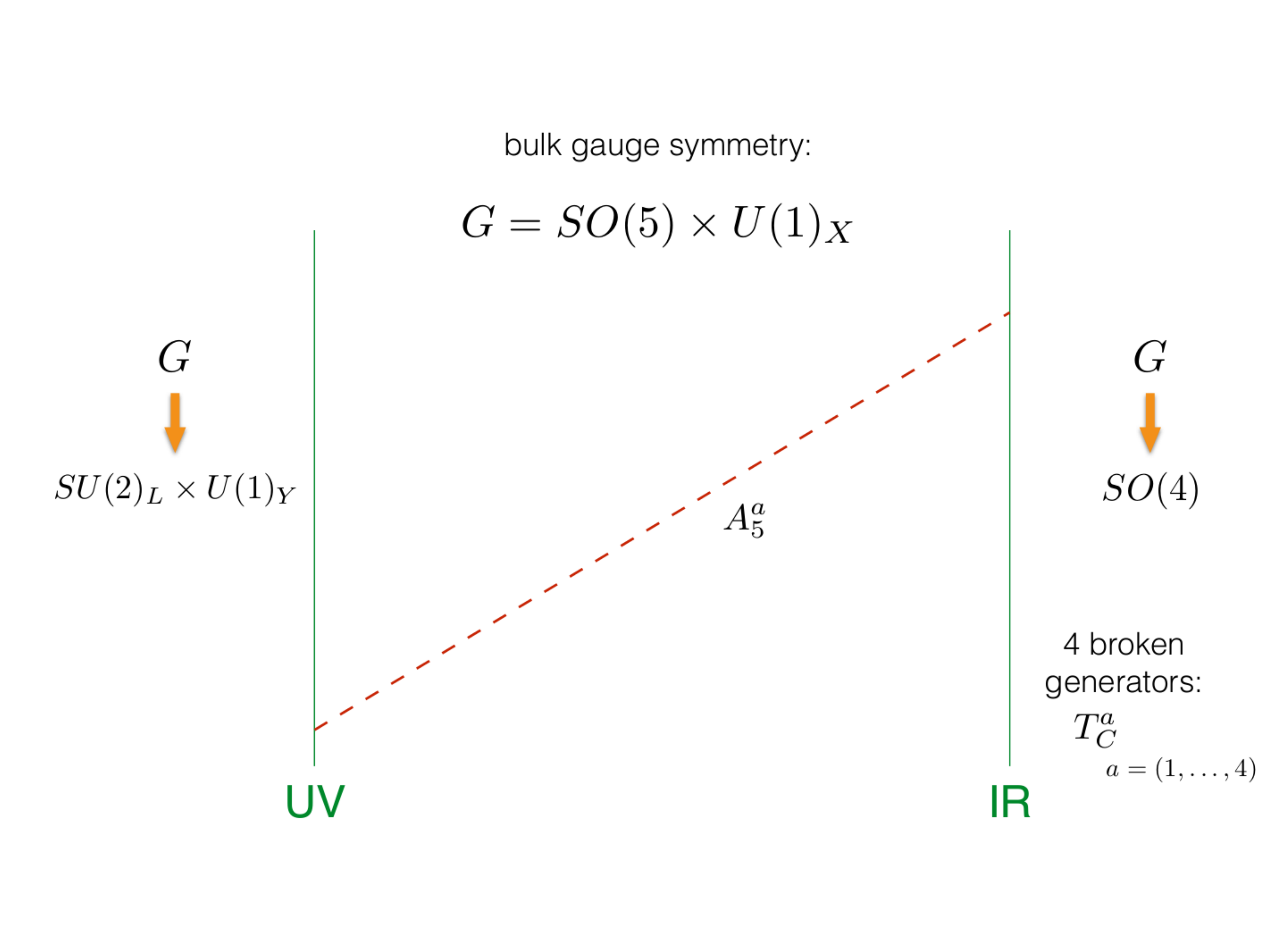}
\end{center}
\caption{The holographic MCH setup in 5D. \label{fig:MCH}}
\end{figure}

We start with a bulk gauge symmetry $G = SO(5) \times U(1)_X$ where the SM $SU(2)_L$ is embedded in $SO(5)$ \cite{MCH}. Notice that $SO(5) \subset SO(4) \sim SU(2)_L \times SU(2)_R$ contains a custodial symmetry. Hypercharge is embedded in a linear combination the diagonal of $SU(2)_R \supset SO(5)$ and $U(1)_X$. $G$ is broken by boundary conditions to the SM gauge groups $SU(2)_L \times U(1)_Y$ on the UV brane to remove the extra gauge symmetries in the low energy theory. Finally, G is broken to $SO(4)$ by boundary conditions on the IR brane. This corresponds to the $SO(5)$ global symmetry in the CFT being spontaneously broken by confinement. 

 There are 4 broken generators $T^a_C$ (the generators corresponding to the coset of $SO(5)/SO(4)$) which are broken on both branes, so there are 4 Goldstone bosons $A_5^a$ which transform as a $(2,2)$ under $SU(2)_L \times SU(2)_R$. After electroweak symmetry breaking 3 are eaten by the $W$ and $Z$ bosons and one remains as the physical Higgs. The Higgs wave function is set by the $A_5$ profile
\begin{equation}
	A_5^a(x,z) = \sqrt{\frac{2}{R}}\frac{z}{R'} T^a_C h^a(x).
\end{equation}
We have chosen the normalization such that the $h^a$'s are canonically normalized in the 4D effective theory. 

There are several relevant scales of the theory. The first is the the KK scale which is set by zeros of the Bessel function solutions of the KK EOMs and is approximately
\begin{equation}
	M_{KK} \simeq \frac{2}{R'}.
\end{equation}
The KK scale sets the mass gap of the strong dynamics, and the lightest KK modes should be an order one number times the KK scale in a natural theory. Another important parameter is the dimensionless gauge coupling
\begin{equation}
	g_* = \frac{g_5}{\sqrt{R}},
\end{equation}
that sets the interaction strength of KK gauge bosons. The Higgs interactions, and thus the SM Yukawas, are also proportional to $g_*$. Then we have the scale of global symmetry breaking
\begin{equation}
	f = \frac{M_{KK}}{g_*} \sim \frac{2}{g_* R'}
\end{equation}
which in the dual theory is the energy scale of the VEV that breaks the global symmetry $SO(5) \rightarrow SO(4)$.

Contributions to the Higgs potential are cutoff at an energy scale\begin{equation}
	g_* f \leq 4\pi f,
\end{equation}
which acts as a compositeness scale for the Higgs boson. The inequality comes from the requirement that the effective theory is perturbative. Notice that the dimensionless gauge coupling controls the cutoff, and for a perturbative scenario ($g_* \leq 4\pi$), the contributions to the Higgs potential are shut off before we lose perturbative calculability at $\Lambda \sim 4 \pi f$. This is the reason the Higgs potential is calculable: the potential is not sensitive to contributions above $g_* f$ which we can not perturbatively calculate. This result can be viewed as a consequence of collective symmetry breaking in the extra dimension. 

How do we get an effective potential for the Higgs? The tree level potential is vanishing since 5D gauge invariance forbids a potential for $A_5$. However, a radiative potential is generated since we have explicitly broken the $SO(5)$ global symmetry by gauging a subgroup and from the fact that the zero modes do not form complete $SO(5)$ representations. In order to determine the potential, we calculate the bulk EOMs for the gauge fields and fermions in the presence of a classical $A_5$ background. Their spectrum will depend on the $A_5$ VEV background and therefore generate a CW potential. However, this calculation is hard! The bulk fermion EOMs are very complicated with the z-dependent Higgs VEV turned on, and the VEV couples the EOM of different components of each fermion multiplet.

The trick is to perform a 5D gauge transformation that completely removes $A_5$ from the bulk action\cite{Falkowski:2006vi}:
\begin{equation}
	\Omega(z) = e^{i g_5 \int_R^z dz' A_5^a T^a}
\end{equation}
where $\Omega$ is just the Wilson line from $R$ to $z$. Notice that $\Omega(R) = 1$, so this is the identity transformation on the UV brane. This gauge transformation removes $A_5$ from the pure gauge action. Under the gauge transformation, the fermions pick up a phase
\begin{equation}
	\psi =  \Omega(z) \tilde{\psi}
\end{equation}
which also removes $A_5$ from the bulk fermion EOMs. Working in terms of the redefined fields $\tilde{\psi}$, the bulk fermion EOMs are simple and decoupled. However, the initial boundary conditions were given as conditions on $\psi$. We must now apply boundary conditions on the rotated fermions $ \Omega(z)  \tilde{\psi}$. Only the IR boundary conditions are affected since $\Omega(R) = 1$. Therefore $A_5$ still shows up in the IR boundary conditions for the fermions in the form of the Wilson line.

The 4D Coleman Weinberg potential generated by a KK tower takes the form 
\begin{equation}
	V = (-1)^F \frac{N}{2} \sum_n \int \frac{d^4p}{(2\pi)^4} \log \left[p^2+m_n^2(h) \right]
\end{equation}
where $n$ runs over the KK modes, $N$ is the number of DOFs at each level of the KK tower (3 for a gauge boson, 4 for a Dirac fermion) and $m_n(h)$'s are the Higgs-dependent masses. 

The easiest way to perform this sum is to find a function that encodes the KK spectrum as simple poles and to enclose the $Re(m^2) >0$ half of the complex $m^2$-plane with a contour integral. The integral picks out the residues of the poles and performs the sum for us at the cost of having to do an integral along the $m = i k$ axis. After the use of dimensional regularization to compute the $d^4p$ integral, the result can be massaged to the form
\begin{equation}
	(-1)^F \frac{N}{(4 \pi)^2}
	\int_0^\infty dk k^{3} \log[\rho(-k^2)]  
\end{equation}
where $\rho(z)$ is a spectral function which must be holomorphic for $Re(z) > 0$ and its zeros encode the KK spectrum by $\rho(m_n^2) = 0$. 

The spectral function is obtained by application of the boundary conditions to the fermion and gauge boson EOM solutions in order to obtain a quantization condition on $m_n(h)$. After applying the UV boundary conditions, there is a solution if and only if the coefficient matrix $\mathcal{M}$ of the IR boundary conditions $\mathcal{M}.\mathcal {A} = 0 $ is non-invertible ($\mathcal{A}$ is the vector of undetermined normalization coefficients). Thus $\rho$ is given by the $det(\mathcal{M})$. One can show that the spectral functions take the form
\begin{equation}
	\rho(-k^2) = 1 + F(-k^2) \sin^2\left(\frac{\lambda_R h}{ f }\right)
\end{equation}
where the form factor $F(-k^2)$ depends on the exact warping and $\lambda_r$ is a numerical factor that depends on the $SO(5)$ representation of the fields contributing to the $A_5$ potential. 

The form factors can be exactly calculated for the AdS$_5$ background. For large momenta, the form factors are warped down as $F \propto e^{-4 k / m_{KK}}$. This shows that $M_{KK} = g_* f$ acts as a momentum cutoff to the Higgs potential since contributions from energies above this scale will not affect the potential. The potential involves contributions from gauge bosons and from fermions
\begin{equation}
	V_{\text{eff}} = V_{\text{gauge}} + V_{\text{fermion}}.
\end{equation}
If we take the fermions to be embedded in the fundamental \textbf{5} of $SO(5)$, the result is~\cite{Falkowski:2006vi}
\begin{align}
	V_{\text{gauge}} &= \alpha \sin^2\left(\frac{h}{f}\right) \nonumber \\
	V_{\text{fermion}} &= \beta_1 \sin^2 \left(\frac{h}{f}\right) + \beta_2 \sin^4 \left(\frac{h}{f}\right)
\end{align}
where $\alpha$, $\beta_i$ are given by momentum integrals of quantities involving the form factors $F(p^2)$. Constraints on Higgs couplings require that the minimum of the potential $v = \langle h \rangle$ satisfies $f /v  \gtrsim 3-5$ since the angle $v/f$ controls how aligned the Goldstone mode is aligned with $SU(2)_L$ and leads to deviations in Higgs couplings if it is too large. This introduces a source of fine-tuning of order $v^2 / f^2$. 

\section{Conclusions\label{sec:Conclusions}}

 So far there is no direct evidence for BSM physics, making the lightness of the Higgs boson ever more puzzling. These lectures were reviewing one of the leading theoretical ideas for new physics that could solve the hierarchy problem around the few TeV scale: the idea that the Higgs is not actually an elementary particle, but rather a composite pNGB. We have outlined the main features of such pNGB's essential for CH model building and highlighted the mechanism of collective symmetry breaking, as the essential tool behind CH/Little Higgs models. We have sketched out how to construct the major versions of such models and also contrasted their properties. We have used the AdS/CFT correspondence to establish the connection between pNGB Higgs models and holographic CH models, and also explained in detail partial compositeness, the modern way of introducing fermion masses into model with strong dynamics and symmetry breaking. While the experimental bounds on the putative top and spin 1 partners are getting ever stronger, the amount of tuning needed for these models is still around a few percent, roughly what was initially implied from the LEP bounds almost 20 years ago.

\section*{Acknowledgements}

We thank Rouven Essig, Ian Low and Tom DeGrand for organizing TASI 2016 and arranging these lectures.    This work is supported in part by the NSF grant PHY-1719877.

\end{document}